\documentclass[longauth]{aa}

\usepackage{graphicx}
\usepackage{natbib}
\usepackage{booktabs}
\usepackage{multirow}
\usepackage{threeparttable}

\usepackage{txfonts}
\usepackage{hyperref}
\hypersetup{
    colorlinks=true,
    linkcolor=blue,
    filecolor=magenta,      
    urlcolor=blue,
    citecolor=blue
}

\makeatletter
\renewcommand*\aa@pageof{, page \thepage{} of \pageref*{LastPage}}
\makeatother

\usepackage[switch, modulo]{lineno}

\usepackage{orcidlink} 
\newcommand{\orcid}[1]{\orcidlink{#1}}

\usepackage{euclid}

\newcommand{\rms}{RMS~}

\newcommand{\ie}{i{.}e{.}~}
\newcommand{\eg}{e{.}g{.}~}
\newcommand{\eq}{Eq{.}~}                 

\newcommand{\tab}{Table~}  
\newcommand{\fg}{Fig{.}~}
\newcommand{\fgs}{Figs{.}~}
\newcommand{\sct}{Sect{.}~}
\newcommand{\scts}{Sects{.}~}

\begin{document}

\title{\Euclid: Galaxy morphology and photometry from bulge-disc decomposition of Early Release Observations\thanks{This paper is published on behalf of the Euclid Consortium}}

\author{L.~Quilley\orcid{0009-0008-8375-8605}\thanks{\email{louis.quilley@univ-lyon1.fr}}\inst{\ref{aff1}}
\and V.~de~Lapparent\orcid{0009-0007-1622-1974}\inst{\ref{aff2}}
\and M.~Bolzonella\orcid{0000-0003-3278-4607}\inst{\ref{aff3}}
\and M.~Baes\orcid{0000-0002-3930-2757}\inst{\ref{aff4}}
\and I.~Damjanov\orcid{0000-0003-4797-5246}\inst{\ref{aff5}}
\and B.~H\"au{\ss}ler\orcid{0000-0002-1857-2088}\inst{\ref{aff6}}
\and F.~R.~Marleau\orcid{0000-0002-1442-2947}\inst{\ref{aff7}}
\and A.~Nersesian\orcid{0000-0001-6843-409X}\inst{\ref{aff8},\ref{aff4}}
\and T.~Saifollahi\orcid{0000-0002-9554-7660}\inst{\ref{aff9}}
\and D.~Scott\orcid{0000-0002-6878-9840}\inst{\ref{aff10}}
\and J.~G.~Sorce\orcid{0000-0002-2307-2432}\inst{\ref{aff11},\ref{aff12},\ref{aff13}}
\and C.~Tortora\orcid{0000-0001-7958-6531}\inst{\ref{aff14}}
\and M.~Urbano\orcid{0000-0001-5640-0650}\inst{\ref{aff9}}
\and N.~Aghanim\orcid{0000-0002-6688-8992}\inst{\ref{aff12}}
\and B.~Altieri\orcid{0000-0003-3936-0284}\inst{\ref{aff15}}
\and A.~Amara\inst{\ref{aff16}}
\and S.~Andreon\orcid{0000-0002-2041-8784}\inst{\ref{aff17}}
\and N.~Auricchio\orcid{0000-0003-4444-8651}\inst{\ref{aff3}}
\and C.~Baccigalupi\orcid{0000-0002-8211-1630}\inst{\ref{aff18},\ref{aff19},\ref{aff20},\ref{aff21}}
\and M.~Baldi\orcid{0000-0003-4145-1943}\inst{\ref{aff22},\ref{aff3},\ref{aff23}}
\and A.~Balestra\orcid{0000-0002-6967-261X}\inst{\ref{aff24}}
\and S.~Bardelli\orcid{0000-0002-8900-0298}\inst{\ref{aff3}}
\and A.~Basset\inst{\ref{aff25}}
\and P.~Battaglia\orcid{0000-0002-7337-5909}\inst{\ref{aff3}}
\and A.~Biviano\orcid{0000-0002-0857-0732}\inst{\ref{aff19},\ref{aff18}}
\and A.~Bonchi\orcid{0000-0002-2667-5482}\inst{\ref{aff26}}
\and D.~Bonino\orcid{0000-0002-3336-9977}\inst{\ref{aff27}}
\and E.~Branchini\orcid{0000-0002-0808-6908}\inst{\ref{aff28},\ref{aff29},\ref{aff17}}
\and M.~Brescia\orcid{0000-0001-9506-5680}\inst{\ref{aff30},\ref{aff14},\ref{aff31}}
\and J.~Brinchmann\orcid{0000-0003-4359-8797}\inst{\ref{aff32},\ref{aff33}}
\and A.~Caillat\inst{\ref{aff34}}
\and S.~Camera\orcid{0000-0003-3399-3574}\inst{\ref{aff35},\ref{aff36},\ref{aff27}}
\and V.~Capobianco\orcid{0000-0002-3309-7692}\inst{\ref{aff27}}
\and C.~Carbone\orcid{0000-0003-0125-3563}\inst{\ref{aff37}}
\and J.~Carretero\orcid{0000-0002-3130-0204}\inst{\ref{aff38},\ref{aff39}}
\and S.~Casas\orcid{0000-0002-4751-5138}\inst{\ref{aff40},\ref{aff41}}
\and M.~Castellano\orcid{0000-0001-9875-8263}\inst{\ref{aff42}}
\and G.~Castignani\orcid{0000-0001-6831-0687}\inst{\ref{aff3}}
\and S.~Cavuoti\orcid{0000-0002-3787-4196}\inst{\ref{aff14},\ref{aff31}}
\and A.~Cimatti\inst{\ref{aff43}}
\and C.~Colodro-Conde\inst{\ref{aff44}}
\and G.~Congedo\orcid{0000-0003-2508-0046}\inst{\ref{aff45}}
\and C.~J.~Conselice\orcid{0000-0003-1949-7638}\inst{\ref{aff46}}
\and L.~Conversi\orcid{0000-0002-6710-8476}\inst{\ref{aff47},\ref{aff15}}
\and Y.~Copin\orcid{0000-0002-5317-7518}\inst{\ref{aff48}}
\and F.~Courbin\orcid{0000-0003-0758-6510}\inst{\ref{aff49},\ref{aff50}}
\and H.~M.~Courtois\orcid{0000-0003-0509-1776}\inst{\ref{aff51}}
\and M.~Cropper\orcid{0000-0003-4571-9468}\inst{\ref{aff52}}
\and J.-C.~Cuillandre\orcid{0000-0002-3263-8645}\inst{\ref{aff53}}
\and A.~Da~Silva\orcid{0000-0002-6385-1609}\inst{\ref{aff54},\ref{aff55}}
\and H.~Degaudenzi\orcid{0000-0002-5887-6799}\inst{\ref{aff56}}
\and G.~De~Lucia\orcid{0000-0002-6220-9104}\inst{\ref{aff19}}
\and A.~M.~Di~Giorgio\orcid{0000-0002-4767-2360}\inst{\ref{aff57}}
\and J.~Dinis\orcid{0000-0001-5075-1601}\inst{\ref{aff54},\ref{aff55}}
\and F.~Dubath\orcid{0000-0002-6533-2810}\inst{\ref{aff56}}
\and C.~A.~J.~Duncan\orcid{0009-0003-3573-0791}\inst{\ref{aff46}}
\and X.~Dupac\inst{\ref{aff15}}
\and S.~Dusini\orcid{0000-0002-1128-0664}\inst{\ref{aff58}}
\and A.~Ealet\orcid{0000-0003-3070-014X}\inst{\ref{aff48}}
\and M.~Farina\orcid{0000-0002-3089-7846}\inst{\ref{aff57}}
\and S.~Farrens\orcid{0000-0002-9594-9387}\inst{\ref{aff53}}
\and F.~Faustini\orcid{0000-0001-6274-5145}\inst{\ref{aff26},\ref{aff42}}
\and S.~Ferriol\inst{\ref{aff48}}
\and S.~Fotopoulou\orcid{0000-0002-9686-254X}\inst{\ref{aff59}}
\and M.~Frailis\orcid{0000-0002-7400-2135}\inst{\ref{aff19}}
\and E.~Franceschi\orcid{0000-0002-0585-6591}\inst{\ref{aff3}}
\and M.~Fumana\orcid{0000-0001-6787-5950}\inst{\ref{aff37}}
\and S.~Galeotta\orcid{0000-0002-3748-5115}\inst{\ref{aff19}}
\and K.~George\orcid{0000-0002-1734-8455}\inst{\ref{aff60}}
\and B.~Gillis\orcid{0000-0002-4478-1270}\inst{\ref{aff45}}
\and C.~Giocoli\orcid{0000-0002-9590-7961}\inst{\ref{aff3},\ref{aff23}}
\and P.~G\'omez-Alvarez\orcid{0000-0002-8594-5358}\inst{\ref{aff61},\ref{aff15}}
\and A.~Grazian\orcid{0000-0002-5688-0663}\inst{\ref{aff24}}
\and F.~Grupp\inst{\ref{aff62},\ref{aff60}}
\and S.~V.~H.~Haugan\orcid{0000-0001-9648-7260}\inst{\ref{aff63}}
\and J.~Hoar\inst{\ref{aff15}}
\and W.~Holmes\inst{\ref{aff64}}
\and F.~Hormuth\inst{\ref{aff65}}
\and A.~Hornstrup\orcid{0000-0002-3363-0936}\inst{\ref{aff66},\ref{aff67}}
\and P.~Hudelot\inst{\ref{aff2}}
\and K.~Jahnke\orcid{0000-0003-3804-2137}\inst{\ref{aff68}}
\and M.~Jhabvala\inst{\ref{aff69}}
\and E.~Keih\"anen\orcid{0000-0003-1804-7715}\inst{\ref{aff70}}
\and S.~Kermiche\orcid{0000-0002-0302-5735}\inst{\ref{aff71}}
\and A.~Kiessling\orcid{0000-0002-2590-1273}\inst{\ref{aff64}}
\and M.~Kilbinger\orcid{0000-0001-9513-7138}\inst{\ref{aff53}}
\and B.~Kubik\orcid{0009-0006-5823-4880}\inst{\ref{aff48}}
\and K.~Kuijken\orcid{0000-0002-3827-0175}\inst{\ref{aff72}}
\and M.~K\"ummel\orcid{0000-0003-2791-2117}\inst{\ref{aff60}}
\and M.~Kunz\orcid{0000-0002-3052-7394}\inst{\ref{aff73}}
\and H.~Kurki-Suonio\orcid{0000-0002-4618-3063}\inst{\ref{aff74},\ref{aff75}}
\and R.~Laureijs\inst{\ref{aff76},\ref{aff77}}
\and D.~Le~Mignant\orcid{0000-0002-5339-5515}\inst{\ref{aff34}}
\and S.~Ligori\orcid{0000-0003-4172-4606}\inst{\ref{aff27}}
\and P.~B.~Lilje\orcid{0000-0003-4324-7794}\inst{\ref{aff63}}
\and V.~Lindholm\orcid{0000-0003-2317-5471}\inst{\ref{aff74},\ref{aff75}}
\and I.~Lloro\orcid{0000-0001-5966-1434}\inst{\ref{aff78}}
\and G.~Mainetti\orcid{0000-0003-2384-2377}\inst{\ref{aff79}}
\and D.~Maino\inst{\ref{aff80},\ref{aff37},\ref{aff81}}
\and E.~Maiorano\orcid{0000-0003-2593-4355}\inst{\ref{aff3}}
\and O.~Mansutti\orcid{0000-0001-5758-4658}\inst{\ref{aff19}}
\and O.~Marggraf\orcid{0000-0001-7242-3852}\inst{\ref{aff82}}
\and K.~Markovic\orcid{0000-0001-6764-073X}\inst{\ref{aff64}}
\and M.~Martinelli\orcid{0000-0002-6943-7732}\inst{\ref{aff42},\ref{aff83}}
\and N.~Martinet\orcid{0000-0003-2786-7790}\inst{\ref{aff34}}
\and F.~Marulli\orcid{0000-0002-8850-0303}\inst{\ref{aff84},\ref{aff3},\ref{aff23}}
\and R.~Massey\orcid{0000-0002-6085-3780}\inst{\ref{aff85}}
\and E.~Medinaceli\orcid{0000-0002-4040-7783}\inst{\ref{aff3}}
\and S.~Mei\orcid{0000-0002-2849-559X}\inst{\ref{aff86}}
\and M.~Melchior\inst{\ref{aff87}}
\and Y.~Mellier\thanks{Deceased}\inst{\ref{aff88},\ref{aff2}}
\and M.~Meneghetti\orcid{0000-0003-1225-7084}\inst{\ref{aff3},\ref{aff23}}
\and E.~Merlin\orcid{0000-0001-6870-8900}\inst{\ref{aff42}}
\and G.~Meylan\inst{\ref{aff89}}
\and A.~Mora\orcid{0000-0002-1922-8529}\inst{\ref{aff90}}
\and M.~Moresco\orcid{0000-0002-7616-7136}\inst{\ref{aff84},\ref{aff3}}
\and L.~Moscardini\orcid{0000-0002-3473-6716}\inst{\ref{aff84},\ref{aff3},\ref{aff23}}
\and R.~Nakajima\orcid{0009-0009-1213-7040}\inst{\ref{aff82}}
\and C.~Neissner\orcid{0000-0001-8524-4968}\inst{\ref{aff91},\ref{aff39}}
\and R.~C.~Nichol\orcid{0000-0003-0939-6518}\inst{\ref{aff16}}
\and S.-M.~Niemi\inst{\ref{aff76}}
\and C.~Padilla\orcid{0000-0001-7951-0166}\inst{\ref{aff91}}
\and S.~Paltani\orcid{0000-0002-8108-9179}\inst{\ref{aff56}}
\and F.~Pasian\orcid{0000-0002-4869-3227}\inst{\ref{aff19}}
\and K.~Pedersen\inst{\ref{aff92}}
\and W.~J.~Percival\orcid{0000-0002-0644-5727}\inst{\ref{aff93},\ref{aff94},\ref{aff95}}
\and V.~Pettorino\inst{\ref{aff76}}
\and S.~Pires\orcid{0000-0002-0249-2104}\inst{\ref{aff53}}
\and G.~Polenta\orcid{0000-0003-4067-9196}\inst{\ref{aff26}}
\and M.~Poncet\inst{\ref{aff25}}
\and L.~A.~Popa\inst{\ref{aff96}}
\and L.~Pozzetti\orcid{0000-0001-7085-0412}\inst{\ref{aff3}}
\and F.~Raison\orcid{0000-0002-7819-6918}\inst{\ref{aff62}}
\and R.~Rebolo\inst{\ref{aff44},\ref{aff97},\ref{aff98}}
\and A.~Renzi\orcid{0000-0001-9856-1970}\inst{\ref{aff99},\ref{aff58}}
\and J.~Rhodes\orcid{0000-0002-4485-8549}\inst{\ref{aff64}}
\and G.~Riccio\inst{\ref{aff14}}
\and E.~Romelli\orcid{0000-0003-3069-9222}\inst{\ref{aff19}}
\and M.~Roncarelli\orcid{0000-0001-9587-7822}\inst{\ref{aff3}}
\and E.~Rossetti\orcid{0000-0003-0238-4047}\inst{\ref{aff22}}
\and R.~Saglia\orcid{0000-0003-0378-7032}\inst{\ref{aff60},\ref{aff62}}
\and Z.~Sakr\orcid{0000-0002-4823-3757}\inst{\ref{aff100},\ref{aff101},\ref{aff102}}
\and D.~Sapone\orcid{0000-0001-7089-4503}\inst{\ref{aff103}}
\and B.~Sartoris\orcid{0000-0003-1337-5269}\inst{\ref{aff60},\ref{aff19}}
\and M.~Schirmer\orcid{0000-0003-2568-9994}\inst{\ref{aff68}}
\and P.~Schneider\orcid{0000-0001-8561-2679}\inst{\ref{aff82}}
\and T.~Schrabback\orcid{0000-0002-6987-7834}\inst{\ref{aff7}}
\and A.~Secroun\orcid{0000-0003-0505-3710}\inst{\ref{aff71}}
\and E.~Sefusatti\orcid{0000-0003-0473-1567}\inst{\ref{aff19},\ref{aff18},\ref{aff20}}
\and G.~Seidel\orcid{0000-0003-2907-353X}\inst{\ref{aff68}}
\and S.~Serrano\orcid{0000-0002-0211-2861}\inst{\ref{aff104},\ref{aff105},\ref{aff106}}
\and C.~Sirignano\orcid{0000-0002-0995-7146}\inst{\ref{aff99},\ref{aff58}}
\and G.~Sirri\orcid{0000-0003-2626-2853}\inst{\ref{aff23}}
\and L.~Stanco\orcid{0000-0002-9706-5104}\inst{\ref{aff58}}
\and J.~Steinwagner\orcid{0000-0001-7443-1047}\inst{\ref{aff62}}
\and P.~Tallada-Cresp\'{i}\orcid{0000-0002-1336-8328}\inst{\ref{aff38},\ref{aff39}}
\and A.~N.~Taylor\inst{\ref{aff45}}
\and I.~Tereno\inst{\ref{aff54},\ref{aff107}}
\and R.~Toledo-Moreo\orcid{0000-0002-2997-4859}\inst{\ref{aff108}}
\and F.~Torradeflot\orcid{0000-0003-1160-1517}\inst{\ref{aff39},\ref{aff38}}
\and I.~Tutusaus\orcid{0000-0002-3199-0399}\inst{\ref{aff101}}
\and L.~Valenziano\orcid{0000-0002-1170-0104}\inst{\ref{aff3},\ref{aff109}}
\and T.~Vassallo\orcid{0000-0001-6512-6358}\inst{\ref{aff60},\ref{aff19}}
\and G.~Verdoes~Kleijn\orcid{0000-0001-5803-2580}\inst{\ref{aff77}}
\and A.~Veropalumbo\orcid{0000-0003-2387-1194}\inst{\ref{aff17},\ref{aff29},\ref{aff28}}
\and Y.~Wang\orcid{0000-0002-4749-2984}\inst{\ref{aff110}}
\and J.~Weller\orcid{0000-0002-8282-2010}\inst{\ref{aff60},\ref{aff62}}
\and G.~Zamorani\orcid{0000-0002-2318-301X}\inst{\ref{aff3}}
\and E.~Zucca\orcid{0000-0002-5845-8132}\inst{\ref{aff3}}
\and C.~Burigana\orcid{0000-0002-3005-5796}\inst{\ref{aff111},\ref{aff109}}
\and V.~Scottez\inst{\ref{aff88},\ref{aff112}}}
                                                                                   
\institute{Centre de Recherche Astrophysique de Lyon, UMR5574, CNRS, Universit\'e Claude Bernard Lyon 1, ENS de Lyon, 69230, Saint-Genis-Laval, France\label{aff1}
\and
Institut d'Astrophysique de Paris, UMR 7095, CNRS, and Sorbonne Universit\'e, 98 bis boulevard Arago, 75014 Paris, France\label{aff2}
\and
INAF-Osservatorio di Astrofisica e Scienza dello Spazio di Bologna, Via Piero Gobetti 93/3, 40129 Bologna, Italy\label{aff3}
\and
Sterrenkundig Observatorium, Universiteit Gent, Krijgslaan 281 S9, 9000 Gent, Belgium\label{aff4}
\and
Department of Astronomy \& Physics and Institute for Computational Astrophysics, Saint Mary's University, 923 Robie Street, Halifax, Nova Scotia, B3H 3C3, Canada\label{aff5}
\and
European Southern Observatory, Alonso de C\'ordova 3107, Vitacura, Santiago de Chile, Chile\label{aff6}
\and
Universit\"at Innsbruck, Institut f\"ur Astro- und Teilchenphysik, Technikerstr. 25/8, 6020 Innsbruck, Austria\label{aff7}
\and
STAR Institute, University of Li{\`e}ge, Quartier Agora, All\'ee du six Ao\^ut 19c, 4000 Li\`ege, Belgium\label{aff8}
\and
Universit\'e de Strasbourg, CNRS, Observatoire astronomique de Strasbourg, UMR 7550, 67000 Strasbourg, France\label{aff9}
\and
Department of Physics and Astronomy, University of British Columbia, Vancouver, BC V6T 1Z1, Canada\label{aff10}
\and
Univ. Lille, CNRS, Centrale Lille, UMR 9189 CRIStAL, 59000 Lille, France\label{aff11}
\and
Universit\'e Paris-Saclay, CNRS, Institut d'astrophysique spatiale, 91405, Orsay, France\label{aff12}
\and
Leibniz-Institut f\"{u}r Astrophysik (AIP), An der Sternwarte 16, 14482 Potsdam, Germany\label{aff13}
\and
INAF-Osservatorio Astronomico di Capodimonte, Via Moiariello 16, 80131 Napoli, Italy\label{aff14}
\and
ESAC/ESA, Camino Bajo del Castillo, s/n., Urb. Villafranca del Castillo, 28692 Villanueva de la Ca\~nada, Madrid, Spain\label{aff15}
\and
School of Mathematics and Physics, University of Surrey, Guildford, Surrey, GU2 7XH, UK\label{aff16}
\and
INAF-Osservatorio Astronomico di Brera, Via Brera 28, 20122 Milano, Italy\label{aff17}
\and
IFPU, Institute for Fundamental Physics of the Universe, via Beirut 2, 34151 Trieste, Italy\label{aff18}
\and
INAF-Osservatorio Astronomico di Trieste, Via G. B. Tiepolo 11, 34143 Trieste, Italy\label{aff19}
\and
INFN, Sezione di Trieste, Via Valerio 2, 34127 Trieste TS, Italy\label{aff20}
\and
SISSA, International School for Advanced Studies, Via Bonomea 265, 34136 Trieste TS, Italy\label{aff21}
\and
Dipartimento di Fisica e Astronomia, Universit\`a di Bologna, Via Gobetti 93/2, 40129 Bologna, Italy\label{aff22}
\and
INFN-Sezione di Bologna, Viale Berti Pichat 6/2, 40127 Bologna, Italy\label{aff23}
\and
INAF-Osservatorio Astronomico di Padova, Via dell'Osservatorio 5, 35122 Padova, Italy\label{aff24}
\and
Centre National d'Etudes Spatiales -- Centre spatial de Toulouse, 18 avenue Edouard Belin, 31401 Toulouse Cedex 9, France\label{aff25}
\and
Space Science Data Center, Italian Space Agency, via del Politecnico snc, 00133 Roma, Italy\label{aff26}
\and
INAF-Osservatorio Astrofisico di Torino, Via Osservatorio 20, 10025 Pino Torinese (TO), Italy\label{aff27}
\and
Dipartimento di Fisica, Universit\`a di Genova, Via Dodecaneso 33, 16146, Genova, Italy\label{aff28}
\and
INFN-Sezione di Genova, Via Dodecaneso 33, 16146, Genova, Italy\label{aff29}
\and
Department of Physics "E. Pancini", University Federico II, Via Cinthia 6, 80126, Napoli, Italy\label{aff30}
\and
INFN section of Naples, Via Cinthia 6, 80126, Napoli, Italy\label{aff31}
\and
Instituto de Astrof\'isica e Ci\^encias do Espa\c{c}o, Universidade do Porto, CAUP, Rua das Estrelas, PT4150-762 Porto, Portugal\label{aff32}
\and
Faculdade de Ci\^encias da Universidade do Porto, Rua do Campo de Alegre, 4150-007 Porto, Portugal\label{aff33}
\and
Aix-Marseille Universit\'e, CNRS, CNES, LAM, Marseille, France\label{aff34}
\and
Dipartimento di Fisica, Universit\`a degli Studi di Torino, Via P. Giuria 1, 10125 Torino, Italy\label{aff35}
\and
INFN-Sezione di Torino, Via P. Giuria 1, 10125 Torino, Italy\label{aff36}
\and
INAF-IASF Milano, Via Alfonso Corti 12, 20133 Milano, Italy\label{aff37}
\and
Centro de Investigaciones Energ\'eticas, Medioambientales y Tecnol\'ogicas (CIEMAT), Avenida Complutense 40, 28040 Madrid, Spain\label{aff38}
\and
Port d'Informaci\'{o} Cient\'{i}fica, Campus UAB, C. Albareda s/n, 08193 Bellaterra (Barcelona), Spain\label{aff39}
\and
Institute for Theoretical Particle Physics and Cosmology (TTK), RWTH Aachen University, 52056 Aachen, Germany\label{aff40}
\and
Institute of Cosmology and Gravitation, University of Portsmouth, Portsmouth PO1 3FX, UK\label{aff41}
\and
INAF-Osservatorio Astronomico di Roma, Via Frascati 33, 00078 Monteporzio Catone, Italy\label{aff42}
\and
Dipartimento di Fisica e Astronomia "Augusto Righi" - Alma Mater Studiorum Universit\`a di Bologna, Viale Berti Pichat 6/2, 40127 Bologna, Italy\label{aff43}
\and
Instituto de Astrof\'{\i}sica de Canarias, E-38205 La Laguna, Tenerife, Spain\label{aff44}
\and
Institute for Astronomy, University of Edinburgh, Royal Observatory, Blackford Hill, Edinburgh EH9 3HJ, UK\label{aff45}
\and
Jodrell Bank Centre for Astrophysics, Department of Physics and Astronomy, University of Manchester, Oxford Road, Manchester M13 9PL, UK\label{aff46}
\and
European Space Agency/ESRIN, Largo Galileo Galilei 1, 00044 Frascati, Roma, Italy\label{aff47}
\and
Universit\'e Claude Bernard Lyon 1, CNRS/IN2P3, IP2I Lyon, UMR 5822, Villeurbanne, F-69100, France\label{aff48}
\and
Institut de Ci\`{e}ncies del Cosmos (ICCUB), Universitat de Barcelona (IEEC-UB), Mart\'{i} i Franqu\`{e}s 1, 08028 Barcelona, Spain\label{aff49}
\and
Instituci\'o Catalana de Recerca i Estudis Avan\c{c}ats (ICREA), Passeig de Llu\'{\i}s Companys 23, 08010 Barcelona, Spain\label{aff50}
\and
UCB Lyon 1, CNRS/IN2P3, IUF, IP2I Lyon, 4 rue Enrico Fermi, 69622 Villeurbanne, France\label{aff51}
\and
Mullard Space Science Laboratory, University College London, Holmbury St Mary, Dorking, Surrey RH5 6NT, UK\label{aff52}
\and
Universit\'e Paris-Saclay, Universit\'e Paris Cit\'e, CEA, CNRS, AIM, 91191, Gif-sur-Yvette, France\label{aff53}
\and
Departamento de F\'isica, Faculdade de Ci\^encias, Universidade de Lisboa, Edif\'icio C8, Campo Grande, PT1749-016 Lisboa, Portugal\label{aff54}
\and
Instituto de Astrof\'isica e Ci\^encias do Espa\c{c}o, Faculdade de Ci\^encias, Universidade de Lisboa, Campo Grande, 1749-016 Lisboa, Portugal\label{aff55}
\and
Department of Astronomy, University of Geneva, ch. d'Ecogia 16, 1290 Versoix, Switzerland\label{aff56}
\and
INAF-Istituto di Astrofisica e Planetologia Spaziali, via del Fosso del Cavaliere, 100, 00100 Roma, Italy\label{aff57}
\and
INFN-Padova, Via Marzolo 8, 35131 Padova, Italy\label{aff58}
\and
School of Physics, HH Wills Physics Laboratory, University of Bristol, Tyndall Avenue, Bristol, BS8 1TL, UK\label{aff59}
\and
Universit\"ats-Sternwarte M\"unchen, Fakult\"at f\"ur Physik, Ludwig-Maximilians-Universit\"at M\"unchen, Scheinerstr.~1, 81679 M\"unchen, Germany\label{aff60}
\and
FRACTAL S.L.N.E., calle Tulip\'an 2, Portal 13 1A, 28231, Las Rozas de Madrid, Spain\label{aff61}
\and
Max Planck Institute for Extraterrestrial Physics, Giessenbachstr. 1, 85748 Garching, Germany\label{aff62}
\and
Institute of Theoretical Astrophysics, University of Oslo, P.O. Box 1029 Blindern, 0315 Oslo, Norway\label{aff63}
\and
Jet Propulsion Laboratory, California Institute of Technology, 4800 Oak Grove Drive, Pasadena, CA, 91109, USA\label{aff64}
\and
Felix Hormuth Engineering, Goethestr. 17, 69181 Leimen, Germany\label{aff65}
\and
Technical University of Denmark, Elektrovej 327, 2800 Kgs. Lyngby, Denmark\label{aff66}
\and
Cosmic Dawn Center (DAWN), Denmark\label{aff67}
\and
Max-Planck-Institut f\"ur Astronomie, K\"onigstuhl 17, 69117 Heidelberg, Germany\label{aff68}
\and
NASA Goddard Space Flight Center, Greenbelt, MD 20771, USA\label{aff69}
\and
Department of Physics and Helsinki Institute of Physics, Gustaf H\"allstr\"omin katu 2, University of Helsinki, 00014 Helsinki, Finland\label{aff70}
\and
Aix-Marseille Universit\'e, CNRS/IN2P3, CPPM, Marseille, France\label{aff71}
\and
Leiden Observatory, Leiden University, Einsteinweg 55, 2333 CC Leiden, The Netherlands\label{aff72}
\and
Universit\'e de Gen\`eve, D\'epartement de Physique Th\'eorique and Centre for Astroparticle Physics, 24 quai Ernest-Ansermet, CH-1211 Gen\`eve 4, Switzerland\label{aff73}
\and
Department of Physics, P.O. Box 64, University of Helsinki, 00014 Helsinki, Finland\label{aff74}
\and
Helsinki Institute of Physics, Gustaf H{\"a}llstr{\"o}min katu 2, University of Helsinki, 00014 Helsinki, Finland\label{aff75}
\and
European Space Agency/ESTEC, Keplerlaan 1, 2201 AZ Noordwijk, The Netherlands\label{aff76}
\and
Kapteyn Astronomical Institute, University of Groningen, PO Box 800, 9700 AV Groningen, The Netherlands\label{aff77}
\and
NOVA optical infrared instrumentation group at ASTRON, Oude Hoogeveensedijk 4, 7991PD, Dwingeloo, The Netherlands\label{aff78}
\and
Centre de Calcul de l'IN2P3/CNRS, 21 avenue Pierre de Coubertin 69627 Villeurbanne Cedex, France\label{aff79}
\and
Dipartimento di Fisica "Aldo Pontremoli", Universit\`a degli Studi di Milano, Via Celoria 16, 20133 Milano, Italy\label{aff80}
\and
INFN-Sezione di Milano, Via Celoria 16, 20133 Milano, Italy\label{aff81}
\and
Universit\"at Bonn, Argelander-Institut f\"ur Astronomie, Auf dem H\"ugel 71, 53121 Bonn, Germany\label{aff82}
\and
INFN-Sezione di Roma, Piazzale Aldo Moro, 2 - c/o Dipartimento di Fisica, Edificio G. Marconi, 00185 Roma, Italy\label{aff83}
\and
Dipartimento di Fisica e Astronomia "Augusto Righi" - Alma Mater Studiorum Universit\`a di Bologna, via Piero Gobetti 93/2, 40129 Bologna, Italy\label{aff84}
\and
Department of Physics, Institute for Computational Cosmology, Durham University, South Road, Durham, DH1 3LE, UK\label{aff85}
\and
Universit\'e Paris Cit\'e, CNRS, Astroparticule et Cosmologie, 75013 Paris, France\label{aff86}
\and
University of Applied Sciences and Arts of Northwestern Switzerland, School of Engineering, 5210 Windisch, Switzerland\label{aff87}
\and
Institut d'Astrophysique de Paris, 98bis Boulevard Arago, 75014, Paris, France\label{aff88}
\and
Institute of Physics, Laboratory of Astrophysics, Ecole Polytechnique F\'ed\'erale de Lausanne (EPFL), Observatoire de Sauverny, 1290 Versoix, Switzerland\label{aff89}
\and
Aurora Technology for European Space Agency (ESA), Camino bajo del Castillo, s/n, Urbanizacion Villafranca del Castillo, Villanueva de la Ca\~nada, 28692 Madrid, Spain\label{aff90}
\and
Institut de F\'{i}sica d'Altes Energies (IFAE), The Barcelona Institute of Science and Technology, Campus UAB, 08193 Bellaterra (Barcelona), Spain\label{aff91}
\and
DARK, Niels Bohr Institute, University of Copenhagen, Jagtvej 155, 2200 Copenhagen, Denmark\label{aff92}
\and
Waterloo Centre for Astrophysics, University of Waterloo, Waterloo, Ontario N2L 3G1, Canada\label{aff93}
\and
Department of Physics and Astronomy, University of Waterloo, Waterloo, Ontario N2L 3G1, Canada\label{aff94}
\and
Perimeter Institute for Theoretical Physics, Waterloo, Ontario N2L 2Y5, Canada\label{aff95}
\and
Institute of Space Science, Str. Atomistilor, nr. 409 M\u{a}gurele, Ilfov, 077125, Romania\label{aff96}
\and
Consejo Superior de Investigaciones Cientificas, Calle Serrano 117, 28006 Madrid, Spain\label{aff97}
\and
Universidad de La Laguna, Dpto. Astrof\'\i sica, E-38206 La Laguna, Tenerife, Spain\label{aff98}
\and
Dipartimento di Fisica e Astronomia "G. Galilei", Universit\`a di Padova, Via Marzolo 8, 35131 Padova, Italy\label{aff99}
\and
Institut f\"ur Theoretische Physik, University of Heidelberg, Philosophenweg 16, 69120 Heidelberg, Germany\label{aff100}
\and
Institut de Recherche en Astrophysique et Plan\'etologie (IRAP), Universit\'e de Toulouse, CNRS, UPS, CNES, 14 Av. Edouard Belin, 31400 Toulouse, France\label{aff101}
\and
Universit\'e St Joseph; Faculty of Sciences, Beirut, Lebanon\label{aff102}
\and
Departamento de F\'isica, FCFM, Universidad de Chile, Blanco Encalada 2008, Santiago, Chile\label{aff103}
\and
Institut d'Estudis Espacials de Catalunya (IEEC),  Edifici RDIT, Campus UPC, 08860 Castelldefels, Barcelona, Spain\label{aff104}
\and
Satlantis, University Science Park, Sede Bld 48940, Leioa-Bilbao, Spain\label{aff105}
\and
Institute of Space Sciences (ICE, CSIC), Campus UAB, Carrer de Can Magrans, s/n, 08193 Barcelona, Spain\label{aff106}
\and
Instituto de Astrof\'isica e Ci\^encias do Espa\c{c}o, Faculdade de Ci\^encias, Universidade de Lisboa, Tapada da Ajuda, 1349-018 Lisboa, Portugal\label{aff107}
\and
Universidad Polit\'ecnica de Cartagena, Departamento de Electr\'onica y Tecnolog\'ia de Computadoras,  Plaza del Hospital 1, 30202 Cartagena, Spain\label{aff108}
\and
INFN-Bologna, Via Irnerio 46, 40126 Bologna, Italy\label{aff109}
\and
Infrared Processing and Analysis Center, California Institute of Technology, Pasadena, CA 91125, USA\label{aff110}
\and
INAF, Istituto di Radioastronomia, Via Piero Gobetti 101, 40129 Bologna, Italy\label{aff111}
\and
ICL, Junia, Universit\'e Catholique de Lille, LITL, 59000 Lille, France\label{aff112}}      

\titlerunning{\Euclid: Galaxy morphology and photometry from bulge-disc decomposition of EROs}
 
\authorrunning{Quilley et al.}
  
\date{Received 21 February 2025 / Accepted 19 February 2026}
\abstract
{The background galaxies in \Euclid Early Release Observation images of the Perseus cluster make up a remarkable sample for the combination of a $0.57$ deg$^2$ area, 25.3 and 23.2 AB mag depth, and angular resolutions in the optical and near-infrared bands of \ang{;;0.1} and \ang{;;0.3}, respectively.
As part of the effort towards characterising the history of the Hubble sequence, we performed a morphological analysis of $2445$ and $12\,786$ galaxies with $\IE\le21$ and $\IE\le23$, respectively.
We used single-S\'ersic profiles and the sums of a S\'ersic bulge and an exponential disc to model these galaxies with \texttt{SourceXtractor++} and analysed their positional, structural, and flux parameters in order to assess their similarities and differences.
The fitted galaxies to $\IE\le21$ span the various Hubble types with ubiquitous bulge and disc components and a bulge-to-total light ratio ($B/T$) that takes all values from 0 to 1. The effective radius of the single-S\'ersic profile is an intermediate estimate of galaxy size (between the bulge and disc effective radii) depending on $B/T$. The axis ratio of the single-S\'ersic profile is higher than the disc axis ratio, and this difference increases with $B/T$. The type of model impacts the photometry with $-0.08$ to 0.01 mag median systematic \IE offsets between single-S\'ersic and bulge-disc total magnitudes and a 0.05 to 0.15 mag dispersion from low to high $B/T$. We measured a median $0.3$ mag bulge-disc colour difference in rest-frame $M_g - M_i$ that originates from the disc-dominated galaxies, whereas bulge-dominated galaxies have median colours similar to those of their components. Remarkably, we also measured redder inside disc colour gradients based on 5 to 10\% systematic variations of disc effective radii between the optical and near-infrared bands.
This analysis demonstrates the usefulness and limitations of single-S\'ersic profile modelling and the power of bulge-disc decomposition for characterising the morphology of lenticulars and spirals in \Euclid images. We make available the catalogues of best-fit parameters for the morphological and SED fits.}

\keywords{Galaxies: evolution -- Galaxies: bulges  -- Galaxies: elliptical and lenticular, cD -- Galaxies: spiral -- Galaxies: structure}

\maketitle
\nolinenumbers

\section{Introduction \label{sct-intro}}

Observations from the \HST (HST) had suggested a large excess of irregularly shaped galaxies over the expected number counts of elliptical and spiral galaxies at apparent magnitudes $I(F814W)\le$ 20--22 in the \textit{Hubble} Deep Field \citep{Abraham-1996-galaxy-morphology-Hubble-Deep-Field}, which were interpreted as the `building blocks' of larger and present-day galaxies. Subsequent studies showed that ellipticals and disc galaxies were also frequent at redshift $z\approx1$, indicating that the Hubble sequence of morphological types was already in place at these redshifts \citep{Brinchmann-1998-HST-CFRS-LDSS,VandenBergh-2000-galaxy-morphology-HDF,Conselice-2005-field-evolution-morphology,Oesch-2010-buildup-Hubble-sequence-Cosmos,Buitrago-2013-early-types,Mortlock-2013-Hubble-sequence-CANDELS-UDS}. Thanks to its improved sensitivity and angular resolution as well as redder wavelengths than HST, the \textit{James Webb} Space Telescope (JWST) has recently revealed that a significant fraction of these HST irregular galaxies were actually among the many flocculent discs with small bulges seen at redshifts 1 to 3 (\citealp{Ferreira-2022-disks-JWST,Ferreira-2023-hubble-seq-JWST};  see also \citealp{Wang-2025-giant-disk-redshift-3}). Together with the detected spheroids (some of which might be face-on large bulge-to-disc ratio galaxies), the various galaxies detected by JWST appear to describe the full Hubble sequence seen in the nearby Universe \citep{Hubble-1926-extragalactic-nebulae}. 

The morphologies of galaxies are a key to unravelling their evolution since the star-formation history is markedly different for different Hubble types as well as within their major morphological components, the bulge, and the disc \citep{Martig-2009-morphological-quenching, Bait-2017-interdep-morph-SF-environment, Eales-2017-galaxy-end-sequence-colors-morphology-continuity,Morselli-2017-bulge-disk-structure-SF-activity, Bremer-2018-GAMA-survey-morph-transf-GV, Sampaio-2022-morph-prior-quenching, Quilley-2022-bimodality}. Through study of the bulge and disc decomposition of thousands of galaxies, \cite{Allen-2006-MGC-BD-decomp} showed that the galaxy colour bimodality is due to a dichotomy in the structure of their bulges and discs, and \cite{Gadotti-2009-bulge-structure-SDSS} explored the scaling relations of pseudo-bulges, classical bulges, and ellipticals whose structural differences point to different formation processes. Subsequent analyses have studied in detail the properties of bulges and discs in the nearby Universe \citep{Simard-2011-BD-decomp-SDSS, Meert-2015-SDSS-BD-catalog, Lange-2016-bulge-disk-decomposition-GAMA, Kim-2016-bulges-SDSS, Casura-2022-bulge-disk-decomposition-GAMA, Robotham-2022-ProFuse, Rigamonti-2024-kinematic-bulge-disk} and at larger redshifts in order to trace their evolution with over a large fraction of the age of the Universe \citep{Bruce-2014-bulge-disk-size-mass-relations-CANDELS, Margalef-Bentabol-2016-bulge-disk-decomposition-CANDELS-z-3, Dimauro-2018-bulge-disk-decomposition-CANDELS, Hashemizadeh-2022-DEVILS-bulge-emergence-since-z-1, Nedkova-2024-bulge-disc-decomposition-UVJ-diagrams-size-mass-relations}.

Reliable bulge-disc decompositions are computationally expensive \citep{Gao-2017-optimal-bulge-disk}, and if improperly constrained, they may be degenerate and biased \citep{Quilley-2023-PhD-thesis}. Therefore, the modelling of large samples of galaxies has predominantly been based on a single S\'ersic component to obtain estimates of galaxy sizes and shapes \citep[e{.}g{.}][among many others]{Trujillo-2004-size-luminosity-mass-z-3, Kelvin-2012-GAMA-sersic-fits, Morishita-2014-size-morpho-GOODS-HST, Mowla-2019-size-mass-CANDELS, Kartatelpe-2023-CEERS-gal-morpho-JWST, Lee-2024-morpho-JWST-fields}. Through bulge and disc decomposition of the Extraction de Formes Id\'ealis\'ees de Galaxies en Imagerie (EFIGI) statistical sample of $4458$ nearby well-resolved galaxies \citep{Baillard-2011-EFIGI}, which densely samples the full Hubble sequence in the local Universe, \citet{Quilley-2022-bimodality} took another step in providing evidence that the morphological sequence is an inverse evolutionary sequence that can be parametrised in terms of shape, size, and star-formation stage by a bulge and a disc. In this scenario, the ageing of galaxies is marked by bulge growth and disc quenching \citep{Huertas-Company-2016-mass-assembly-morpho-transformations-z-3-CANDELS, Quilley-2022-bimodality, Martorano-2024-size-mass-JWST}, with specific size-luminosity relations for both components \citep{Bruce-2014-bulge-disk-size-mass-relations-CANDELS, Quilley-2023-scaling-bulges-and-disks, Nedkova-2024-bulge-disc-decomposition-UVJ-diagrams-size-mass-relations}. Direct observations of the varying morphologies of galaxies with look-back time are necessary to confirm this scenario.

With its half-square-degree field of view, the \Euclid space telescope \citep{EuclidSkyOverview} offers an unprecedented opportunity to observe millions of galaxies with sufficient angular resolution to study their morphology from cosmic noon ($z\sim2$, \citealt{Madau-Dickinson-2014-cosmic-SFH}) to the present epoch \citep{Bretonniere-EP13}. The \Euclid morphology challenge (EMC hereafter) has evaluated the ability of five model fitting codes to obtain reliable photometry \citep{Merlin-EP25} and structural parameters \citep{Bretonniere-EP26} of galaxy samples extracted from synthetic images in the \IE band\footnote{See \url{https://www.euclid-ec.org/science/overview/} for a summary of Euclid instruments, bands and surveys.} from the VISible instrument \citep[VIS,][]{EuclidSkyVIS} with the characteristics of the Euclid Wide Survey (EWS; \citealp{Scaramella-EP1}). The codes tested were \texttt{DeepLeGATo} \citep{Tuccillo-2018-DeepLeGATo}, \texttt{Galapagos-2} \citep{Haussler-2022-separating-spheroid-disk}, \texttt{Morfometryka} \citep{Ferrari-2015-Morfometryka}, \texttt{ProFit} \citep{Robotham-2017-ProFit}, and \texttt{SourceXtractor++} \citep{Bertin-2020-SourceXtractor-plus-plus, Kummel-2020-SourceXtractor-plus-plus}. Some of the conclusions of the EMC were that \texttt{SourceXtractor++} was one of the best codes to perform such tasks (hence our decision to use it in the current study) and that the input structural parameters of synthetic galaxies could be reliably recovered (\ie the difference between the true and measured structural parameters was compatible with null bias, with a dispersion lower than $10\%$) down to $\IE\le23$ for single-S\'ersic profiles and down to $\IE\le21$ for two-component profiles (bulge-disc decomposition). The EMC studies prepared the implementation of single-S\'ersic fits in the \Euclid pipeline \citealp[see]{Q1-TP004, Q1-SP040}.

With the \IE median depth of $25.3$ mag at a signal-to-noise ratio of ten for galaxies \citep{EROData}, the \Euclid images of the Perseus cluster of galaxies obtained within the \citet[ERO]{EROcite} represent a unique opportunity to perform a study of the morphology of distant galaxies in the background of the cluster. We therefore performed on these galaxies single Sérsic model fitting as well as bulge-disc decomposition. This allowed us to compare the performance of these configurations and their parameters in characterising distant \Euclid galaxies. We demonstrate some limitations in modelling galaxies as single-S\'ersic profiles and how the resulting flux and structural parameters relate to the measurements of their bulges and discs. We also measured the colours of the bulges and discs of galaxies, which may be symptomatic of distinct histories of star formation when different, as well as colour gradients within these galaxies. Morphological analyses based on the multi-component profile-fitting of galaxies in the Perseus cluster are also being performed (\citealp{Mondelin-2025-ERO-outskirts-SB-and-color-profiles}; Tarsitano et al. in prep{.}). From these ERO images, analyses were also performed of the Perseus cluster luminosity and mass functions \citep{EROPerseusOverview}, of the intracluster light \citep{Kluge-2025-ERO-Perseus-ICL}, and of the cluster's dwarf galaxies \citep{Marleau-2025-ERO-Perseus-dwarf}. 

In this work, after presenting the observations (\sct\ref{sct-imaging}), the calculations of the point spread functions (PSFs; \sct\ref{sct-psf}) for VIS \citep{EuclidSkyVIS} and the Near-Infrared Spectrometer and Photometer \citep[NISP,][]{EuclidSkyNISP}, and the source selection approach (\sct\ref{sct-source-selection}), we describe the single-S\'ersic and bulge-disc decomposition approaches applied to all galaxies in the images (\sct\ref{sct-methodo}). We then compare the derived structural galaxy parameters from the VIS images using both modelling methods (\sct\ref{sct-quality-check}), namely, the centres of the models (\sct\ref{sct-check-position}), their effective radii (\sct\ref{sct-check-Re}), their S\'ersic indices (\sct\ref{sct-check-n}), their axis ratios (\sct\ref{sct-check-elong}), and the position angles of their major axes (\sct\ref{sct-check-angle}), and we focus on similarities and differences between the single-S\'ersic profile parameters and those of the bulge and disc. We then examine the impact of both modelling methods on the derived total photometry of galaxies as well as the differences among them and compare to the adaptive aperture photometry (\sct\ref{sct-model-photo}). Lastly, we use the derived model parameters in the VIS and NISP bands to detect a bulge-disc colour difference (\sct\ref{sct-bulge-disk-color}) as well as the colour gradients within single-S\'ersic and disc profiles (\sct\ref{sct-bulge-disk-gradient}). We also discuss the implications of the various detected biases and colour effects on the weak lensing and spin alignment measurements planned with \Euclid (\sct\ref{sct-check-shear-spin}). Sections \ref{sct-summary} and \ref{sct-conclusions} provide a summary of the results and the conclusions and perspectives that we draw from them, respectively. In the appendix, we provide data, model, and residual images for both the single-S\'ersic fits and bulge-disc decompositions in the \IE band for 18 galaxies that illustrate successful and problematic bulge-disc decompositions.

\section{Data \label{sct-data}}

\subsection{Imaging\label{sct-imaging}}

The \Euclid ERO images of the Perseus cluster were obtained in all its bands: VIS \IE (\ang{;;0.1} pixels); and NISP \YE, \JE, and \HE (\ang{;;0.3} pixels). These images were produced from four Reference Observation Sequences (ROSs, see \citealp{Scaramella-EP1}), while the Euclid Wide Survey (EWS; \citealt{Scaramella-EP1}) and Euclid Deep Survey (EDS) will be observed at the depth of one ROS and more than 40 ROSs (depending on the field), respectively. The total exposure time for the ERO-Perseus images is 7360 seconds in \IE, 1472 seconds in \YE, and 1582 seconds in \JE and \HE, over a field of view of $0.57$ deg$^2$.

The ERO-Perseus images were reduced (removal of instrumental signatures, astrometric and photometric calibration, and stacking) in a specific ERO pipeline \citep{EROData}, distinct from the \Euclid Science Ground segment that will process all the \Euclid fields \citep{Frailis-2019-Euclid-SGS}. We use the compact-source stacks provided by \citet{EROData}, which are optimised for compact-source photometry: the sky background of these stack images was modelled and subtracted, therefore removing the extended and low surface brightness signal.\footnote{The extended emission stacks optimised for low-surface brightness analyses, also provided by the authors, should be used for studying the Perseus cluster galaxies.} As a result, the outskirts of the brightness distributions of 120 Perseus cluster galaxies with large angular diameters ($\gtrsim$ \ang{;;5}, and $\IE\lesssim16.5$) are unusable for modelling (84 out of 120 of them are excluded from the present statistical analysis, see \sct\ref{sct-source-selection}).

\subsection{Point spread function\label{sct-psf}}

\begin{figure}
\includegraphics[width=\columnwidth]{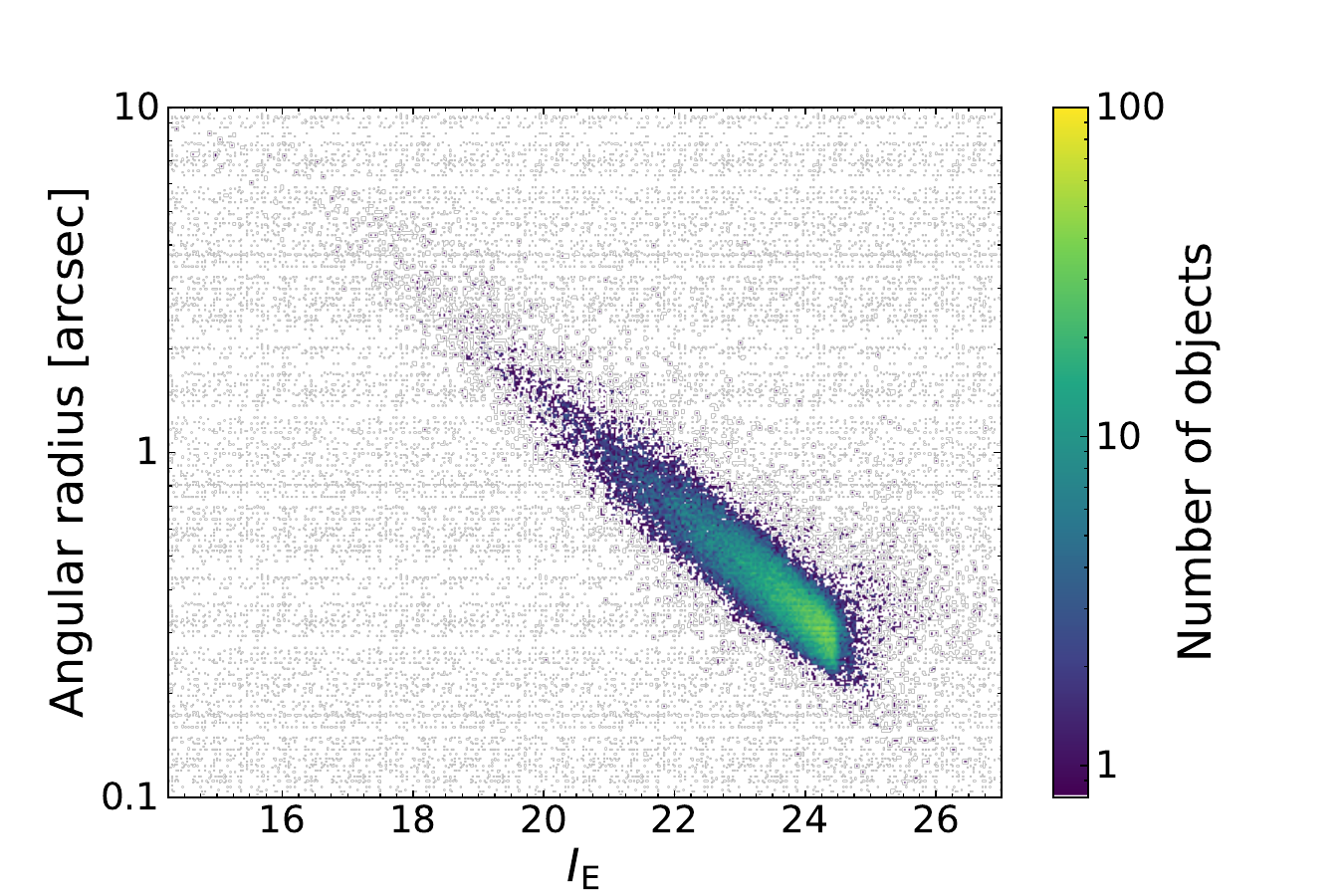}
\caption{Distribution of the angular radii corresponding to the isophotal areas calculated by \texttt{SourceXtractor++} on all objects identified as galaxies in the ERO-Perseus field by \citet{EROPerseusOverview} versus their \texttt{SExtractor}'s \texttt{MAG\_AUTO} \IE magnitude. All $38\,032$ modelled sources with $\IE\le24.5$ are plotted, as described in \sct\ref{sct-source-selection}.}
\label{area-vs-mag}
\end{figure}

Precise measurement of the PSF is essential for reliable profile modelling of galaxies. Figure \ref{area-vs-mag} shows the square root of the isophotal area measured by \texttt{SourceXtractor++} \citep{Bertin-2020-SourceXtractor-plus-plus, Kummel-2022-SE-use} in the VIS image at a $2.0\sigma$ detection threshold, for all sources identified as galaxies (see \sct\ref{sct-source-selection}), therefore representing the approximate angular diameter of the sources. Indeed, the majority of galaxies have an angular radius in the \ang{;;0.2}--\ang{;;1.0} interval, and an \IE apparent magnitude in the 19--24.5 interval.

To measure the PSF, we used the fully automated \texttt{PSFex} software \citep{Bertin-2011-PSFEx} in combination with \texttt{SExtractor}\footnote{See \url{https://sextractor.readthedocs.io/en/latest/} for documentation.} \citep{Bertin-1996-Sextractor}, which first extracts $51\times51$ and $35\times35$ pixel vignettes of the stars in the fields, under saturation limits of $12\,000$ and $5000$ for the VIS and NISP images, respectively. We then limit the calculation of the PSF to the $15\,245$, $8091$, $8053$, and $7501$ stars with a signal-to noise larger than 500 and adopt for the \texttt{PSFex} a \texttt{FWHM\_RANGE} parameter of $2.0$--$2.7$, $1.7$--$2.1$, $1.8$--$2.2$, and $1.9$--$2.5$ pixels for the VIS \IE, and NISP \YE, \JE, and \HE bands, respectively (in this section, all quadruplets of values correspond to this sequence of bands). This limits the calculation of the PSF to galaxies in the \texttt{MAG\_AUTO} $16.5$--22 mag \IE interval, and to a value of \texttt{FLUX\_RADIUS} (half-light radius) in the intervals corresponding to half the \texttt{FWHM\_RANGE} parameter. 

The resulting PSFs were modelled as Moffat functions at a sub-pixel resolution of $0.489$, $0.404$, $0.426$, and $0.471$ pixels. The derived PSF mean FWHM is \ang{;;0.16}, \ang{;;0.47}, \ang{;;0.49}, and \ang{;;0.50}. Therefore, the PSF is undersampled given the pixel sizes of \ang{;;0.1} and \ang{;;0.3} in the VIS and NISP images, respectively. We measure a mean $\chi^2$ per degree of freedom of approximately $9$, $12$, $5$ and $2$, with variations of these values by $\sim50\%$ across the field of view. The PSFs are calculated in circular apertures with diameters of 51 pixels (\ang{;;5.1}) in \IE, and 35 pixels (\ang{;;10.5}) in the three NISP bands. The $\chi^2$ value is sensitive to the details of the PSF in its wings, where the residuals per pixel are nevertheless less than $10^{-3}$. Lower $\chi^2$ values near unity with identical central PSF are obtained for other versions of the \IE reduced image. The FWHM of the resulting PSF varies across the field of view by $\lesssim5\%$ in all bands, and we adopt a polynomial of degree 5 for modelling these angular variations, as this corresponds to a minimum of the reduced $\chi^2$.
The ellipticities of the PSF FWHM have $(a-b)/(a+b)=0.03$, $0.04$, $0.003$, and $0.02$ (where $a$ and $b$ are the major and minor axes of the ellipses, respectively), with variations of the same order of magnitude across the field of view as the FWHM.

We caution that the PSF with a $1.6$ pixel full width at half maximum (FHWM) in the \IE band may bias the effective radii and S\'ersic indices of the 20\% smallest of the \IE$\le21$ bulges with an effective diameter (twice the effective radii) in the $0.1$ to $4.8$ pixels interval (that is at most three times the PSF FWHM). The three times lower resolution in each of the NISP bands increases the fraction of galaxies with an effective diameter smaller than 1.47 arcsec (three times the NISP \JE PSF FHWM) to 55\%. The flat bulge residual in \fg\ref{fig:exple-2} and the low-level residual structure in \fg\ref{fig:exple-1} (with effective bulge radii of 1.6 and 2.2 pixels, respectively) provide a visual illustration of the modelling in the central pixels of galaxies and the possible impact of the PSF.

For the \IE$\le21$ discs, 90\% of their effective diameters are larger than $8.1\times 2=16.2$ pixels (that is ten times the \IE PSF FWHM). In the NISP bands, 87\% of the discs even have their effective diameters larger than three times the \JE PSF FWHM. In contrast, 99.1\% and 94.4\% of the VIS single-S\'ersic diameters are larger than three times the PSF FWHM for \IE$\le21$, and \IE$\le23$, respectively; these fractions decrease to 73.7\% and 45.2\%, respectively, in the NISP bands (but the values of effective radii in the 21--23 \IE interval are not used in the current analysis). These various values indicate that the disc effective radii should only be weakly affected by the PSFs and hence have a limited impact on the extracted colour gradients (see \sct\ref{sct-bulge-disk-gradient}), and even less so on the fluxes (see \sct\ref{sct-model-photo}). Statistics on the general impact of the undersampled \Euclid PSF on the various models used with \texttt{SourceXtractor++} are provided in \citet{Merlin-EP25} for the photometry and in \citet{Bretonniere-EP26} for the structural parameters (as well as the improvement in using sub-pixel sampling for the PSF Moffat model, as done here).

\subsection{Source selection
\label{sct-source-selection}}

\begin{figure}
\includegraphics[width=\columnwidth]{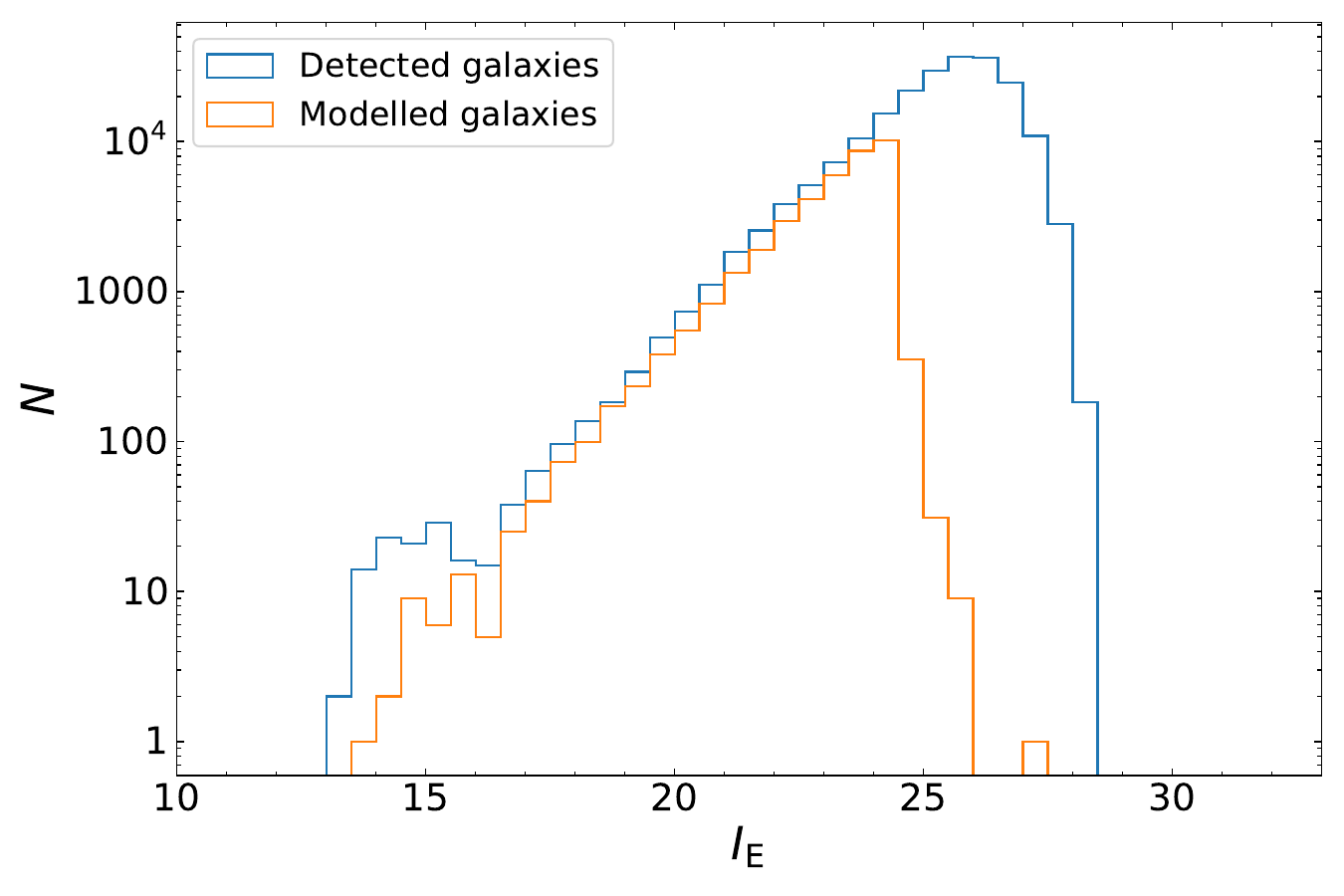}
\caption{Distribution of \IE magnitude for the $212\,975$ sources classified as galaxies in the ERO-Perseus field by \citet{EROPerseusOverview} and labelled as `Detected' compared to the $38\,082$ galaxies fit with the single-S\'ersic profile and the bulge-disc decomposition using \texttt{SourceXtractor++}, labelled as `Modelled'. The magnitudes of the `Detected' objects are from the photometric redshift catalogue based on \texttt{MAG\_AUTO} photometry using \texttt{SExtractor}, whereas those for the `Modelled' objects are \texttt{auto\_mag} photometry calculated with \texttt{SourceXtractor++}.}
\label{number-counts}
\end{figure}

After a \ang{;;1} cross-matching of the VIS and NISP coordinates to discard most spurious sources coming from cosmic rays and edge effects ($263\,196$ sources), a thorough star-galaxy separation was performed by \citet{EROPerseusOverview} using six criteria: matching with \textit{Gaia} DR3 stars; $g-z$ versus $z-\HE$ colour-colour diagram based on complementary observations with MegaCam at the Canada-France-Hawaii Telescope (CFHT); the \texttt{CLASS\_STAR} neural-network-based stellarity classifier of \texttt{SExtractor}; compactness using a peak surface brightness-magnitude criterion; and a \texttt{SExtractor} \texttt{KRON\_RADIUS} smaller than 3.5 pixels on the \IE image; a \texttt{SExtractor} \texttt{SPREAD\_MODEL} parameter indicating that the spread of the source is comparable to the PSF. Objects satisfying at least four of the above criteria are considered as stars ($49\,922$). This yields a catalogue of $212\,975$ galaxies, whose number counts as a function of \IE magnitude are plotted in \fg\ref{number-counts} (blue curve). Here, for consistency with the reduced images on which the SourceExtractor++ modelling was performed, we used the magnitude zero-points of 30.132 in the \IE band, and $30.000$ in all three NISP bands, which where provided at an early stage of the project by J.-C. Cuillandre. In the rest of the article we updated the values to those published by \citet{EROData}. 

We then selected sources with \texttt{MAG\_AUTO} limited to $\IE\leq24.5$ to be modelled, using the ASSOC mode of \texttt{SourceXtractor++}, and a \ang{;;1} cross-matching. This leads to $38\,082$ galaxies whose profiles were modelled with the various configurations of the current analysis (see \sct\ref{sct-model-fitting}), and whose number-counts as a function of \IE magnitude are plotted in orange in \fg\ref{number-counts}. At this stage, the apparent magnitudes are not yet corrected for Milky Way extinction, since what matters to evaluate our ability to model galaxies is the observed flux and the spatial resolution of the objects. This correction is performed in \sct\ref{sct-bulge-disk-color} before examining bulge and disc colours.

Figure \ref{number-counts} shows that at the bright end ($\IE\sim15$), only about one-third of the detected galaxies in the Perseus cluster are modelled, because the profiles of bright and nearby cluster galaxies are biased in the compact-source stacks at large radii. At the faint end, completeness in the photometry drops sharply at $\IE\sim27$ in the number of detected galaxies, whereas profiles are modelled out to $\IE=24.5$. Although only `Detected' galaxies with $\IE\le24.5$ are modelled, the `Modelled' histogram contains fainter galaxies than this limit because the cross-match can either erroneously select a brighter or fainter neighbouring source or not find any source since the partition performed by \texttt{SourceXtractor++} can be different from that with \texttt{SExtractor}. This nevertheless has no impact on the galaxies with $\IE\le23$ onto which the present analysis is focused. 

\section{Methodology
\label{sct-methodo}}

\subsection{Luminosity model fitting\label{sct-model-fitting}}

\subsubsection{Generalities \label{sct-model-fitting-generalities}}

We use the \texttt{SourceXtractor++}\footnote{See \url{https://sourcextractorplusplus.readthedocs.io/en/latest/} for documentation.} software to model the sky projected brightness distributions of all the galaxies identified in the \IE image. To this end, the code performs a least-squares fit using the differences between the resampled convolved model with the local PSF model (see \sct\ref{sct-psf}) and the pixel values, weighted by the pixel uncertainty. The differences are modified by a derivable monotonic function that reduces the influence of large deviations from the model, such as the contamination by neighbours. The loss function also includes a penalty term acting as a Gaussian prior on the profile parameters, which contributes to regularisation (and is among the strengths of \texttt{SourceXtractor++} compared to the original \texttt{SExtractor}).

Several configurations of model fitting were performed on the ERO-Perseus field galaxies, which are described below, and which are all based on the generalisation to elliptical isophotes of the S\'ersic profile \citep{Sersic-1963-sersic-model}, which describes the variation of the light intensity $I$ as a function of the angular radius $r$ (hence in the case of circularly symmetric isophotes):
\begin{equation}
    I(r) = I_\mathrm{e} \exp\left\{ -b_n\left[\left( \frac{r}{R_\mathrm{e}}\right) ^{1/n} -1\right]\right\}.
    \label{eq-sersic}
\end{equation}
The $R_\mathrm{e}$ parameter is the effective radius that encloses half of the total light of the profile, $I_\mathrm{e} = I(R_\mathrm{e})$ is the intensity at $R_\mathrm{e}$, $n$ is the S\'ersic index that defines the steepness of the profile, with higher values of $n$ corresponding to steeper central profiles and less steep outskirts, and $b_n$ is a normalisation parameter depending on $n$ only. For discs, we also used an exponential profile, which is equivalent to a S\'ersic profile with an index $n=1$. For the bulges, the choice of a de Vaucouleurs profile, with $n=4$, is common \citep[see e.g.][]{Simard-2011-BD-decomp-SDSS, Lackner-Gunn-2012-bulge-disc-decomposition-SDSS, Nedkova-2024-mass-size-z-3} and so is the choice made here of a free S\'ersic index \citep[see e.g.][]{Allen-2006-MGC-BD-decomp, Dimauro-2018-bulge-disk-decomposition-CANDELS}. We preferred the second option, as the bulge S\'ersic index presents the advantage of encoding information regarding the nature of the bulge, namely, whether it is dominated by a disky pseudo-bulge structure or by a spheroidal classical one \citep{Kormendy-Kennicutt-2004-bulges-disk-galaxies-review, Quilley-2023-scaling-bulges-and-disks}, whereas the use of a de Vaucouleurs profile assumes (a priori) the latter.

The different model fitting configurations that were performed in this study are the elliptical generalisations of either a single-S\'ersic profile to model the whole galaxy light distribution (single-S\'ersic hereafter) or of the sum of a S\'ersic profile and an exponential profile designed to model the bulge and the disc components of each galaxy, respectively (bulge-disc decomposition hereafter). The details of the configurations can be found in \sct\ref{sct-model-fitting-config}.
For each fitted profile or model component {\texttt{SourceXtractor++} returns as parameters the position of the centre of the 2D elliptically symmetric model, the values of its integrated flux, the effective radius along the semi-major axis, the position angle $\theta$,\footnote{This is the position angle between the major axis of the ellipse and the CCD $x$-axis, measured counter-clockwise between $-\pi/2$ and $\pi/2$ radians.} the axis ratio between the minor and major axis effective radii of the fitted elliptical profile $b/a$, and the S\'ersic index $n$ (unless fixed to 1 in the exponential profile).

The model fitting runs were performed using three approaches (unless mentioned otherwise, no prior relating a given parameter between the bulge and disc component or between different bands was used):

\begin{enumerate}
    \item All four \Euclid VIS and NISP images, using the VIS image as the detection image,\footnote{The detection image is the single image from which the groups of pixels defining each source are extracted, and from which estimates of positions and shapes, as well as initial guesses for the model fitting parameters, are defined. Using the higher angular resolution VIS image as the detection image when the measurement images also include those in the NISP bands (on which the models are fitted) ensures that the same galaxies are modelled in the various bands.} with each band having its own independent set of structural parameters for either the single-S\'ersic profile or the bulge-disc decomposition.
    \item All four \Euclid  VIS and NISP images, using VIS for the detection image$^6$, but with two sets of structural parameters for either the single-S\'ersic profile or the bulge-disc decomposition, one for the VIS image and a common one for all three NISP images, yet independent photometry in the \YE, \JE, and \HE bands.
    \item Only on the VIS image with a unique centre, axis ratio, and position angle for both the bulge and disc components in the bulge-disc decomposition (see \sct\ref{sct-model-fitting-config}).
    \end{enumerate}
The results of approach 1 are those that are analysed and discussed throughout \scts \ref{sct-quality-check}, \ref{sct-color-gradients} and \ref{sct-model-photo},  whereas those of approach 2 are only used in Appendix \ref{appendix-common-profile}, and those of approach 3 are discussed at the end of \sct\ref{sct-check-position}. 

\subsubsection{Configurations\label{sct-model-fitting-config}}

We list here the initial values and allowed ranges that are provided to \texttt{SourceXtractor++} for both the single-S\'ersic model and the bulge-disc decomposition:

\begin{itemize}
    \item The $x$ and $y$ coordinates (in pixel units of the detection image) of the centre of the single-S\'ersic, bulge, and disc models are defined by the built-in function \texttt{get\_pos\_parameters()} that is initialised at the isophotal centroid coordinates and with a range equal to the isophotal radius on both axes.
    \item The total flux $f$ of each galaxy is computed with the built-in function \texttt{get\_flux\_parameter()} initialised as the isophotal flux, and that can vary by a factor of $10^3$.
    \item The bulge-to-total light ratio $B/T$ of the bulge-disc decomposition (\ie for each galaxy, the ratio between the luminosity enclosed in its bulge and the total luminosity) is initialised at 0.5 and varies exponentially in the $[0.001,1]$ interval, with the fluxes of the bulge and disc components then being defined as $f_\mathrm{B} = B/T\,f$ and $f_\mathrm{D} = (1 - B/T)\,f$.
    \item The S\'ersic indices of either the single-S\'ersic profile or the bulge component are initialised at 4 and vary linearly in the $[0.5,10]$ interval.
    \item The effective radii (in pixel units of the detection image) of the different profiles are initialised at the isophotal radius (given in pixels by the \texttt{o.radius} built-in function) and vary exponentially in a range defined in terms of a factor times the initial value, with $[0.01,2]$ for the bulge and $[0.1,10]$ for the single-S\'ersic or disc radii.
\end{itemize}

The position angle $\theta$ (in radian) and axis ratio $b/a$ (of the minor to major axis effective radii of the fitted elliptical profile) are usually defined so that the angle spans the $[-\pi,\pi]$ range and is then re-projected onto $[-\pi/2,\pi/2]$, whereas the axis ratio parameter $b/a$ spans $[0,1]$ on a linear scale or $[0.01,1]$ in exponential scale.
However, we adopted a Cartesian parametrisation (see the appendix of \citealt{Tessore-2023-appendix-e1e2-def}; also used in \citealt{Atek-2025-ERO-mag-lens}) that defines the $e_1$ and $e_2$ parameters varying linearly in the range $[-1,1]$ so that the position angle is computed as
\begin{equation}
    \theta = \frac{1}{2} \mathrm{atan}(e_1/e_2),
\end{equation}
and the axis ratio as
\begin{equation}
    b/a = \frac{\left\vert{1- \sqrt{( {e_1}^2 + {e_2}^2 )}}\right\vert}{{1+\sqrt{( {e_1}^2 + {e_2}^2 )}}}.
\end{equation}
We checked that this parametrisation yields similar profile parameters as the straightforward configuration by running tests using both of them (with all other options left unchanged), and it scans more uniformly through the phase space.

One of the new features of \texttt{SourceXtractor++} compared to \texttt{Sextractor} is the ability to define priors related to different parameters or between different bands for a given parameter (in multi-band fits). As our goal is to explore the similarities and differences between the various profiles, we nevertheless used minimal priors (except in Appendix \ref{appendix-common-profile}), and at the end of \sct\ref{sct-check-position} we further discuss the implications of using more constrained fits in their centring.

The modelling approach 1 is used throughout this article. It treats the bands independently and hence returns best-fit parameters in all four \Euclid bands. In the following analysis, we mainly focus on the parameters for the \IE band, due to the higher resolution it provides (with a pixel scale and PSF FWHM 3 times lower than in the NISP bands). NISP parameters are leveraged in \sct\ref{sct-bulge-disk-gradient} to study the variation of structural parameters with wavelengths. 

Generally, we use $B/T$ as an indicator of galaxy morphology, and specifically the $B/T$ fitted in the \IE band, which differs from those in the other bands. This choice is supported by the higher precision on its measure thanks to the better angular resolution of VIS, which leads to overall smaller uncertainties on $B/T(\IE)$ than on the $B/T$ fitted in the NISP bands: it is the case for $92\%$, $84\%$, and $78\%$ of the \IE$<21$ galaxies, for the \YE, \JE, and \HE bands, respectively, and the median ratio of the uncertainties on $B/T$ between \IE and a NISP band is $0.37$, $0.48$, $0.53$ for the \YE, \JE and \HE, respectively. We also examined the variations of $B/T$ with the observing band. For instance, we measure an increase in the \JE-to-\IE median ratio of the $B/T$ values, which are $0.97$, $1.13$, $1.40$ and $2.18$, as well as in the dispersions of 0.22, 0.37, 0.69 and 1.58 for decreasing $B/T(\IE)$ ranges of $[0.5,1.0]$, $[0.3,0.5]$, $[0.1,0.3]$ and $[0.05,0.1]$, respectively. We also emphasize that the variations are smaller for bands closer to each other in wavelength.

\subsection{Spectral energy distribution fitting of model photometry \label{sct-methodo-SED}}

To estimate absolute magnitudes, we used the photometric redshifts derived in \cite{EROPerseusOverview}, based on the Kron adaptive aperture photometry in the CFHT $u$, $g$, $r$, $i$, and $z$ bands, as well as in the \Euclid \IE, \YE, \JE, and \HE bands, serving as input to \texttt{Phosphoros}, the Bayesian spectral energy distribution (SED) fitting code developed for the \Euclid pipeline \citep{Q1-TP005}. The observed photometry was compared with a grid of expected values computed from 31 COSMOS templates \citep{Ilbert2009}, optimised for distant galaxies.

To compute the absolute model magnitudes in the $ugriz$ \IE \YE \JE \HE bands, we performed a new SED fitting with the photometric redshift fixed to that of the corresponding galaxy. We used the \IE, \YE, \JE, and \HE model magnitudes measured for whole galaxies fitted with single-S\'ersic profiles or decomposed into bulge and disc components, as well as the separate bulge and disc magnitudes, as input to \texttt{hyperz} \citep{hyperz}. This allows the derivation of absolute magnitudes simultaneously in several bands using the preferred set of templates. Dust attenuation was modelled using the \citet{Calzetti2000} law, with $V$-band attenuation $A_V$ ranging from $0$ to $3$ in steps of 0.1. The cosmological parameters adopted to compute the distances were $\Omm = 0.3$, $\OmLa=0.7$, $h=0.7$ \citep{Planck-collaboration-2016-planck-2015-results-overview}.

To obtain absolute magnitudes while minimising the uncertainties from the best-fit SED, we used the observed band with ${\rm S/N}>3$ closest to the rest-frame band at the assumed photometric redshift (called `pivot' band), then applied a $k$-correction to that band, plus a colour correction to the rest-frame band, both derived from the best-fit template SED. Given the limited number of input photometric bands, we tested the robustness of the results by comparing the absolute magnitudes obtained with the COSMOS templates used for photometric redshifts to those derived with \cite{BC03} models, which include additional parameters such as age, star formation history, and metallicity. The two estimates agree with a dispersion of about $0.05$ magnitudes in the $g$-band and $0.02$ in $i$. In the following, we adopt the estimates based on the COSMOS templates.

Since the SED fitting was carried out independently for the bulge, the disc, and the total photometry, we also verified that the sum of the bulge and disc rest-frame fluxes reproduces the total rest-frame flux. We found good agreement for galaxies with $\IE\le21$: the mean(median) of the difference between the total absolute magnitude and the one from the sum of the bulge and disc fluxes is equal to $0.027(0.009)$ in the $g$-band, with a dispersion of $0.069$, and  $-0.008(-0.002)$ in the $i$-band, with dispersion of $0.053$.
In order to estimate the uncertainties in the absolute magnitudes, we used the uncertainties in the observed magnitudes in the `pivot' bands, to which we add in quadrature the \rms dispersion in the difference of absolute magnitudes between the values obtained with the COSMOS and \cite{BC03} templates: $0.06$ for both the bulges and the discs in the $g$ band, and $0.03$ for both components in the $i$ band. Because we only consider colours of bulges and discs in \sct\ref{sct-bulge-disk-color}, hence differences in absolute magnitudes of a given component of an object, the uncertainties in the distance moduli are not considered.

\section{Comparison of structural parameters between single-S\'ersic and bulge-disc modelling \label{sct-quality-check}}

In this section, we assess the reliability of the \texttt{SourceXtractor++} single S\'ersic and bulge-disc modelling configurations described in \sct\ref{sct-model-fitting} on the $2445$ modelled galaxies in the ERO-Perseus field at \IE band \texttt{auto\_mag} $\le21$. To this end, we examine in the next subsections, the relations between the bulge and disc parameters in order to ensure that the S\'ersic and exponential profiles of the bulge-disc decomposition provide physically meaningful descriptions of the bulge and disc components of galaxies. We emphasise that the ERO-Perseus field is deeper than the EWS, with a point-source detection limit of $\IE=28.0$ (and \YE, \JE, \HE = 25.3) at $5\sigma$ for ERO images \citep{EROPerseusOverview} compared to $\IE=26.2$ (and \YE, \JE, \HE = 24.5) for the EWS \citep[and surface brightness limits of $30.1$ and $29.2$ mag.arcsec$^{-2}$ in the optical and near infrared for the ERO field compared to $29.8$ and $28.4$ mag.arcsec$^{-2}$ for the EWS;][]{Scaramella-EP1}, due to about four times longer exposures in all bands. Therefore, by using the EMC limit (\IE$<21$) based on the EWS for validity of the bulge-disc decompositions in the ERO-Perseus images, we adopt a conservative approach. The following checks are performed on the central positions of the single-S\'ersic, bulge, and disc components, on their effective radii, their axis ratios, and major axis position angles, and on the S\'ersic indices of the single-S\'ersic and the bulge component. All derived parameters are apparent, hence in angular units. The uncertainties for individual galaxies are either those estimated by \texttt{SourceXtractor++} for the output parameters of each object or result from the propagation of these uncertainties. Bootstrap uncertainties were also computed for median values of parameters within sub-samples of galaxies.

The structural parameters derived from profile-fitting are also useful to estimate the morphological types of galaxies, which are related to their evolutionary stage and present and past star-formation rates \citep{Strateva-2001-color-bimodality,Bluck-2014-bulge-mass,Lang-2014-bulge-growth-quenching-CANDELS,Bremer-2018-GAMA-survey-morph-transf-GV,Dimauro-2022-bulge-growth,Quilley-2022-bimodality}. When extending the characterisation of the Hubble sequence \citep{Hubble-1926-extragalactic-nebulae}, in addition to the extent to which the spiral arms are unwound, \citet{De-Vaucouleurs-1959-class-morph} maintained the bulge prominence as a criterion to differentiate among the morphological types: $B/T$ increases from irregulars or `bulge-less' types, through spirals from late- to early-types, lenticulars with very prominent bulges, and ultimately ellipticals that can be considered as `bulge-only' galaxies. Therefore, throughout the analysis we examine the parameters of the fitted galaxy models as a function of $B/T$, which has been shown to vary continuously with the Hubble type \citep{de-Lapparent-2011-EFIGI-stats,Kim-2016-bulges-SDSS,Gao-2019-demographics-bulge,Quilley-2023-scaling-bulges-and-disks}.

Furthermore, visual examination of the data, model, and residual images in the \IE band was performed on all outliers from the trends identified in the following subsections, and critical examples are shown in the appendix. Labelling a fit as visually successful is based on the success in modelling any visible bulge and disc components, as well as obtaining low-intensity residual images compared to the model (except for some regions corresponding to flocculence, spiral arms, bars, and rings, which are not modelled here (see \fgs \ref{fig:exple-1} to \ref{fig:exple-7}). This allowed us to confirm that many of the outliers are galaxies that cannot be modelled successfully by an elliptically symmetric profile with a single central concentration. This can be due to Milky Way stellar contamination, dust lanes in edge-on galaxies splitting the disc into two, or ongoing mergers causing strongly distorted isophotes (see for instance \fg\ref{fig:exple-10}).
During the process of visual inspection of the outliers, and various initial tests of the profile modelling performance,  several hundreds of randomly selected galaxies with \IE total galaxy magnitude in the 17--23 interval were also examined. This examination confirmed that galaxies located inside the major sequences in the various projections of the parameter spaces described below were indeed successfully fitted by the single-S\'ersic model and the bulge-disc decomposition.

\subsection{Effective radii  \label{sct-check-Re}} 

The definitions of the bulge and disc components of galaxies based on galaxies observed in the nearby Universe, are a central concentration of stellar light enclosed within a flattened lower surface brightness profile and more extended stellar system. When the physical bulge and disc are successfully decomposed in nearby galaxies, the effective radius of the bulge is therefore smaller than the effective radius of the disc (see Fig. 17 of \citealt{Quilley-2023-scaling-bulges-and-disks} for example). The single-S\'ersic effective radius is then expected to take values between the bulge and the disc effective radii. Indeed, the total galaxy contains more light than the bulge only, so the radius containing half of the light is larger for the galaxy than for the bulge. The bulge creates an over-concentration of light in the centre of the disc, leading to a single-S\'ersic profile with an effective radius smaller than that of the disc. Altogether, these conditions can be written as $R_\mathrm{e,bulge} \le R_\mathrm{e,1p} \le R_\mathrm{e,disc}$, with equality for $B/T=1$ and $0$, respectively.

We examine in \fg\ref{all-radii-related-cmap-BT} these expected properties of the ERO-Perseus field galaxies, by plotting the disc-to-bulge ratio of effective radii versus the corresponding disc to single-S\'ersic $R_{\rm e}$ ratio. The colour of the points represents the $B/T$ value of the bulge-disc decomposition measured in the \IE image (with the best angular resolution), notably to highlight extreme cases for which either the disc or the bulge component is weak, and therefore the values of their radii should not be over-interpreted. These galaxies lie along the vertical $x=1$ axis and the diagonal $x=y$ axis, respectively. Galaxies along the  $x=1$ line (corresponding to $R_\mathrm{e,1p} = R_\mathrm{e,disc}$), have $B/T \approx 0$  (blue points). These objects are well fitted by a pure exponential profile, and therefore their bulge component is negligible and the disc component and single-S\'ersic profile have similar effective radii (we also note in \sct\ref{sct-check-n} that for such galaxies the single-S\'ersic index is close to 1, as for the exponential profile). Galaxies near the $x=y$ line (corresponding to $R_\mathrm{e,1p} = R_\mathrm{e,bulge}$) have the opposite behaviour with $B/T \sim 1.0$ (red points). Therefore, their disc component is negligible, and their light profiles are well described by a single-S\'ersic function. As a result, they have similar effective radii of their bulge component and single-S\'ersic profile.

\begin{figure}
\includegraphics[width=\columnwidth]{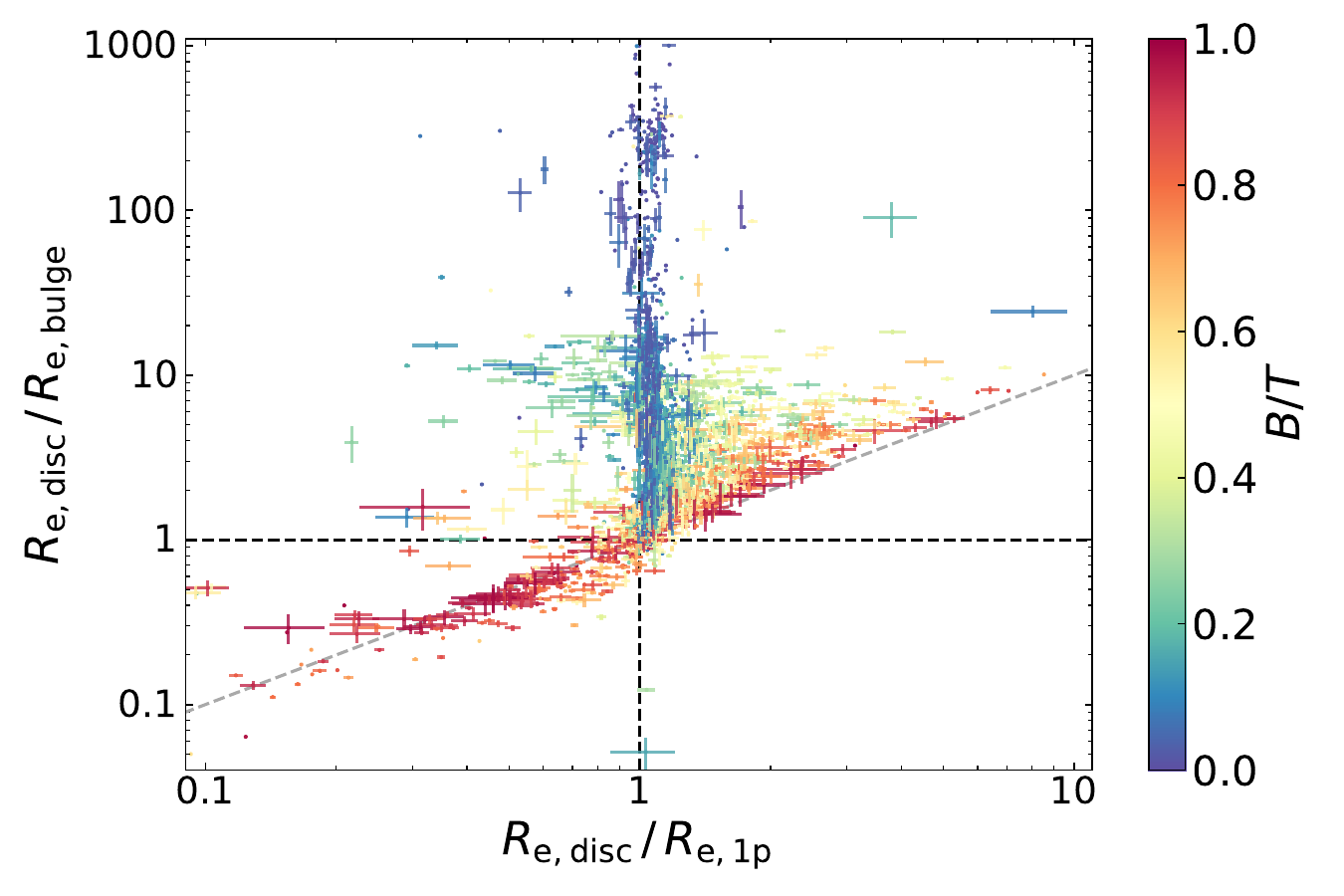}
\caption{Ratios of the disc-to-bulge effective radii as a function of the ratios of the disc-to-single-S\'ersic effective radii, all in the \IE band, for the $2445$ galaxies with $\IE\le21$. In the upper vertical concentration of disc-dominated galaxies (in blue, $B/T\approx0$) and on the right diagonal concentration of bulge-dominated galaxies (in red, $B/T\approx1$), the single-S\'ersic effective radius is consistent with that of either a dominating disc or bulge component. The single-S\'ersic effective radius of the $74.7\%$ of galaxies in the top-right cone is intermediate between the disc effective radius (the largest) and the bulge effective radius (the smallest), with a smooth gradient in the ratio of the disc to single-S\'ersic effective radius, while $B/T$(\IE) varies from zero to one. These galaxies are visually indistinguishable from the types in the present-time Hubble sequence, whereas objects in other regions of the diagram are identified as either non-physical bulge-disc modelling or biased bulge fits due to bars.}
\label{all-radii-related-cmap-BT}
\end{figure}

The top-right quadrant of \fg\ref{all-radii-related-cmap-BT} is the region of galaxies satisfying both aforementioned conditions that $R_\mathrm{e,disc} \ge R_\mathrm{e,bulge}$ and $R_\mathrm{e,disc} \ge R_\mathrm{e,1p}$ and contains 1829 galaxies (that is $74.8\%$ of galaxies with $\IE\le21$). The fact that most galaxies in this quadrant ($98.2\%$) are in a triangle above the diagonal line results from the condition $R_\mathrm{e,1p}\ge R_\mathrm{e,bulge}$, and suggests that the bulge-disc decomposition is successful in modelling both components of the objects. Moreover, in this triangle, the bulge-disc decomposition displays a continuous $B/T$ gradient from 0 to 1 (with colours from blue to red), hence describing the full range of types from disc-only to bulge-only, with a position on the graph depending on the fraction of flux contained by each of the bulge and disc components. These galaxies also take all values of the ratio of disc to bulge effective radii up to about $20$.
The distribution of bulge-to-disc $R_\mathrm{e}$ ratios is consistent with the $B/T$ and size distributions of bulges and discs measured by \citet{Quilley-2023-scaling-bulges-and-disks} for the EFIGI sample, which fully and densely samples the Hubble sequence of morphological types in the nearby Universe ($z\lesssim0.05$; \citealp{Baillard-2011-EFIGI}). Nearby galaxies possessing both a bulge and a disc are either lenticular or spiral types of the Hubble sequence \citep{Quilley-2022-bimodality}. However, some profile fits of nearby elliptical types also include a significant disc component when the single-S\'ersic profile is not sufficient for modelling both the inner and outer parts of the objects. 

Visual inspection in the \IE band of a random sub-sample of $\sim300$ galaxies in the top-right quadrant of \fg\ref{all-radii-related-cmap-BT} confirms their irregular, lenticular, spiral, and elliptical morphologies, indicating that the Hubble sequence is also present at the higher redshifts reached by this \Euclid image. We emphasise that among the hundreds of galaxies with $\IE\le23$ that were visually examined across the field (see \sct\ref{sct-quality-check}), there was no object that looked markedly different from the EFIGI nearby morphological sequence in their bulge and disc properties. The EFIGI sample can be used as a reference because its galaxies have similar or better spatial resolution than galaxies in the current \Euclid field (depending on the object distances in both samples), and take all possible values and configurations of disc inclination, axis ratio, dynamical features (spiral arms, bars, rings), and texture (from stellar formation and dust), as shown in \cite{de-Lapparent-2011-EFIGI-stats}. In particular, a significant fraction of galaxies with $\IE\le21$ are seen to host bars in the ERO-Perseus field (this is confirmed with the first `Quick' \Euclid data release (Q1), \citealp{Q1-SP043}). If the detected morphological types in the ERO-Perseus $\IE\lesssim21$ sample (corresponding to mostly $z\leq0.54$) are indistinguishable from the present-time morphological types in the Hubble sequence \citep{Hubble-1926-extragalactic-nebulae}, fainter galaxies with $21\lesssim \IE\lesssim23$ are mostly too poorly resolved to draw detailed conclusions about them. 

In contrast, the 281 galaxies ($11.5\%$ of the sample) in \fg\ref{all-radii-related-cmap-BT} below the $y=1$ black horizontal dashed line correspond to bulges larger than their discs. The majority of these galaxies lie near the $y=x$ diagonal (described above) and have $B/T\ge0.8$. However, we note in the lower part of this graph that 101 galaxies ($4.1\%$ of galaxies with $\IE\le21$) have $R_\mathrm{e,disc} < R_\mathrm{e,bulge}$ while showing a non-negligible disc component ($B/T \leq 0.7$), therefore leaving 180 well-modelled bulge-dominated galaxies. Visual examination of the \IE images of the 101 outliers shows that they correspond to non-physical bulge-disc models: in many cases, the S\'ersic profile intended for the bulge models the galaxy at radii well beyond the visual extent of the bulge (see \fg\ref{fig:exple-9}); and some cases result from bulge and disc components modelling nearly disjoint parts of galaxies, as already highlighted in \sct\ref{sct-check-position} (see \fgs\ref{fig:exple-11} and \ref{fig:exple-13}).

Lastly, the top left quadrant of \fg\ref{all-radii-related-cmap-BT} includes 305 galaxies ($12.5\%$ of the sample) with discs whose effective radii are larger than their bulges, but also smaller than the single-S\'ersic effective radius. Discarding the vertical sequence of disc-like galaxies (in blue) with some dispersion, that is only counting galaxies with ${R_\mathrm{e,disc}/R_\mathrm{e,1p} \le 0.9}$, yields 117 objects, hence $4.8\%$ of the sample. Visual inspection of these outliers shows that many of them are symptomatic of the difficulty to model both a peaked bulge and a low-surface-brightness disc with a single-S\'ersic profile: such fits underestimate the central flux of the bulge, hence expand the effective radius beyond that of the disc (see \fgs\ref{fig:exple-3} and \ref{fig:exple-4}). Other outliers are due to a bar that is fitted as if there was a very elongated bulge within a lower axis ratio disc. The single-S\'ersic fit for these objects is dominated by the bar which artificially boosts its axis ratio and its effective radius, then defined along the major axis. In these cases, the minor axis radius of the single-S\'ersic fits is nevertheless closer to the disc effective radius (see \fg\ref{fig:exple-5}). Adding to the $1859$ galaxies of the top-right quadrant, the left part of the dispersed vertical sequence of disc-dominated galaxies with $0.9\le R_\mathrm{e,disc}/R_\mathrm{e,1p}\le1$, that is 188 objects, plus the 180 bulge-dominated galaxies ($B/T > 0.7$) from the bottom half, leads to $2227$ galaxies with both single-S\'ersic models and bulge-disc decompositions that are likely to be reliable, as their effective radii are consistent with each other. This leaves 218 likely outlier fits (101 with spurious combinations of effective bulge and disc radii, and 117 with overestimated $R_\mathrm{e,1p}$), corresponding to 9\% of the 2445 modelled galaxies with $\IE\le21$. In the following, we exclude the 398 galaxies with $R_\mathrm{e,disc}/R_\mathrm{e,bulge}<1$ or $R_\mathrm{e,disc} / R_\mathrm{e,1p} \le 0.9$ when examining bulge and disc properties (leaving $2047$ objects), whereas we exclude the 117 galaxies with $R_\mathrm{e,disc}/R_\mathrm{e,bulge}\ge1$ and $R_\mathrm{e,disc} / R_\mathrm{e,1p} \le 0.9$ when examining single-S\'ersic profiles (leaving $2328$ objects).

\begin{figure}
\includegraphics[width=\columnwidth]{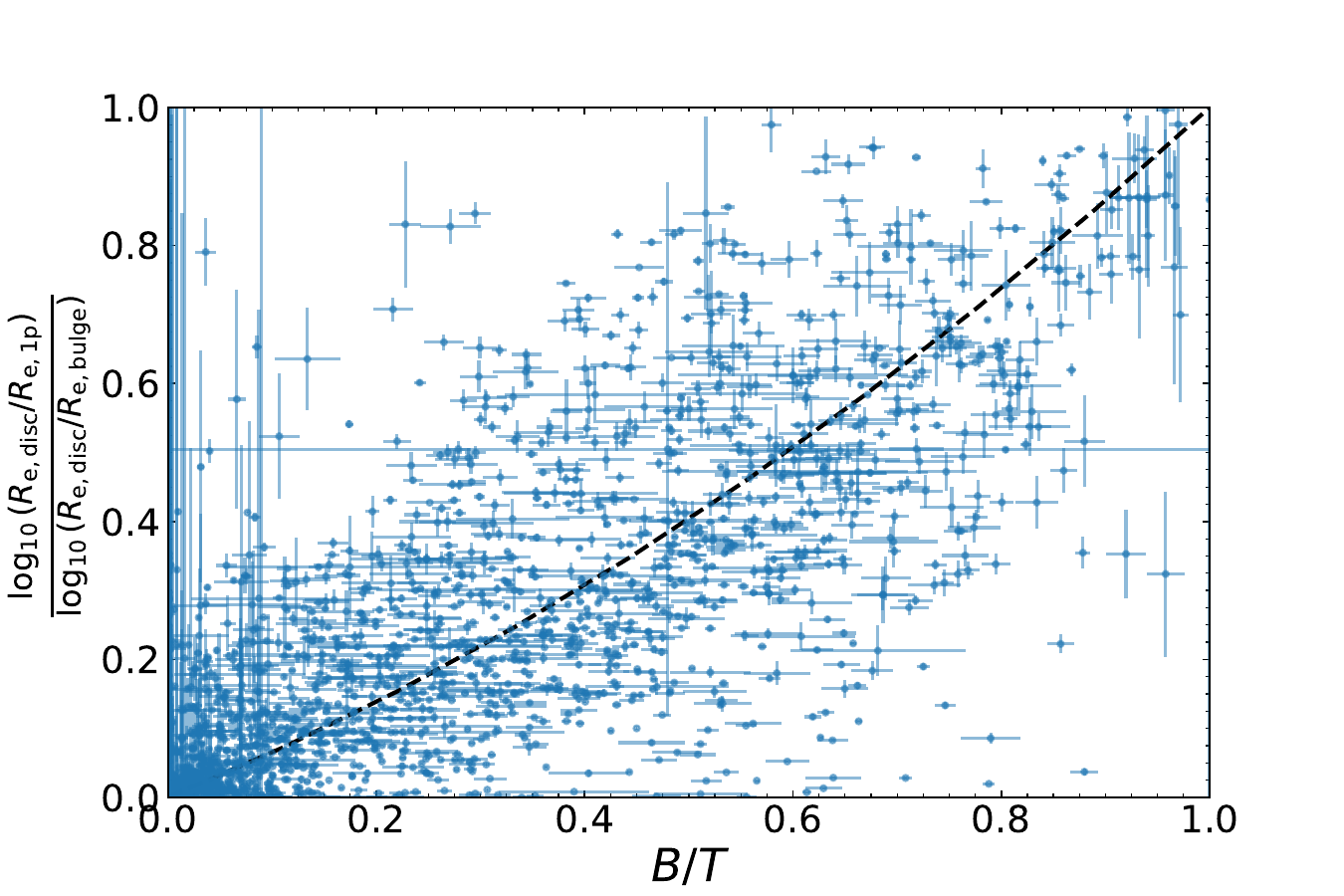}
\caption{Correlation between the disc-to-single S\'ersic ratio of effective radii relative to the disc-to-bulge ratio of radii (in logarithmic scale) and $B/T$ in the \IE band for the $1826$ galaxies with $\IE\le21$, verifying $R_\mathrm{e,bulge} < R_\mathrm{e,1p} < R_\mathrm{e,disc}$.}
\label{radii-1p-from-BD-values}
\end{figure}

We further explore in \fg\ref{radii-1p-from-BD-values} the relationship between the various effective radii and $B/T$ for galaxies located in the top-right wedge of \fg\ref{all-radii-related-cmap-BT} by examining, for all galaxies satisfying $R_\mathrm{e,bulge} < R_\mathrm{e,1p} < R_\mathrm{e,disc}$, how the disc-to-bulge ratio of effective radii relative to the disc to single-S\'ersic ratio of effective radii (in logarithmic scale) varies as a function of $B/T$. A correlation between the combination of effective radii ratios and $B/T$ is observed in \fg\ref{radii-1p-from-BD-values}, with a Pearson correlation coefficient of $r=0.77$\footnote{We also computed the location of the single-S\'ersic effective radius within the interval between the bulge and disc effective radii (hence in linear scale) $\frac{R_\mathrm{e,disc} - R_\mathrm{e,1p}}{R_\mathrm{e,disc} - R_\mathrm{e,bulge}}$, and obtained similar results, but we present here the best polynomial model according to both the Akaike and Bayesian information criterion.}.
We performed a second-degree polynomial fit (dashed line) under the constraints that $R_\mathrm{e,1p} = R_\mathrm{e,bulge}$ for $B/T = 1$ and $R_\mathrm{e,1p} = R_\mathrm{e,disc}$ for $B/T = 0$, leading to the following polynomial:
\begin{equation}
    \frac{\log_{10}(R_\mathrm{e,disc} / R_\mathrm{e,1p})}{\log_{10}(R_\mathrm{e,disc} / R_\mathrm{e,bulge})} = (0.39\pm 0.02)\; (B/T)^2 + (0.61 \pm 0.02) B/T
    \label{eq-R1p-BT-log}
.\end{equation}
The dispersion around this polynomial is 0.16. 

The polynomial fit in \fg\ref{radii-1p-from-BD-values} may be useful to produce synthetic galaxies with bulges and discs of realistic size, which match observational samples for which only a single-S\'ersic radii has been measured. We emphasise that in making mock galaxy fields (see for example \citealt{EuclidSkyFlagship}), the use of measured single-S\'ersic radii distributions to generate disc radii creates an underestimation of the latter, which is dependent on $B/T$, and therefore creates a bias in the disc radii with the morphological type of galaxies. 

Figure \ref{radii-1p-from-BD-values} also highlights the relevance of bulge-disc decomposition when discussing the size evolution of spiral and lenticular galaxies, and samples including these types of galaxies. Indeed, because the single-S\'ersic effective radius relates to both the extent and the concentration of a galaxy's light distribution, one should be careful in interpreting a variation of this radius which does not imply a variation in disc size, since it can also arise from a variation of $B/T$ at fixed disc radius.

\subsection{Centre positions \label{sct-check-position}}

\begin{figure}
\includegraphics[width=\columnwidth]{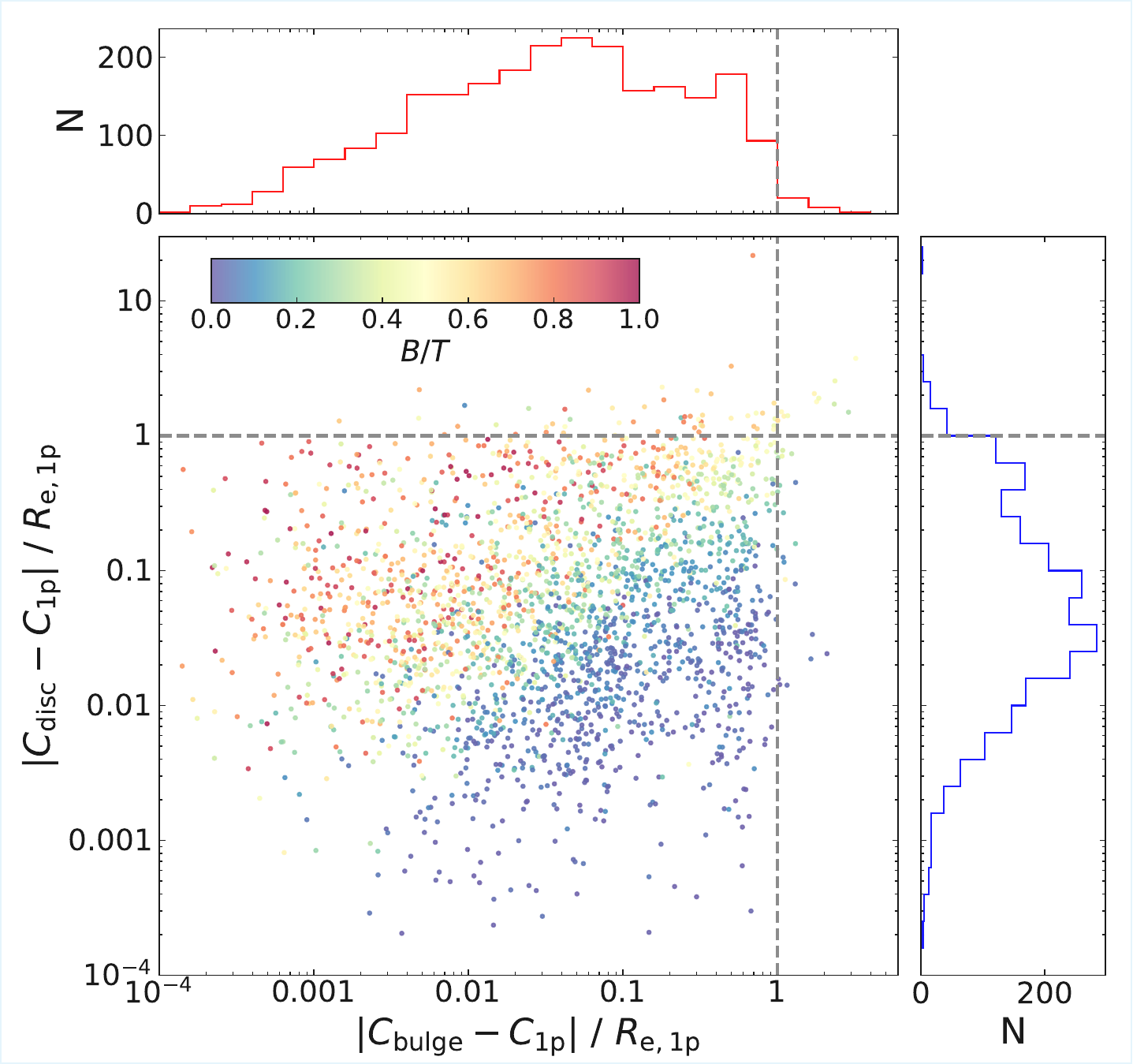}
\caption{Distances between the centres of the single-S\'ersic profile and those of the disc and bulge components, normalised by the effective radius of the single-S\'ersic component $R_\mathrm{e,1p}$ (left), and distances between the bulge and disc centres, normalised by the bulge and disc effective radii (right). Points are  colour-coded by $B/T$(\IE) of the bulge-disc decomposition for the $2445$ ERO-Perseus field galaxies with $\IE\le21$. As expected, in galaxies with a dominant bulge or disc, the brighter component is also the closer to the centre of the single-S\'ersic profile.}
\label{position-offset-BD-1p}
\end{figure}

Because we did not apply a constraint on the relative centring of the two components of the bulge-disc decomposition nor on the centring of the components or the single-S\'ersic profile in different bands, any offset between the centring of the components is indicative of some anomaly in the fits or in the object (atypical galaxies), which may allow one to detect unreliable fitted profiles. Figure \ref{position-offset-BD-1p} shows the distance between the position of the centre of the single-S\'ersic profile, denoted $C_\mathrm{1p}$, and those of the bulge or disc, denoted $C_\mathrm{B}$ and $C_\mathrm{D}$, respectively. In both cases, the distance is normalised by the effective radius of the single-S\'ersic profile $R_\mathrm{e,1p}$. The histograms along each axis are shown, and the colour of the points represent $B/T$ of the bulge-disc decomposition. A majority of the galaxies ($97\%$) have their bulges and discs centred consistently with the single-S\'ersic component within $10^{-4}$ to 1 effective radius, with peaks at $0.05 R_\mathrm{e,1p}$ and $0.03 R_\mathrm{e,1p}$. There are only 72 galaxies that either have a bulge or disc centre offset larger than $R_\mathrm{e,1p}$ (see fractions below). There is a clear trend that distances to the bulge centre are smaller than to the disc centre when the galaxy is bulge dominated ($B/T > 0.5$), and vice versa.

Four areas are delimited in \fg\ref{position-offset-BD-1p} by the grey dashed-lines. The bottom-left quadrant corresponds to single-S\'ersic profiles whose centre aligns closely to those of both the bulge and disc ($2363$ out of $2445$ galaxies, \ie $96.6\%$). The top-right quadrant indicates a single-S\'ersic profile offset larger than the single-S\'ersic $R_\mathrm{e}$ from both the bulge and disc centres (13 galaxies, $0.5\%$), hence one of the two types of fits can be considered as spurious. The top-left (50 galaxies, $2.0\%$) and bottom-right (19 galaxies, $0.8\%$) quadrants, on the other hand, show an offset of the single-S\'ersic centre from that of the disc (or bulge) but not from the bulge (disc, respectively), meaning that the bulge and disc centres differ and that the single S\'ersic centre is aligned with either one (predominantly the brighter component). 

One can also directly examine the offset between the bulge and disc centres. Visual inspection of the images in the \IE band was performed on the 49  ($2\%$ of the $\IE<21$ sample) and 225 ($9\%$) galaxies with bulge-disc offsets larger than 1 and 0.5 $R_\mathrm{e,disc}$, respectively. This revealed various configurations in which the light profile of the galaxy is well fit by the sum of a S\'ersic (bulge) and exponential (disc) profiles, but these have a significant offset in their centres, hence can be considered as adjacent: edge-on discs with no visible bulge but dust or surface brightness irregularities (flocculence) creating two separate flux maxima along their length or width, respectively (see \fg\ref{fig:exple-8}); spiral galaxies with a bright bulge and a luminous secondary feature (arm, \ion{H}{ii} region), but a comparatively faint disc so the two profiles model a bulge and the secondary component, respectively (see \fg\ref{fig:exple-11}). Therefore, offsets larger than the disc $R_\mathrm{e}$ between the centres of the bulge and disc components are successful at indicating that the profile-fitting failed to model the physical bulge and disc. Visual examination showed that decreasing the offset threshold to a fraction of $R_\mathrm{e,disc}$ would eliminate more objects that have a spurious fits, but would also have the disadvantage of discarding asymmetric galaxies whose fitted parameters are usable (mentioned in \sct\ref{sct-quality-check}).

With the goal to decrease the fraction of outlier fits (that do not successfully model both the physical bulge and disc), and can be identified by offsets in their relative centrings, one could add some constraints to the current `free' configuration in the bulge and disc centring (see \sct\ref{sct-methodo}): one could either constrain the centres of both components to be less distant from each other by some margin or require that both of them be less distant than some fixed coordinates (\eg the isophotal centroid) by some other margin. Both types of prior would nevertheless require us to define the positional margin in terms of the object size, hence one of the three effective radii (bulge, disc, single-S\'ersic) or a combination of them (as well as possibly the object inclination, see \sct\ref{sct-quality-check}), which would require some more extensive tests. To circumvent these difficulties, one could consider the extreme case of a null margin, therefore defining a unique position for the centres of both the bulge and disc components. However, since most galaxies do not have their bulge perfectly centred on the disc (due to inherent asymmetries), not allowing for any flexibility in the bulge and disc centring would ultimately bias the other bulge and disc parameters. We tested this option of concentric bulge and disc on the ERO-Perseus VIS image (along with a common angle and axis ratio, see approach 3 in \sct\ref{sct-model-fitting}), and obtained $34.7\%$ of spurious fits for which the bulge $R_\mathrm{e}$ is larger than the disc $R_\mathrm{e}$ by factor of 2, compared to a more likely fraction of $4.1\%$ in the `free' configuration, for galaxies with $\IE\le21$ (see \sct\ref{sct-check-Re}).

\subsection{S\'ersic indices \label{sct-check-n}}

In single-S\'ersic fits, the S\'ersic index $n$ of a galaxy depends on its morphology, with the specific cases of $n \sim 1$ (\ie exponential discs) derived for late-type galaxies and $n\sim 4$ (the de Vaucouleurs profile) derived for elliptical galaxies \citep{De-Vaucouleurs-1948-r14}. In between these cases, galaxy morphologies are better defined by their $B/T$, and \citet{Quilley-2022-bimodality} confirmed with visual morphological types that $B/T$ traces the Hubble sequence with some intrinsic dispersion. But simulated images show that the measured value of $B/T$ can differ significantly from the true values \citep{Haussler-2022-separating-spheroid-disk} in particular for $B/T\lesssim0.2$ and $B/T\gtrsim0.8$. Using prior fits of the bulges to ensure their physical identification in the bulge-disc decomposition and limit the degeneracies in $B/T$, \cite{Quilley-2023-scaling-bulges-and-disks} showed that the bulge S\'ersic index $n_\mathrm{B}$ and $B/T$ vary continuously along the Hubble sequence, as well as with the bulge effective radius. The location of bulges along the size-luminosity and size-surface brightness \citep{Kormendy-1977-II-kormendy-relation} planes also depends on $B/T$ and $n_\mathrm{B}$ \citep{Quilley-2023-scaling-bulges-and-disks}. This analysis also showed that there is a continuous sequence from pseudo-bulges (with small $R_\mathrm{e}$, $B/T$, and $n_\mathrm{B}$) hosted by late-type spiral galaxies to classical bulges (with large $R_\mathrm{e}$, $B/T$ and $n_\mathrm{B}$) hosted by early-type spirals and lenticulars.

Within this context, we analyse in \fg\ref{nsersic-nsersicB-BT-3d-space} the 3D parameter space $n$, $n_\mathrm{B}$, and $B/T$ for galaxies with $\IE\le21$ by plotting every pair of parameters with the third as a colour map. The goal was to examine how these three parameters may describe galaxy morphologies. In the top panel of \fg\ref{nsersic-nsersicB-BT-3d-space}, galaxies in the ERO-Perseus field display a correlation between the single-S\'ersic index $n$ and $B/T$ for the extreme values: There are very few or no objects with a very large $B/T$ and a small $n$ or the inverse. For intermediate values of $B/T$ (in the 0.1 to 0.8 interval), there is a weak diagonal excess of points, but the distribution of $n$ values for a given $B/T$ remains wide, with wings up to the $0.5$ and $10$ boundaries. To systematically check models that do not lie along this diagonal trend, we empirically defined the dashed black lines to exclude a majority of visually identified spurious fits. Both lines have a slope of 0.8 and intercepts of 0.5 and 2.5, respectively. This gives a total width along the vertical axis of $0.7$ dex (for comparison, the \rms dispersion around a fitted line is $0.26$ dex). The bottom threshold in $n$ as a function of $B/T$ only applies to $B/T > 0.25$, to keep disc-dominated galaxies even with very low $n<1$. This leads to the identification of 253 and 146 outliers (that is $10.3\%$ and $6.0\%$ of the $\IE\le21$ sample), above and below the delineated region. If we compare these 399 outliers to the 218 identified in \fg\ref{all-radii-related-cmap-BT} using the effective radii (see \sct\ref{sct-check-Re}), 88 galaxies are found in common, leaving 529 sources with inconsistencies between the bulge-disc decomposition and the single-S\'ersic models.

\begin{figure}
\centering
\includegraphics[width=0.85\columnwidth]{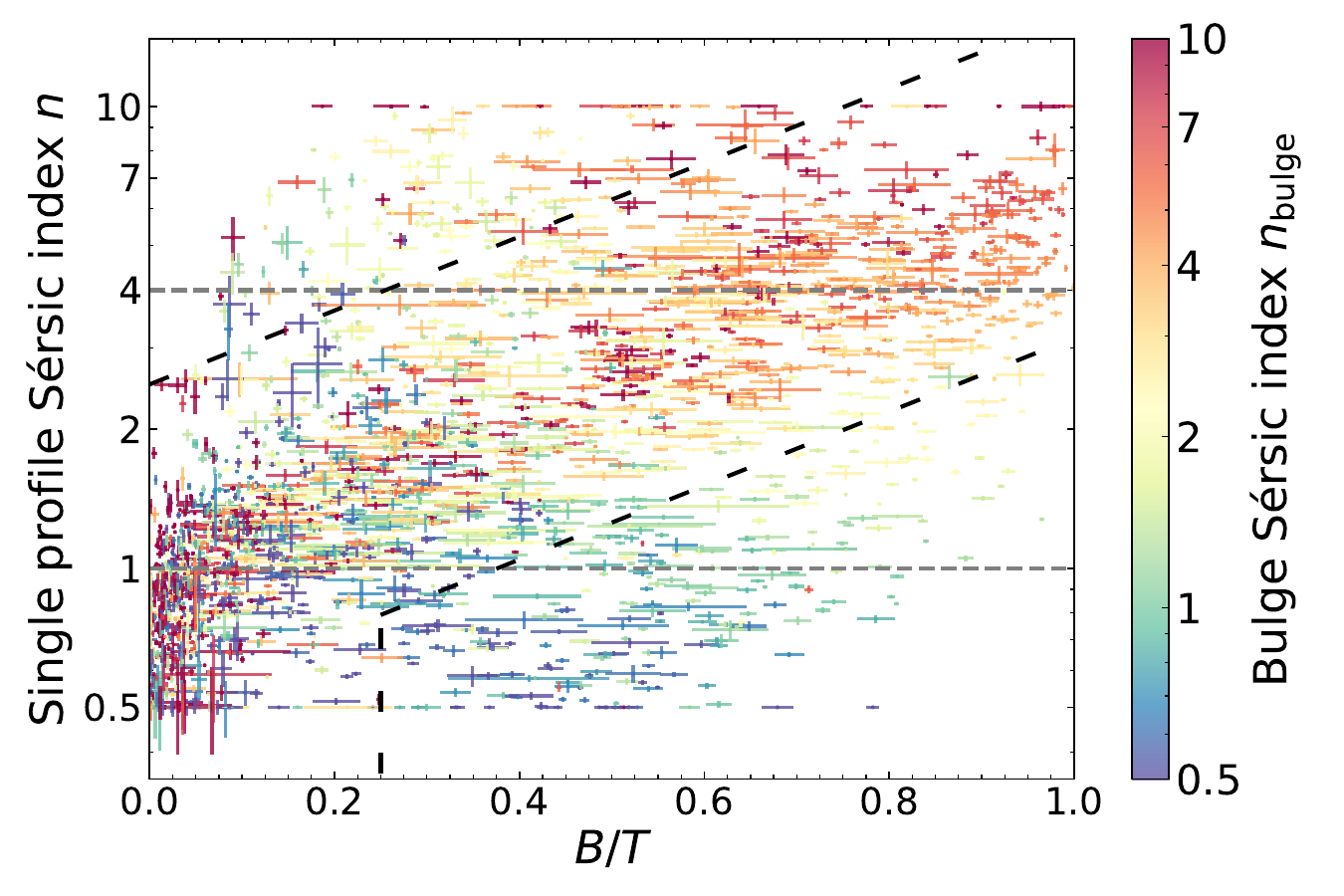}
\includegraphics[width=0.85\columnwidth]{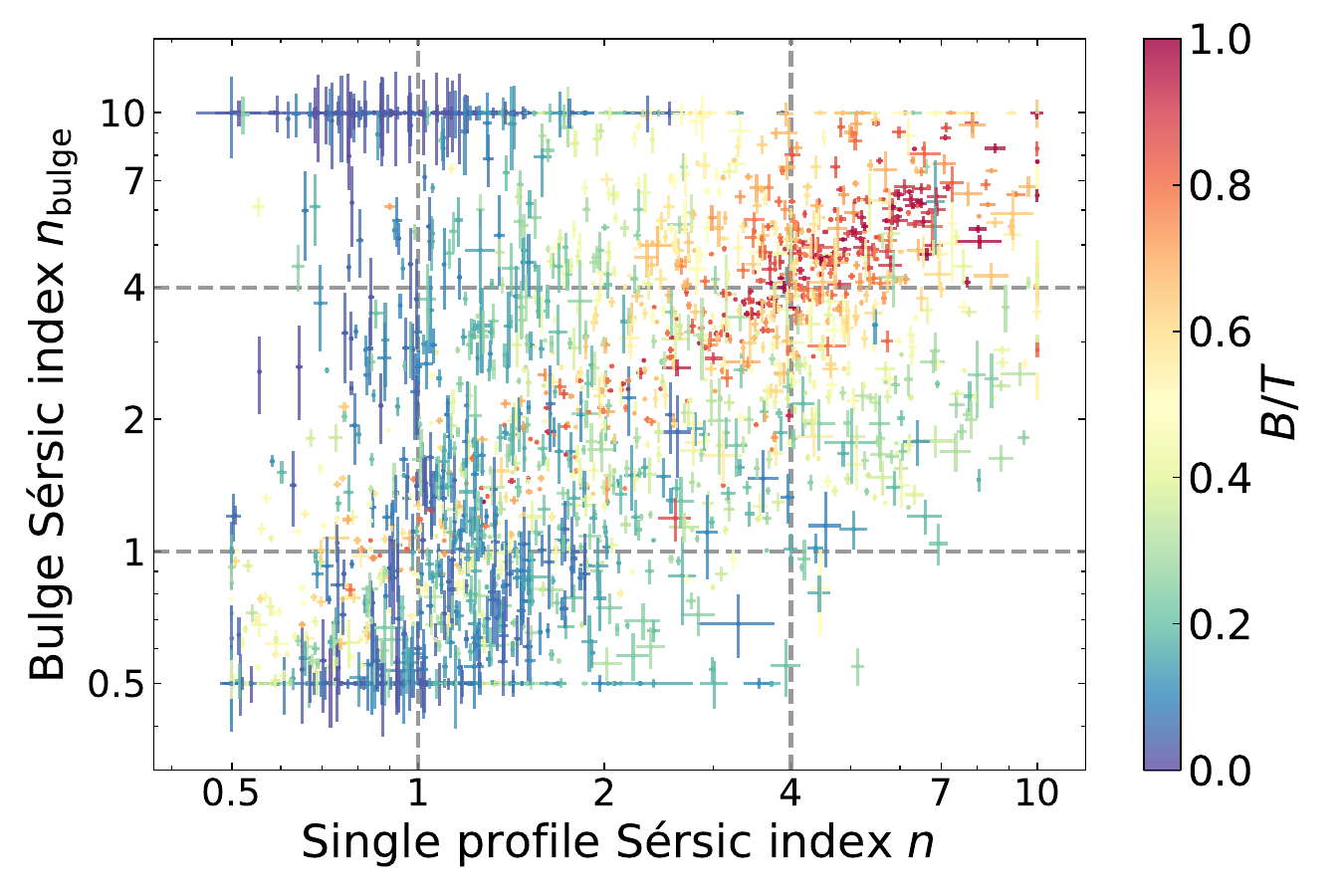}
\includegraphics[width=0.85\columnwidth]{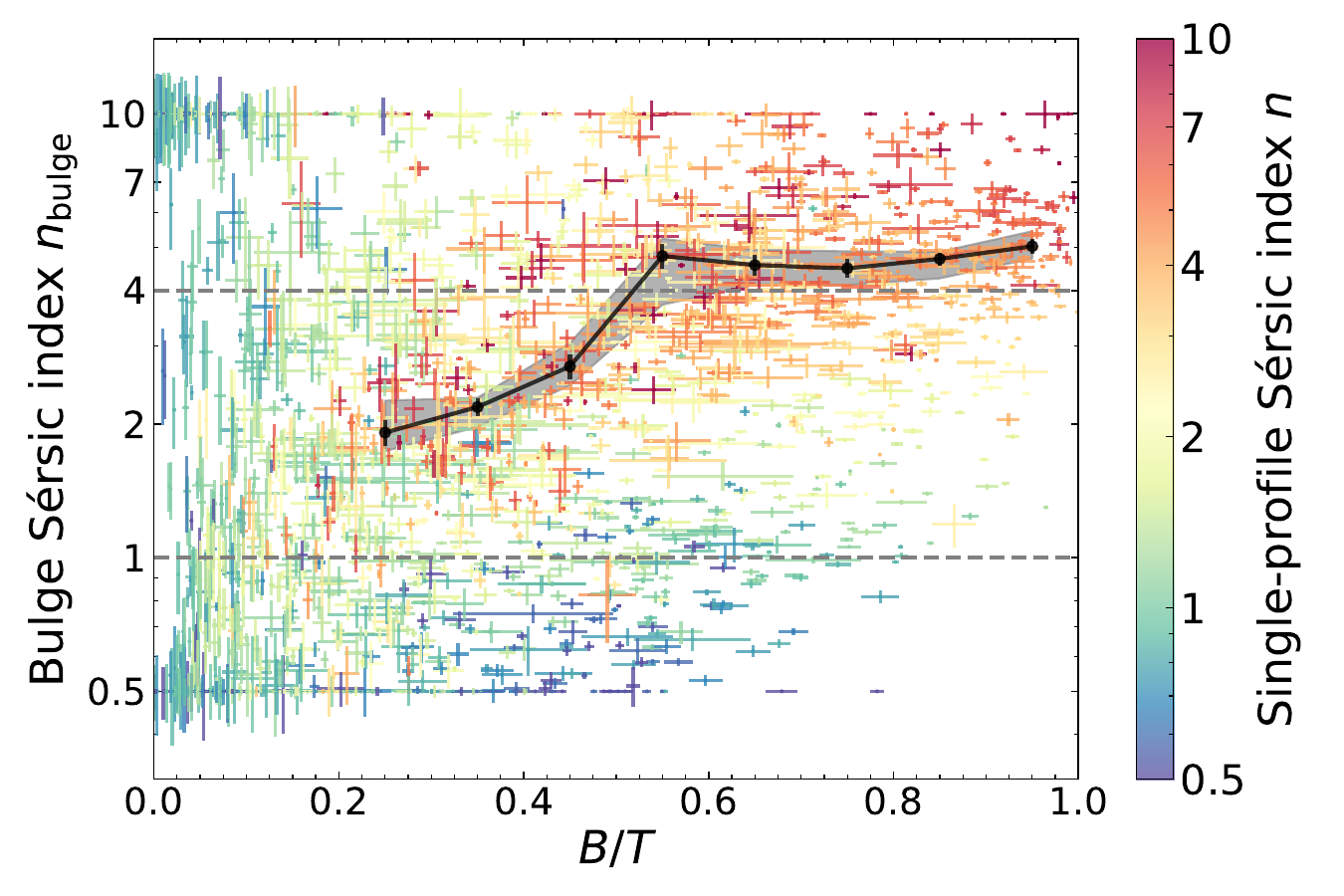}
\caption{Relations between the S\'ersic index $n$ from the single-S\'ersic modelling, the bulge-to-total light ratio $B/T$(\IE), and the bulge S\'ersic index $n_\mathrm{B}$ from the bulge-disc decomposition for the $2445$ galaxies with $\IE\le21$. Each plot shows one projection of this 3D parameter space and incorporates the third parameter as a colour map on the points. The inclined dashed black lines in the top panel delimit above and below the visually inspected samples (see text). Once spurious fits and bulges that are too small or too faint to be modelled were identified and discarded, an overall correlation between $n$, $B/T$, and $n_\mathrm{B}$ remained, which all increase jointly as galaxy types change from late to earlier types along the Hubble sequence and as their bulges grow from small pseudo-bulges to more prominent classical bulges. In this latter regime, the increase in $n_\mathrm{B}$ with $B/T$ flattens, as shown by the solid black line displaying the median values and bootstrap uncertainties of $n_\mathrm{bulge}$ in $B/T$ intervals of $0.1$ in width, from $0.2$ to $1.0$ (the grey shaded area shows the 10--90\% percentile range around these values)}
\label{nsersic-nsersicB-BT-3d-space}
\end{figure}

Visual inspection of these selected objects indicated that those appearing below the bottom dashed black line in the top panel of \fg\ref{nsersic-nsersicB-BT-3d-space} are not properly modelled as the sum of a bulge and a disc. Rather, they are modelled as two non-concentric light profiles that do not correspond to the physical bulge or disc (see \sct\ref{sct-check-position}). Some of these objects are distorted asymmetric galaxies in which the S\'ersic profile models the bulge and its immediate surroundings in the disc, whereas the exponential profile models a sub-region of the asymmetric part of the galaxy disc further from the bulge (see ID 18b in \fg\ref{fig:exple-15}). This inclusion of disc light in the profile designed for the bulge leads to both marked overestimation and underestimation of $B/T$ and $n_\mathrm{B}$, respectively (see \tab\ref{tab-values-model-exples}). Another type of object identified below the bottom dashed black line in the top panel of \fg\ref{nsersic-nsersicB-BT-3d-space} are edge-on or highly inclined galaxies with no visible bulge. These are modelled by two adjacent light profiles along the disc, therefore leading to an intermediate value of $B/T$ and a bulge profile close to exponential ($n_\mathrm{B} \approx 1$), whereas the single-S\'ersic profile appears centred on the galaxy and its index $n$ is correctly estimated (see \fgs\ref{fig:exple-8}, \ref{fig:exple-11}, and \fg\ref{fig:exple-12}; these galaxies were already identified and discussed in \sct\ref{sct-check-position}). Visual examination of the fits above the top dashed black line indicated that a majority of these galaxies have neighbouring stars or galaxies that \texttt{SourceXtractor++} failed to segment out from the galaxy under study, hence including it as part of the disc flux in the profile fitting. These contaminations lead to an overestimation of the modelled disc flux and hence an underestimation of $B/T$.

The middle panel of \fg\ref{nsersic-nsersicB-BT-3d-space} shows a linear sequence of $n_\mathrm{B} \approx n$ for galaxies with $B/T \geq 0.7$ as for these galaxies the bulge component dominates, hence its parameters are close to those for the single-S\'ersic fit. 
There are also 703 galaxies with $B/T \leq 0.1$ and $n\lesssim2$ for which $n_\mathrm{B}$ can take all possible values in the entire $[0.5,10]$ range, with larger uncertainties than for other points. For these objects, both $B/T$ and $n$ indicate that the modelled galaxies are late types, and therefore their bulges are faint and hence difficult to model. This may lead to incorrect, hence more dispersed values for the bulge parameters (in the top-left corner of the graph -- for $n_\mathrm{B} > 4$ and $n < 2$ -- these objects include the fits of contaminating stars or \ion{H}{ii} regions in very weak or bulge-less late-type disc galaxies, which are seen as red points in the bottom-left end of the diagonal in the top panel of \fg\ref{nsersic-nsersicB-BT-3d-space}).

The bottom panel of \fg\ref{nsersic-nsersicB-BT-3d-space} shows the third projection of the $n$--$n_\mathrm{B}$--$B/T$ parameter space. The outliers identified in the description of the previous panels are also visible here. Regarding the remaining data points, that is the red to orange points defined by $n\ge2.5$, the relation between $\logten(n_\mathrm{B})$ and $B/T$ flattens for $B/T>0.5$ (as confirmed by the median values for $n\ge2.5$ displayed as a black line), after a steep increase from 2.7 to 4.8 in $n_\mathrm{B}$ (the curve is only plotted for $B/T\ge0.25$, by lack of $n\ge2.5$ points below).This contrasts with both the $\logten(n)$--$B/T$ and $\logten(n_\mathrm{bulge})$--$\logten(n)$ relations appearing linear when removing outliers.

The continuous variations of bulge properties measured in \cite{Quilley-2023-scaling-bulges-and-disks} and summarised at the beginning of this subsection, led them to interpret the  increase of $n_\mathrm{B}$ with $B/T$ as the transition from pseudo-bulges to classical bulges. Pseudo-bulges with typically $n_\mathrm{B} \lesssim 2$ \citep{Kormendy-Kennicutt-2004-bulges-disk-galaxies-review, Fisher-Drory-2008-bulges-n-sersic} are likely to be the yellow and green points in the bottom panel of \fg\ref{nsersic-nsersicB-BT-3d-space}, representing the continuation of the $n_\mathrm{B}$--$B/T$ trend at $B/T\lesssim0.2$, whereas the orange points in the $0.2 \lesssim B/T \lesssim 0.5$ interval are interpreted as intermediate bulges between pseudo and classical \citep{Mendez-Abreu-2014-composite-bulges, Erwin-2015-composite-bulges, Breda-2023-continuous-seq-bulges, Quilley-2023-scaling-bulges-and-disks}. Still, the change from mostly rotation-supported pseudo-bulges to more pressure-supported classical bulges leads to a significant increase in $n_\mathrm{B}$ with $B/T$. We suggest here that the observed flattening of the $n_\mathrm{B}$--$B/T$ relation in the bottom panel of \fg\ref{nsersic-nsersicB-BT-3d-space} occurs for spheroidal bulges that are dynamically similar to elliptical galaxies found at higher $B/T$, inducing a rather stable $n_\mathrm{B}$ for $B/T\gtrsim0.5$. 

Altogether, the three projections in \fg\ref{nsersic-nsersicB-BT-3d-space} show that once spurious fits are removed, $B/T$, $n$, and $n_\mathrm{B}$ are complementary indicators of the morphology of galaxies, with rough correlations between them (the correlation between $n$ and $B/T$ is further explored in \sct\ref{sct-model-photo} and \fg\ref{mag-diff-BD-1p-explain}). Indeed, a value of $B/T$ corresponds to wide intervals in $n$ and $n_\mathrm{B}$ ($\approx 0.3$ dex). Similarly, the top panel of \fg\ref{nsersic-nsersicB-BT-3d-space} show that splitting early and late morphological types of galaxies at $n=2.5$, as often performed when there is no other indicator of morphology \citep{Barden-2005-GEMS, Vulcani-2014-GAMA-color-gradients}, yields varying fractions of galaxies with $B/T$ in almost the full 0 to 1 range. This analysis demonstrates that the single-S\'ersic index is only an approximate indicator of galaxy types, more in-depth characterisation of morphology requiring bulge-disc decomposition.

\subsection{Axis ratios \label{sct-check-elong}}

The axis ratios of disc galaxies projected onto the sky provides an estimate of their inclination angle, if one assumes that they are circular and takes into account their scale height \citep{Hubble-1926-extragalactic-nebulae, Giovanelli-1994-extinction-opacity-elongation-disk-Sc}, which is necessary for analyses of spin alignments with the cosmic web for example (see \sct\ref{sct-check-shear-spin}). Spiral galaxies were nevertheless shown to have a face-on ellipticity in the $\sim0.1$--$0.2$ interval \citep{Padilla-2008-shape-galaxies-SDSS,Unterborn-2008-inclination-extinction-disks-SDSS,Rodriguez-2013-shape-galaxies-SDSS-Zoo}, which should be taken into account to determine their spatial orientation (complemented by their position angles, see \sct\ref{sct-check-angle}). The apparent axis ratios of elliptical galaxies were extensively studied and result from the projection of oblate spheroids with a wider range of axis ratios \citep{Padilla-2008-shape-galaxies-SDSS,Rodriguez-2013-shape-galaxies-SDSS-Zoo}. In contrast, the face-on elongation of the discs of lenticular galaxies have been poorly explored. Regardless of the intrinsic shape of all these types of galaxy, accurate measurement of their apparent axis ratios are crucial for the cosmic shear measurements of the \Euclid mission (see \sct\ref{sct-check-shear-spin}). Therefore, this parameter of their modelling is examined in this subsection.

In \fg\ref{aspect-ratio-vs}, we compare the measured axis ratio for the galaxies modelled as single-S\'ersic profiles with that for the disc derived from the two-component decomposition, with colour-coding by the value of $B/T$ in the decomposition. Galaxies with $B/T\lesssim0.1$ gather along the diagonal, indicating, as expected, similar axis ratios for both models, whereas galaxies with larger $B/T$ ($\gtrsim0.6$) have a distribution that is more dispersed and skewed above the diagonal. Actually, 196 out of 215 galaxies ($91.2\%$) with $B/T\geq0.8$ are above the diagonal with a ratio of $(b/a)_\mathrm{1p}/(b/a)_\mathrm{disc}$ mostly in the 1.5 to 5 interval (see \fg\ref{aspect-ratio-bias} below). For the small number of models with a significantly more elongated single S\'ersic fit than the disc component, hence below the identity line (dashed grey line), that is $b/a_\mathrm{1p} < (b/a)_\mathrm{disc} - 0.1$, either the disc model does not represent a physical component (galaxy with high $B/T$ or failed decomposition) or the galaxy is a barred spiral seen close to face-on, for which the single-S\'ersic fit mostly models the bar, therefore leading to an underestimated $b/a_\mathrm{1p}$ (see \fg\ref{fig:exple-5}). This graph illustrates how the single-S\'ersic model would provide a biased estimate of the disc axis ratio, hence their inclination and that this bias is larger for more prominent bulges (larger $B/T$).

\begin{figure}
\includegraphics[width=\columnwidth]{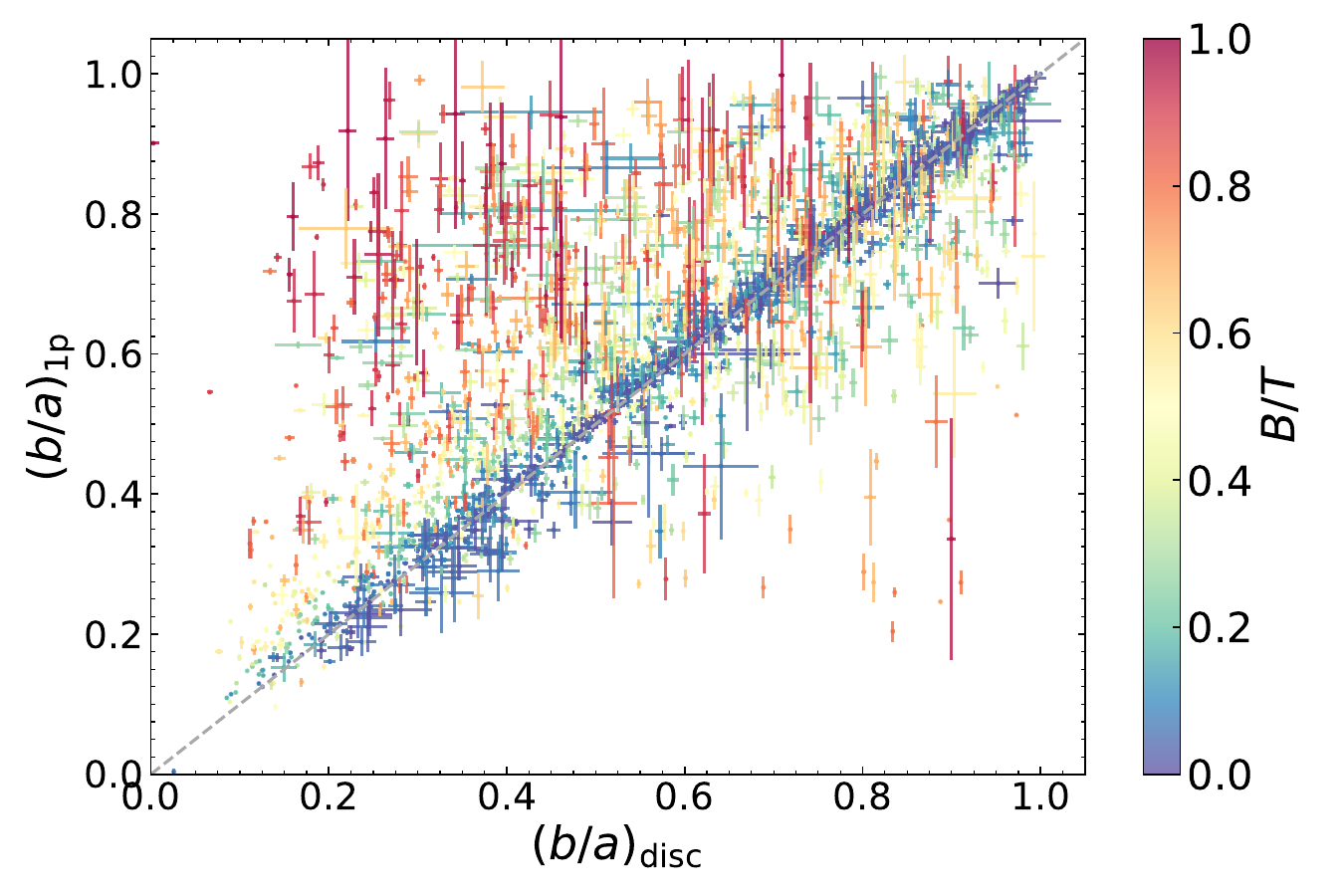}
\caption{Comparison of the axis ratio for the $2445$ galaxies with $\IE\le21$ modelled as single-S\'ersic profiles with that of the disc derived from the bulge-disc decomposition, colour-coded by the overall morphologies of galaxies, parametrised with $B/T$(\IE). This graph shows agreement in the axis ratio of the single-S\'ersic profile and the disc component for low $B/T$ galaxies. The axis ratio is significantly larger for the single-S\'ersic profile than for the disc component for galaxies with prominent bulges, therefore providing an underestimation of their disc inclinations.}
\label{aspect-ratio-vs}
\end{figure}

\begin{figure*}
\includegraphics[width=0.5\textwidth]{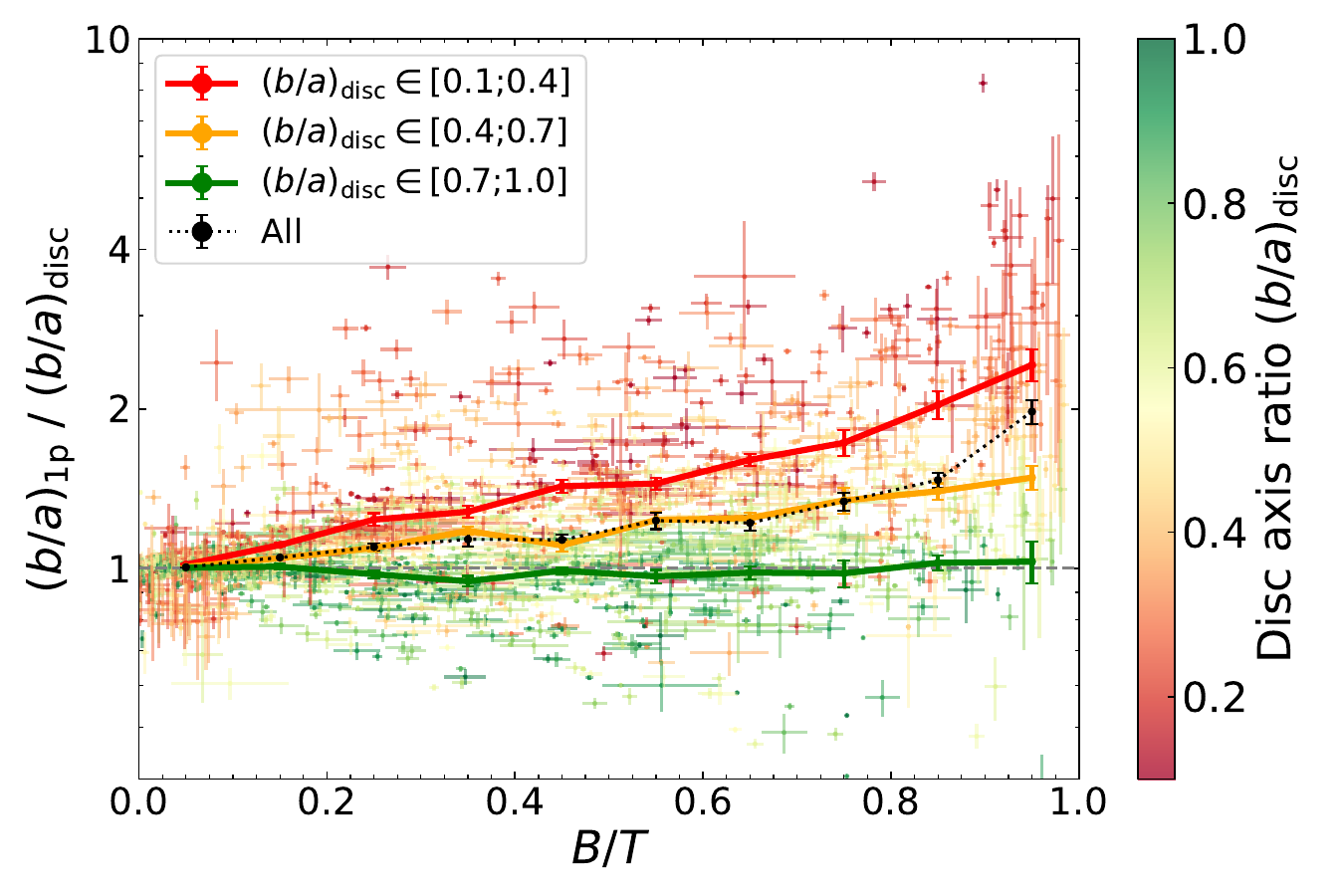}
\includegraphics[width=0.5\textwidth]{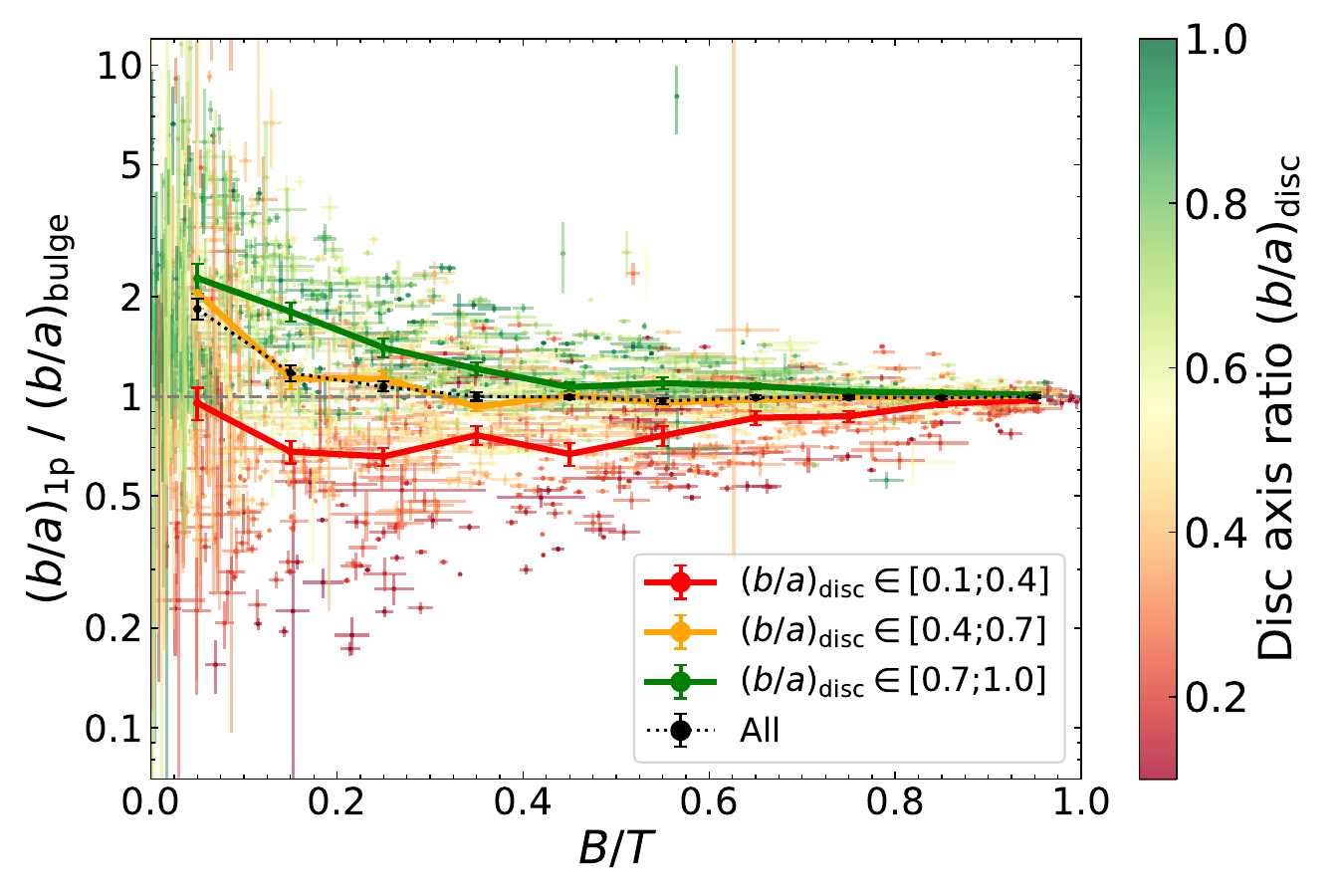}
\caption{Ratio of the axis ratio of the single-S\'ersic profile to that of the disc (left panel) or bulge (right panel) in the bulge-disc decomposition for the $2445$ galaxies with $\IE\le21$ as a function of $B/T$, colour-coded by the axis ratio of the disc. These graphs show biases in estimating the galaxy disc or bulge axis ratio using the single-S\'ersic profile axis ratio, which vary both with $B/T$ and with the axis ratio of the other component. The left panel shows that the single-S\'ersic axis ratio of galaxies with more prominent bulges tends to overestimate that of the disc and hence underestimates its inclination. In contrast, the single-S\'ersic axis ratio either underestimates the bulge axis ratios for weakly inclined discs or overestimates them for highly inclined galaxies, with a vanishing bias as the bulge becomes dominant. These biases should be taken into account in the cosmic shear and spin alignment measurements. Statistics for all curves are listed in \tab\ref{tab-ratio-asp-ratio}.}
\label{aspect-ratio-bias}
\end{figure*}

\begin{table*}[ht]
\caption{Median values of the ratios of axis ratios plotted in \fg\ref{aspect-ratio-bias} and the sample sizes over which these medians are computed.}
\resizebox{\linewidth}{!}{%
\begin{tabular}{ c c c c c c c c c c c c c }
\hline\hline
\multirow{2}{*}{} & $(b/a)_\mathrm{disc}$ & \multicolumn{10}{c}{$B/T$ interval} \\
& interval & $[0,0.1]$ & $[0.1,0.2]$ & $[0.2,0.3]$ & $[0.3,0.4]$ & $[0.4,0.5]$ & $[0.5,0.6]$ & $[0.6,0.7]$ & $[0.7,0.8]$ & $[0.8,0.9]$ & $[0.9,1.0]$ \\
\hline
\multirow{4}{*}{${N_\mathrm{gal}}^{(a)}$} & $[0.1,0.4]$ & 157 & 79 & 83 & 55 & 65 & 68 & 59 & 50 & 52 & 53 \\
& $[0.4,0.7]$ & 279 & 126 & 97 & 108 & 73 & 81 & 70 & 55 & 42 & 30 \\
& $[0.7,1.0]$ & 286 & 87 & 76 & 57 & 64 & 53 & 53 & 37 & 20 & 9 \\
& All & 722 & 294 & 257 & 221 & 205 & 203 & 182 & 142 & 115 & 92 \\

\multirow{4}{*}{ $\frac{(b/a)_\mathrm{1p}}{(b/a)_\mathrm{disc}}^{(b)}$ } & $ [0.1,0.4]$ & $1.02\pm0.00$ & $1.11\pm0.01$ & $1.24\pm0.03$ & $1.28\pm0.03$ & $1.43\pm0.04$ & $1.45\pm0.04$ & $1.60\pm0.04$ & $1.73\pm0.1$ & $2.03\pm0.12$ & $2.42\pm0.16$ \\
& $ [0.4,0.7]$ & $1.01\pm0.00$ & $1.05\pm0.01$ & $1.09\pm0.02$ & $1.18\pm0.02$ & $1.10\pm0.03$ & $1.23\pm0.04$ & $1.24\pm0.03$ & $1.34\pm0.08$ & $1.40\pm0.05$ & $1.48\pm0.08$ \\
& $ [0.7,1.0]$ & $1.00\pm0.00$ & $1.01\pm0.01$ & $0.97\pm0.02$ & $0.95\pm0.02$ & $0.99\pm0.02$ & $0.97\pm0.03$ & $0.98\pm0.03$ & $0.98\pm0.06$ & $1.02\pm0.03$ & $1.03\pm0.09$ \\
& All & $1.00\pm0.00$ & $1.05\pm0.01$ & $1.10\pm0.02$ & $1.13\pm0.03$ & $1.13\pm0.03$ & $1.23\pm0.04$ & $1.22\pm0.04$ & $1.34\pm0.05$ & $1.47\pm0.05$ & $1.98\pm0.10$ \\

\multirow{4}{*}{ $\frac{(b/a)_\mathrm{1p}}{(b/a)_\mathrm{bulge}}^{(b)}$ } & $ [0.1,0.4]$ & $0.95\pm0.11$ & $0.68\pm0.05$ & $0.66\pm0.04$ & $0.76\pm0.05$ & $0.67\pm0.05$ & $0.76\pm0.05$ & $0.86\pm0.04$ & $0.87\pm0.03$ & $0.95\pm0.02$ & $0.97\pm0.02$ \\
& $ [0.4,0.7]$ & $2.05\pm0.22$ & $1.13\pm0.08$ & $1.13\pm0.06$ & $0.93\pm0.03$ & $1.01\pm0.02$ & $0.96\pm0.03$ & $0.98\pm0.02$ & $1.00\pm0.02$ & $0.99\pm0.02$ & $1.01\pm0.01$ \\
& $ [0.7,1.0]$ & $2.27\pm0.24$ & $1.80\pm0.12$ & $1.40\pm0.09$ & $1.21\pm0.05$ & $1.07\pm0.03$ & $1.10\pm0.04$ & $1.07\pm0.02$ & $1.03\pm0.02$ & $1.03\pm0.01$ & $1.01\pm0.02$ \\
& All & $1.84\pm0.14$ & $1.18\pm0.06$ & $1.07\pm0.04$ & $1.00\pm0.03$ & $0.99\pm0.01$ & $0.97\pm0.02$ & $0.99\pm0.01$ & $0.99\pm0.01$ & $0.99\pm0.01$ & $1.00\pm0.01$ \\
\hline
\vspace{1pt}
\end{tabular}}
\small\textbf{Notes.}
$^{(a)}$ Number of galaxies in the considered sample.
$^{(b)}$ Median values with bootstrap errors. 
\label{tab-ratio-asp-ratio}
\end{table*}

In \fg\ref{aspect-ratio-bias}, we directly examine the ratio of the single-S\'ersic axis ratio to that of the disc (left panel), and of the bulge (right panel), as a function of $B/T$. The left panel shows that the single-S\'ersic profiles have mostly larger axis ratios than the corresponding disc components, hence appearing less elongated than the disc, due to the presence of the bulge. For $B/T < 0.1$, the profiles are equally elongated, with an axis ratio of $ 1.005 \pm 0.009$ and a dispersion of $0.047$ dex. For $B/T>0.1$, the dispersion in the ratio of single-S\'ersic to disc axis ratios increases to $0.243$ dex. The colour-coding of the points and the three median curves over indicated intervals, shows that this increasing dispersion results from systematic variations in $(b/a)_\mathrm{1p}/(b/a)_\mathrm{disc}$ with the disc axis ratio. For face-on or weakly inclined discs (in green), the single-S\'ersic model is expected to be as round or elongated as the model disc component, except if spiral arms or a bar bias any of the two types of fit (this may cause the dispersion around the median curve). While discs get more inclined and hence more elongated (yellow, orange and red points), the ratio of single-S\'ersic to disc axis ratio increases for decreasing disc axis ratio, reaching values up to 2--3, with a stronger effect for larger values of $B/T$. The median curves for three intervals of $(b/a)_\mathrm{disc}$ illustrate both trends. Across the $0.1\le B/T\le0.9$ range that corresponds to intermediate and early spiral types as well as lenticulars (the types with the most prominent bulges), modelling non face-on galaxies with single-S\'ersic profiles would make the estimation of their disc axis ratio (hence disc inclination) biased by an amount that increases with both $B/T$ and the disc inclination itself. Even the average curve over all values of the disc axis ratio shows a significant overestimation by $10$ to $50$ \% for $B/T$ from $0.2$ to $0.9$. However, the bias values at $B/T \gtrsim 0.9$ must be taken with caution, since the parameters of such weak discs are poorly measured, hence the large uncertainties for these points in the graph. \tab\ref{tab-ratio-asp-ratio} show the various median values and associated uncertainties\footnote{Estimated as the standard error $\sigma/\sqrt{N}$, where $\sigma$ and $N$ are the standard deviation and number of galaxies of each sample, respectively} in the ratios of axis ratio between the single-S\'ersic profile and those of the bulge and disc components plotted in \fg\ref{aspect-ratio-bias}.

\begin{figure*}
\includegraphics[width=0.48\textwidth]{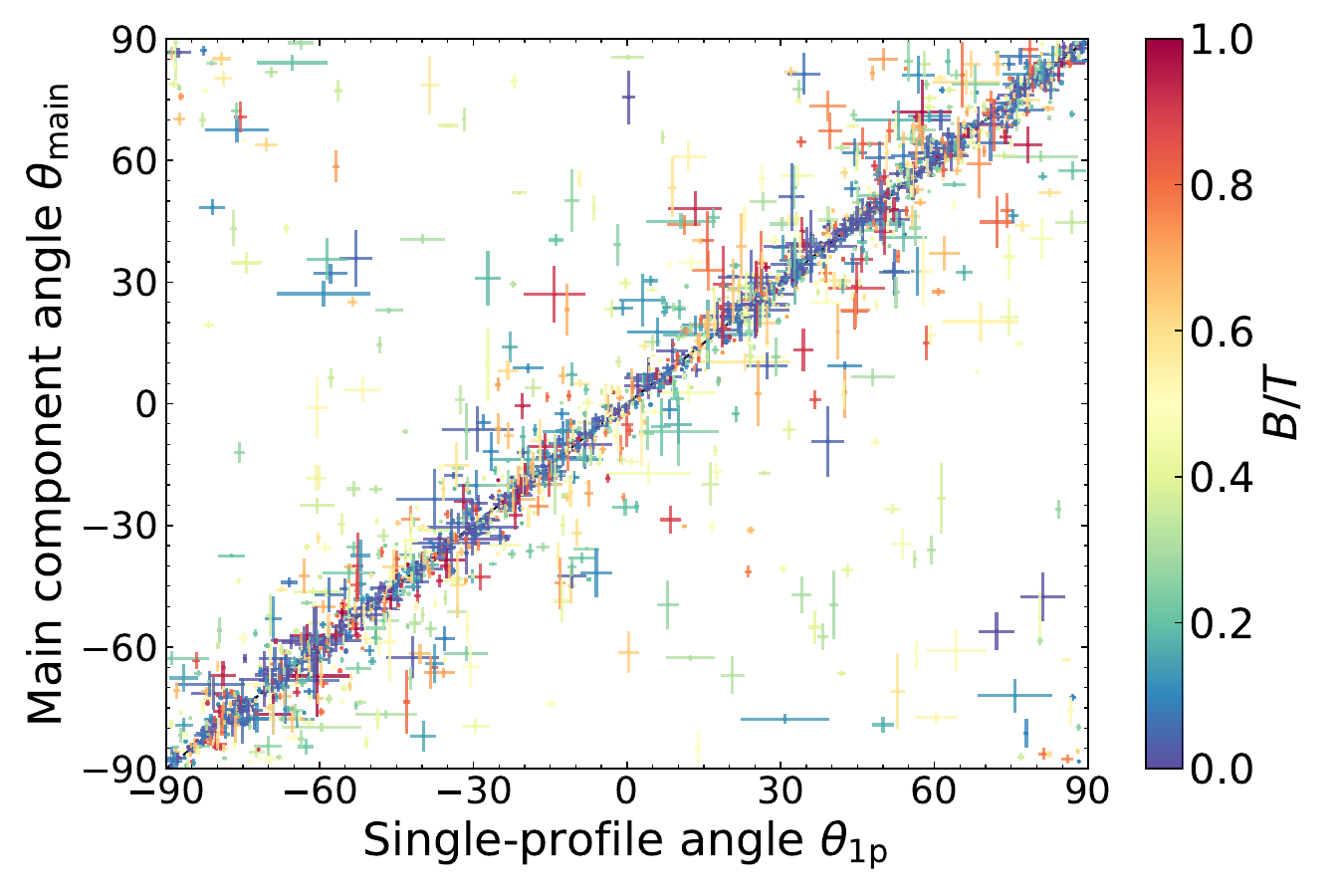}
\includegraphics[width=0.48\textwidth]{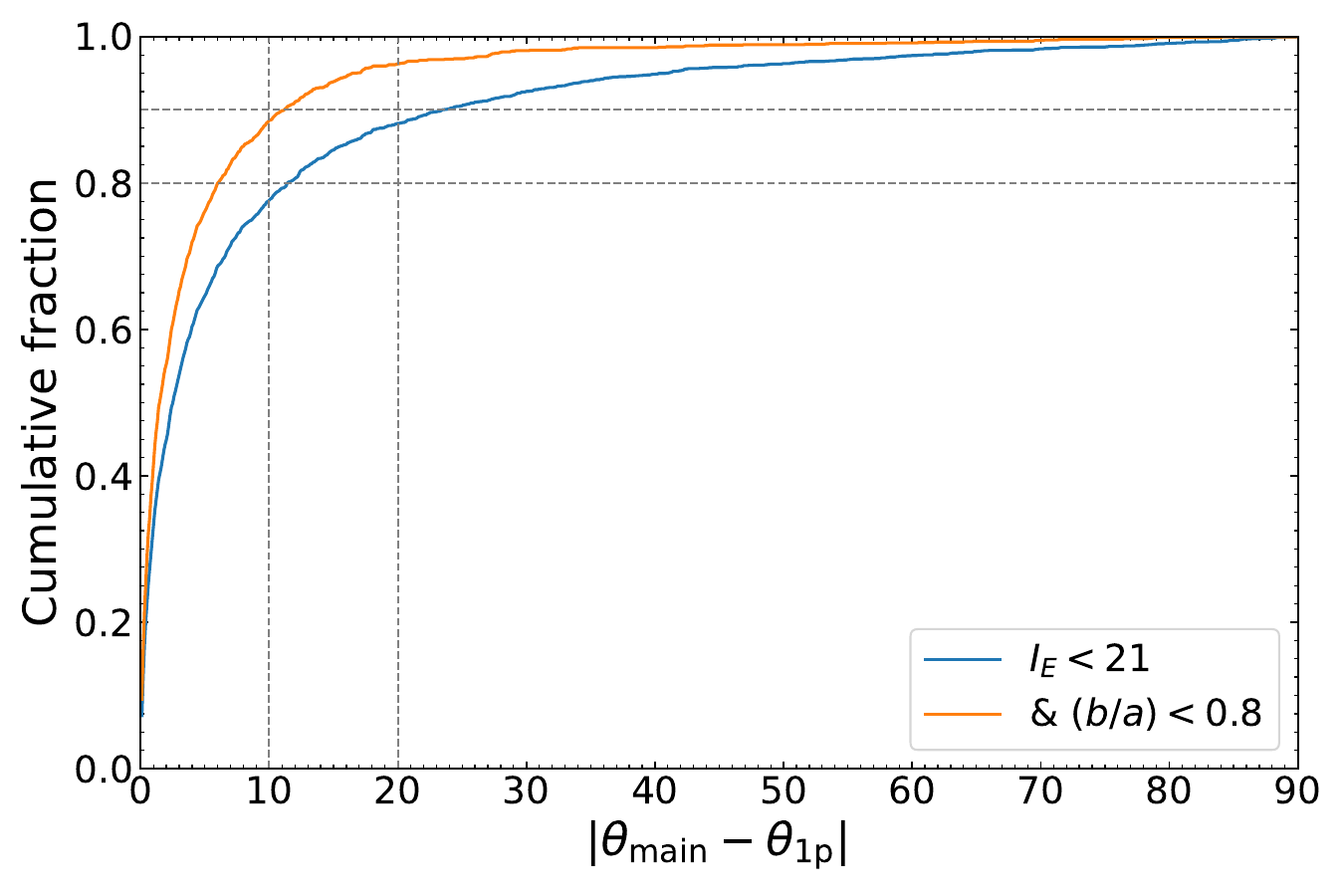}
\caption{Comparison between the position angle of the major axis of the single-S\'ersic profile $\theta_\mathrm{1p}$ and of the dominant profile in the bulge-disc decomposition $\theta_\mathrm{main}$ (see \eq\ref{eq-def-theta-main}) for the $2445$ galaxies with $\IE\le21$. The left panel compares both angles, with a $B/T$(\IE) colour map for the points, highlighting an agreement for most of the galaxies, as well as outliers reaching up to tens of degrees of angle difference. No trend with $B/T$ is noticeable. The right panel shows the cumulative fraction of galaxies for which the difference of $\theta_\mathrm{main} - \theta_\mathrm{1p}$ is less than the value of the threshold (in degrees on the $x$-axis), for all galaxies with $\IE\le21$ (in blue) and limited to $b/a<0.8$ for both models (in orange, corresponding to 1721 galaxies, hence $70.4\%$ of the sample), given that the position angle of a model with elliptical symmetry can take any value when fitted to a face-on galaxy. Vertical dashed lines indicate the $10^\circ$ and $20^\circ$ thresholds, while horizontal dashed lines show the $80\%$ and $90\%$ fractions. The right panel confirms quantitatively the agreement between the two position angles observed in the left panel, with $89\%$ and $96\%$ of non-face-on galaxies having a difference in their measured angles below $10^\circ$ and $20^\circ$, respectively.}
\label{angle-diff}
\end{figure*}

The right panel of \fg\ref{aspect-ratio-bias} shows the analogous graph to the left panel, for the ratio of the single-S\'ersic axis ratio to that of the bulge. For $B/T\lesssim0.1$, the bulge axis ratio is poorly measured, hence the large dispersion and uncertainties. For galaxies with prominent bulges ($B/T\gtrsim0.8$) the axis ratio of the single-S\'ersic ratio is a good estimate of that of the bulge, since it dominates the galaxy profile. For intermediate values of $B/T$ (hence significant discs components), there is a vertical gradient with the disc axis ratio, as indicated by the colour-coding of the points, with evidently a steadily decreasing dispersion for more prominent bulges. There is, however, a null average bias between the bulge and single-S\'ersic axis ratio as shown by the dotted line labelled `All', with a median ratio (black dashed line) being always distant from 1 by less than $1\sigma$. The null bias is also measured for galaxies with a disc axis ratio in the $0.4$--$0.7$ range (intermediate inclinations), for all $B/T\gtrsim0.1$. Nevertheless, the median curve for the galaxies with face-on or weakly inclined discs (in green) shows only a moderate overestimation (see \tab\ref{tab-ratio-asp-ratio}) of the single-S\'ersic axis ratio over that of the bulge: indeed, the isophotes of the bulge of these objects are blended into those of the disc. In contrast, the single-S\'ersic axis ratio underestimates by a larger amount (see \tab\ref{tab-ratio-asp-ratio}) the bulge axis ratio for highly inclined galaxies (in red), because of the inclined and hence elongated disc isophotes that the single-S\'ersic profile must take into account. The implications of these various biases for \Euclid's cosmic shear and spin alignment measurements are discussed in \sct\ref{sct-check-shear-spin}.

\subsection{Position angles \label{sct-check-angle}}

Here we examine the impact of the profile modelling method on the position angle of the major axes of galaxies projected onto the sky (PA hereafter), converted into degrees. To this end, we compared the single-S\'ersic profile PA $\theta_\mathrm{1p}$ to that of the main model component of the bulge-disc decomposition, $\theta_\mathrm{main}$, defined as
\begin{equation}
    \theta_\mathrm{main} =
\begin{cases}
    \theta_\mathrm{disc} & \text{if } B/T < 0.5,\\
    \theta_\mathrm{bulge} & \text{if } B/T \geq 0.5,
\end{cases}
\label{eq-def-theta-main}
\end{equation}
where $\theta_\mathrm{disc}$ and $\theta_\mathrm{bulge}$ are the disc and bulge PA, respectively. The left panel of \fg\ref{angle-diff} shows the direct comparison between $\theta_\mathrm{main}$ and $\theta_\mathrm{1p}$ for all galaxies in the sample. A large majority of points lie along the identity line, indicating that the measured angle for the bulge or disc component containing most of the light is similar to that for the single-S\'ersic fit (it is also the case for the points at the top-left and bottom-right corners). The $B/T$ colour map shows no correlation between the distance of the points from the diagonal and their $B/T$ value.

The tight correlation in the left panel of \fg\ref{angle-diff} is further quantified in the right panel, showing the cumulative fraction of galaxies for which the difference between $\theta_\mathrm{main}$ and $\theta_\mathrm{1p}$ is below a given value in the 0 to 90$^\circ$ interval. For the whole sample of galaxies with $\IE\le21$ (blue histogram), the difference is less than $10^\circ$ for $77.7\%$ of objects and less than $20^\circ$ for $88.1\%$ of them. Because for nearly circular profiles PA is poorly defined, we also examine the cumulative fraction for galaxies whose main component in the bulge-disc decomposition has an axis ratio $(b/a)_\mathrm{main}$ (defined similarly as $\theta_\mathrm{main}$ in \eq\ref{eq-def-theta-main}) below 0.8 (orange curve). For this sub-sample, the fraction of galaxies whose angle difference is below $10^\circ$ and $20^\circ$ rises up to $88.6\%$ and $96.3\%$, respectively. Visual examination of the \IE fits for the objects with $|\theta_\mathrm{main}-\theta_\mathrm{1p}|\ge20^\circ$ shows some spurious fits, such as discs encompassing some flux from a neighbouring source (previously mentioned in \sct\ref{sct-check-n}), or bulge components modelling foreground contaminating stars. This sub-sample also contains barred galaxies with an estimated $B/T<0.5$ (hence $\theta_\mathrm{main}=\theta_\mathrm{disc}$) in which both the single-S\'ersic profile and the bulge component are biased by a bar onto which they are fitted, hence are elongated at an angle that differs from the disc component (see \fg\ref{fig:exple-14}).

\section{Photometry: Impact of measurement method and galaxy morphology  \label{sct-model-photo}}

\begin{figure}
\includegraphics[width=\columnwidth]{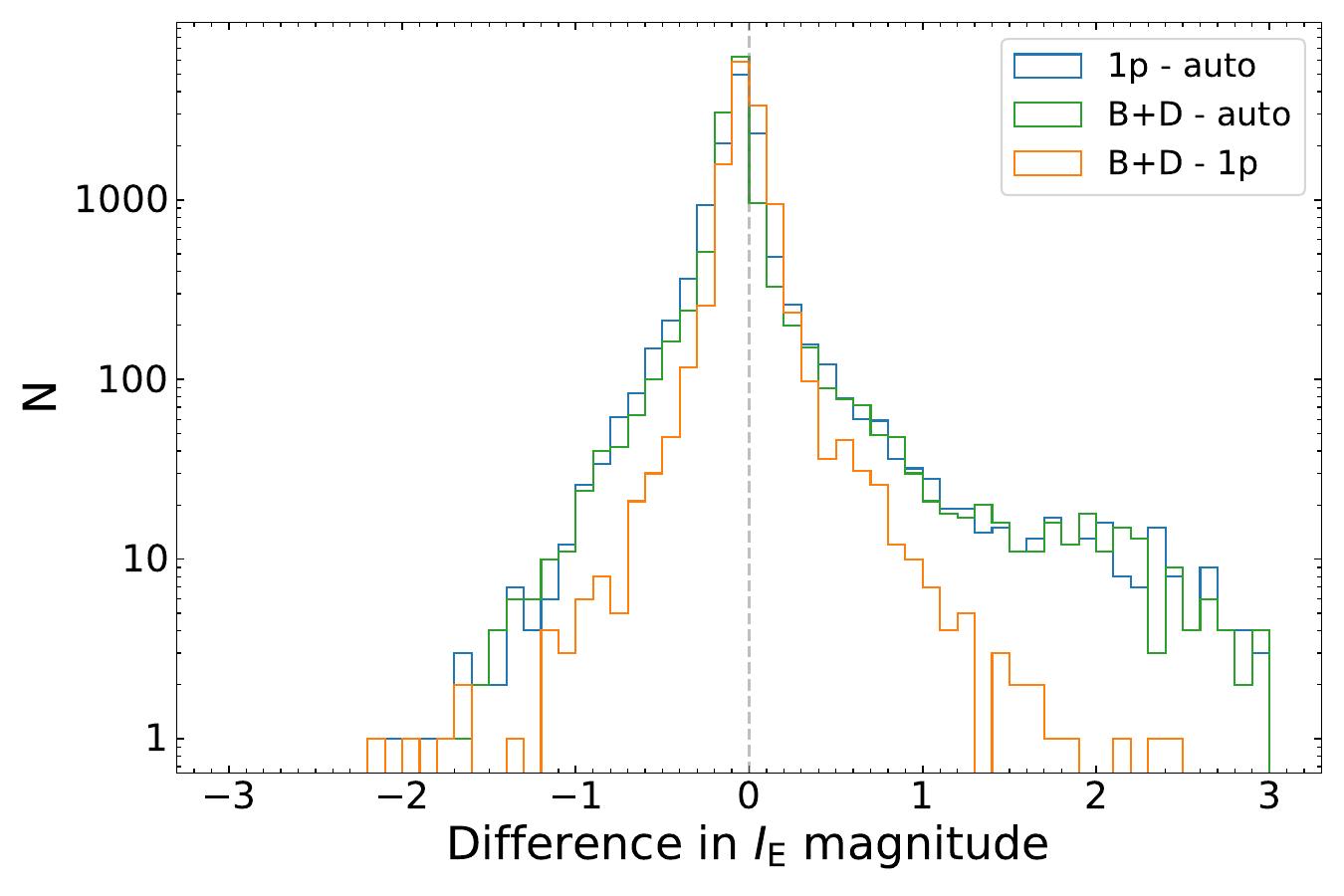}
\caption{Histograms of the differences in the \IE band between the adaptive aperture photometry ($m_\mathrm{auto}$) and the model photometry fitted with a single-S\'ersic profile ($m_\mathrm{1p}$) or with a bulge-disc decomposition ($m_\mathrm{B+D}$) as well as the difference between the two models for the $12\,786$ galaxies with $m_\mathrm{auto}(\IE)\le23$.}
\label{mag-diff-histo}
\end{figure}

\begin{table}[ht]
\caption{Statistics of the distributions of \IE magnitude differences between the various photometry methods.}
\resizebox{\linewidth}{!}{%
\begin{tabular}{ c r r r r r r r r }
\hline\hline
 & \multirow{2}{*}{Mode} & \multirow{2}{*}{Mean} & \multirow{2}{*}{Median} & Std. & \multirow{2}{*}{${q_{5}}^{(b)}$} & \multirow{2}{*}{${q_{25}}^{(b)}$} & \multirow{2}{*}{${q_{75}}^{(b)}$} & \multirow{2}{*}{${q_{95}}^{(b)}$} \\
&  &  &  & err.$^{(a)}$ &  & &  &  \\
\hline
$m_\mathrm{1p} - m_\mathrm{auto}$ & $-0.025$ & $-0.021$ & $-0.039$ & $0.003$ & $-0.381$ & $-0.128$ & $0.014$ & $0.396$ \\
$m_\mathrm{B+D} - m_\mathrm{auto}$ & $-0.075$ & $-0.034$ & $-0.071$ & $0.003$ & $-0.328$ & $-0.120$ & $-0.029$ & $0.367$ \\
$m_\mathrm{B+D} - m_\mathrm{1p}$ & $-0.025$ & $-0.013$ & $-0.024$ & $0.001$ & $-0.176$ & $-0.074$ & $0.036$ & $0.178$ \\
\hline
\vspace{1pt}
\end{tabular}}
\small\textbf{Notes.}
The compared magnitudes are the adaptive aperture photometry ($m_\mathrm{auto}$) and the model photometry fitted with either a single-S\'ersic profile ($m_\mathrm{1p}$) or a bulge-disc decomposition ($m_\mathrm{B+D}$), and the comparison is made for the $12\,786$ galaxies with $m_\mathrm{auto}(\IE)\le23$.
$^{(a)}$ Standard error calculated as $\sigma/\sqrt{N}$ where $\sigma$ is the \rms dispersion and $N=12\,786$ galaxies in the sample.
$^{(b)}$ $q_n$ quantiles corresponding to the $n^\mathrm{th}$ percentile.
\label{tab-histo-mag-diff}
\end{table}

We now explore the effects of the specific choice of photometry method. Specifically, we compare the magnitudes obtained from adaptive aperture photometry, using the \texttt{SourceXtractor++} \texttt{auto\_mag} magnitude (denoted here $m_\mathrm{auto}$), to the PSF-convolved modelled magnitudes from the various runs of luminosity profile model fitting (denoted here $m_{1p}$ for the single-S\'ersic profile, and $m_\mathrm{B+D}$ for the bulge-disc decomposition). The \texttt{SourceXtractor++} \texttt{auto\_mag} uses a Kron elliptical aperture with reduced pseudo-radius $k r_\mathrm{Kron}$ with $r_\mathrm{Kron}$ the Kron radius and $k$ a scale factor set by default to 2.5, while model magnitudes are intended to provide a total magnitude integrated to infinity. In this section we use all fits to the $12\,786$ galaxies with \texttt{auto\_mag} $\IE\le23$, because if the EMC shows that structural parameters may suffer large biases in the bulge-disc decomposition in the $21$ to $23$ \IE interval, the total magnitudes are stable and weakly affected \citep{Bretonniere-EP26,Merlin-EP25}. 

To estimate the average offsets and dispersion between the different flux measurements methods, we plot in \fg\ref{mag-diff-histo} the histograms of the magnitude differences between the model magnitudes and the adaptive aperture ones, as well as the differences between the two types of model magnitudes, in the \IE band. Figure \ref{mag-diff-histo} shows systematic offsets from about $-0.07$ to $-0.02$ between the three distributions of magnitude differences, as listed in \tab\ref{tab-histo-mag-diff}. The difference between both model magnitudes and $m_\mathrm{auto}$ also exhibit a tail at magnitude differences in the $1.5$--3 range, which does not exist for the magnitude difference between both models. We checked that this is due to contaminating stars that are not separated in the aperture magnitude.

\begin{figure}
\includegraphics[width=\columnwidth]{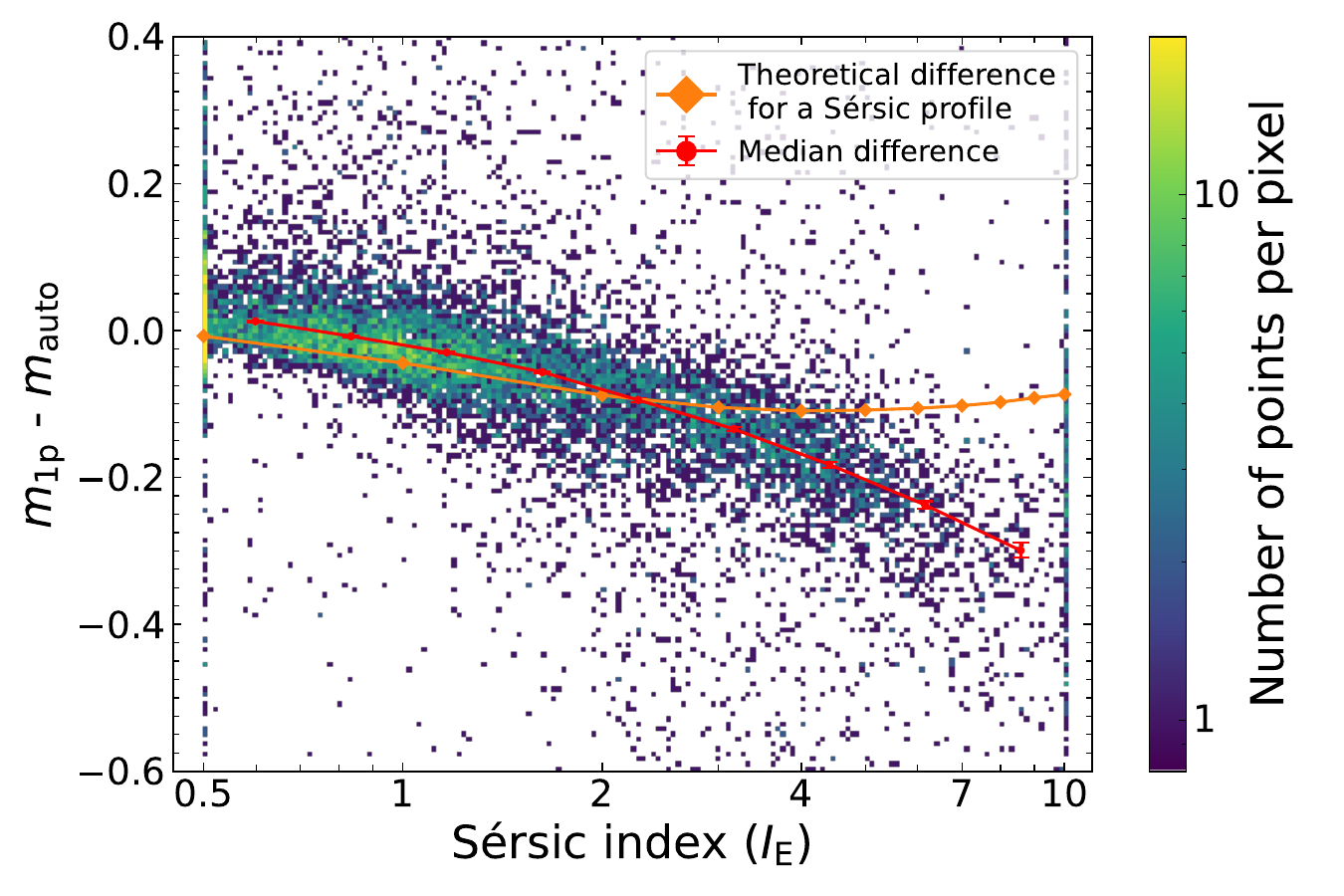}
\caption{Difference between the \IE single-S\'ersic model magnitude $m_\mathrm{1p}$ and the adaptive aperture magnitudes $m_\mathrm{auto}$ as a function of the corresponding S\'ersic index fitted in the \IE band for the $12\,786$ galaxies with $\IE\le23$. Red points show the median value of that difference in bins of $n$ as well as the associated bootstrap uncertainty. The solid orange line represents the expected magnitude difference for a pure S\'ersic luminosity profile. The difference is smaller than expected for disc-like galaxies ($n \lesssim 1.5$), and hence models are fainter than expected, which could be due to the model missing the excess of flux from a small and faint bulge or secondary features (e.g. spiral arms, bars, rings, flocculence). For higher S\'ersic indices, the trend inverts itself, with model photometry being brighter than the adaptive aperture measurement.}
\label{mag-diff-1p-auto-vs-nsersic}
\end{figure}

All magnitude offsets shown in \fg\ref{mag-diff-histo} and \tab\ref{tab-histo-mag-diff} are negative in their mode, mean, and median, as expected for model magnitudes that extrapolate the flux to infinity, contrary to aperture magnitudes that lose some flux. However, the lost flux in the wings of the objects when using $m_\mathrm{auto}$ instead of model magnitudes may be overestimated for spiral galaxies, since a dominant fraction of their discs have been observed to have a down-turn in their light profiles at visible and near-infrared wavelengths \citep{Pohlen-2006-spiral-outer-disks-bending, Mondelin-2025-ERO-outskirts-SB-and-color-profiles}. The flux underestimation of the aperture magnitudes is almost twice as large when compared to the bulge-disc decomposition because the latter better models simultaneously the central peak of flux in the bulges and the wings in the discs than the single-S\'ersic profile. The same effect explains the lost flux in $m_\mathrm{1p}$ compared to $m_\mathrm{B+D}$. 

We further examine in \fg\ref{mag-diff-1p-auto-vs-nsersic} the difference between the single-S\'ersic profile magnitude and $m_\mathrm{auto}$ in the \IE band, as a function of the S\'ersic index $n$ of the fitted profile. Despite the brighter single-S\'ersic magnitude compared to the adaptive aperture magnitude, hence the negative $m_\mathrm{1p}-m_\mathrm{auto}$ difference (see \tab\ref{tab-histo-mag-diff}), the median difference per $0.33$ dex interval of $n$ is positive at $0.013$ in the first bin located at $[0.505,0.698]$. It then decreases steadily all the way to $-0.3$ in the last bin at $[7.2,9.9]$. The \rms dispersion for $m_\mathrm{1p}-m_\mathrm{auto}$ is $0.34$ mag for $n\in[0.5,0.7]$, and it decreases to $0.23$ mag for $n\in[1.4, 1.9]$ and increases again to $0.39$ mag for $n\in[7.2, 10]$. This dispersion results from the fact that galaxies do not all have smooth profiles as most ellipticals and barless lenticulars do. Lenticular galaxies may also have bars and rings, whereas in addition to bars and rings, spiral galaxies display spiral arms and scattered \ion{H}{ii} regions causing flocculence. None of these features can be modelled by the single-S\'ersic profile or by the bulge-disc decomposition. The steadily decreasing $m_\mathrm{1p}-m_\mathrm{auto}$ means that the relative flux lost by the adaptive aperture magnitude is larger for higher S\'ersic indices, with the model magnitude being increasingly brighter. One could try to explain this trend by the fact that a high S\'ersic index corresponds to a centrally steep profile with also more extended wings at large radii,  leading to an increasingly larger part of the source light being located beyond the aperture in which $m_\mathrm{auto}$ is calculated. This effect can be precisely estimated and accounted for.

\begin{figure}
\includegraphics[width=\columnwidth]{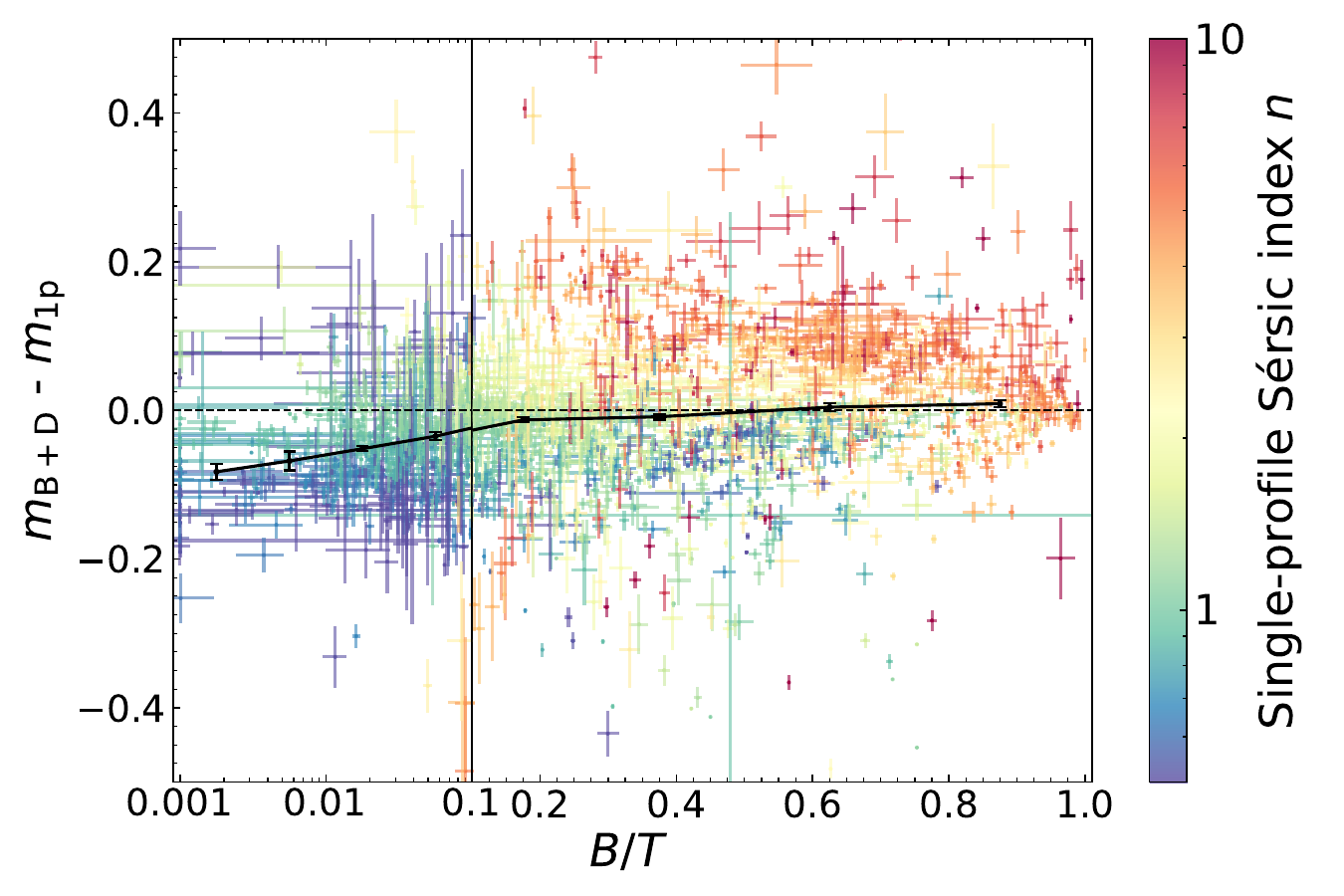}
\caption{Difference between the model magnitudes obtained from the bulge-disc decomposition and the single-S\'ersic fit as a function of $B/T$ for the $2445$ galaxies with $\IE\le21$. The solid black line separates two areas of the graph, with logarithmic and linear scales to the left and to the right, respectively, whereas black dots correspond to median and bootstrap errors in bins of $B/T$. For small $B/T$, the bulge-disc decomposition leads on average to brighter magnitudes, but this median offset vanishes for more prominent bulges ($B/T\gtrsim0.2$). A strong vertical colour gradient appears at all $B/T$, with higher (smaller) $n$ corresponding to more (less) flux in the single-S\'ersic model than in the bulge-disc one.}
\label{mag-diff-BD-1p-vs-BT}
\end{figure}

Indeed, to better understand the magnitude differences in \fg\ref{mag-diff-1p-auto-vs-nsersic} we have added in this graph a black line that shows the magnitude difference $m_\mathrm{1p}-m_\mathrm{auto}$ when both magnitudes are measured on a pure S\'ersic profile, using the values provided by \citet{Graham-2005-sersic-considerations} in their table 1, and calculated for a circular source. This line should be used as a reference for interpreting the observed $m_\mathrm{1p}-m_\mathrm{auto}$, because it allows one to account for the magnitude difference arising from the different definitions, rather than from specifics of the modelling and intrinsic morphological effects. For S\'ersic indices from 0.5 to 2, the observed median $m_\mathrm{1p}-m_\mathrm{auto}$ is closer to 0 and above the expected theoretical value. This means that the loss of flux in the single-S\'ersic model compared to the aperture magnitude is not as large for the irregular or disc-dominated spiral galaxies in the data as for the S\'ersic theoretical curve. This may be due to the strong flocculence in irregular and very late spirals, in which \texttt{SourceXtractor++} might treat the \ion{H}{ii} regions as contaminating sources that are partitioned and not taken into account in the model fits, whereas these bright spots would be included in the aperture magnitude, making it brighter. The low-density tail of positive $m_\mathrm{1p}-m_\mathrm{auto}$ up to about $1$ mag (and for all values of $n$) could correspond to galaxies with unusually bright spots of star formation or contaminating stars that are not partitioned from the galaxy, whereas more common flocculence creates the dispersion of the data points around its median value. 

\begin{figure*}
\includegraphics[width=\columnwidth]{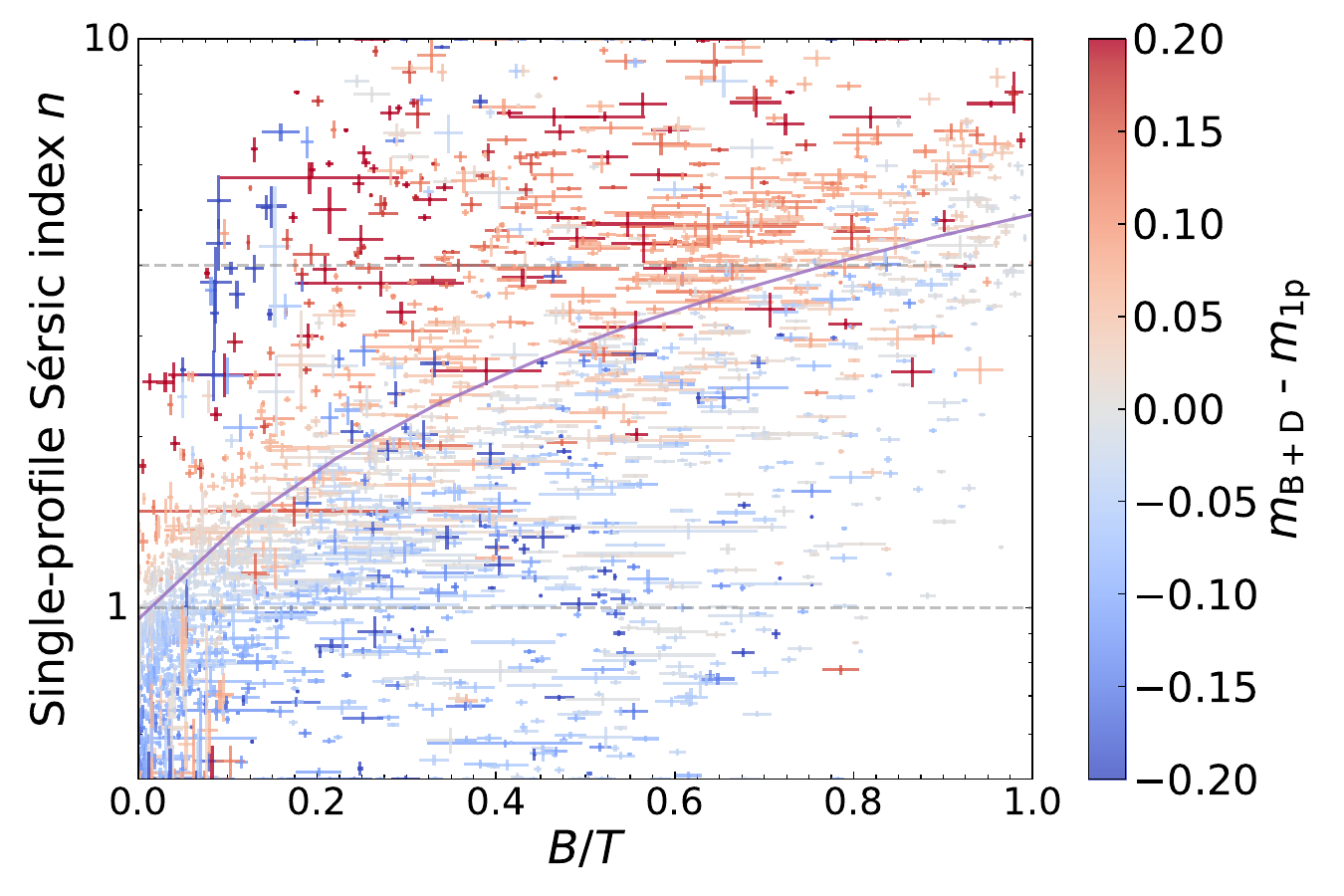}
\includegraphics[width=\columnwidth]{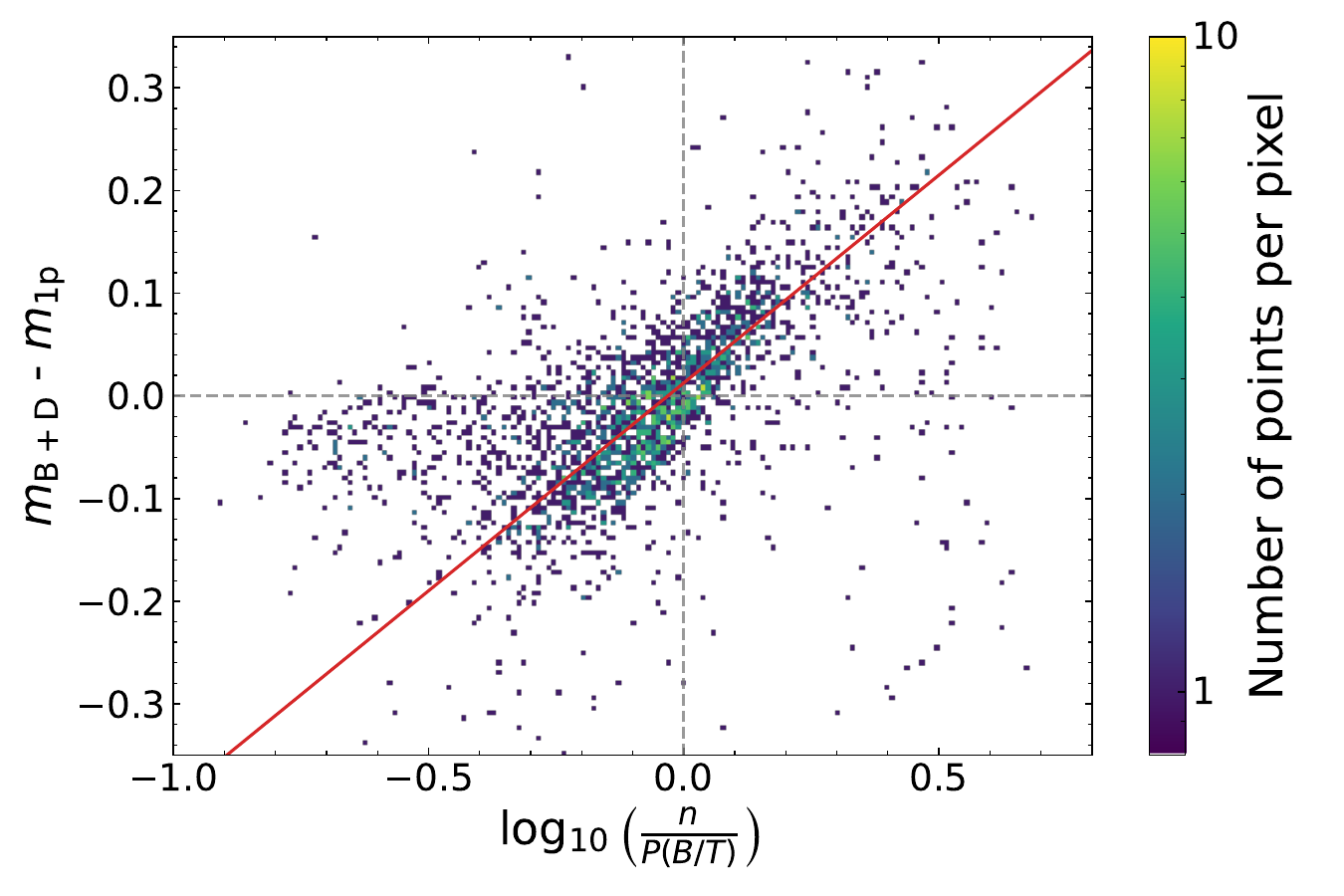}
\caption{Relations between the difference in the model magnitudes of both model configurations and the difference in the S\'ersic index $n$ for the single-S\'ersic fit and $B/T$ of the bulge-disc decomposition for the $2445$ galaxies to $\IE\le21$. Left: Value of $n$ as a function of $B/T$. The \IE magnitude difference between the two models is shown as the colour of the points. The first-degree polynomial best-fitting $P(B/T)=n$ appears as a purple line and matches the separation between the positive and negative magnitude differences, which appear mostly above and below it, respectively. Right: Magnitude difference plotted as a function of the distance between points in the left panel and the best-fit relation computed vertically and in logarithmic scale. The vertical $m_\mathrm{1p}-m_\mathrm{auto}$ offset in \fg\ref{mag-diff-BD-1p-vs-BT} can therefore be explained as the result of a discrepancy between the measured $n$ and $B/T$, with less flux within the bulge-disc decomposition than the single-S\'ersic model of a galaxy since its $n$ is larger than what should be expected from its $B/T$ through their fitted linear relation (and inversely). A difference in model photometry is therefore correlated to a discrepancy between the two parameters, which, in each configuration, better describe the overall galaxy morphology.}
\label{mag-diff-BD-1p-explain}
\end{figure*}

For larger values of $n$, the theoretical difference in magnitude stops decreasing at $n\approx4$, and even slightly increases for $n \geq 5$, to reach $-0.09$ at $n=10$. In contrast, the measured $m_\mathrm{1p}-m_\mathrm{auto}$ continues to decrease. This effect is due to the fact that a single-S\'ersic profile is insufficient to model the complexity of the whole galaxy's light distribution when there is a prominent bulge, causing the fitted profile to underestimate the light either at the centre or in the outskirts of the galaxy and cannot fit both parts well. More generally, when the surface brightness distribution of a galaxy is too complex to be modelled by a single-S\'ersic profile, it leads to offsets in its magnitude estimation and calls for more complex modelling methods, involving more parameters, such as the bulge-disc decomposition.

Finally, we examine in \fg\ref{mag-diff-BD-1p-vs-BT} the impact of the chosen luminosity profile-fitting configuration on the model photometry by comparing the total \IE magnitudes measured with a single-S\'ersic profile and through a bulge-disc decomposition. The difference between these two magnitudes is plotted as a function of $B/T$ in logarithmic scales for $B/T\leq0.1$, and linear scales for $B/T > 0.1$ in order to better highlight the observed trends (the two parts of the graph are separated by a black solid  vertical line). In this graph, the colour of the points represent the single-S\'ersic index $n$, which we showed in \fg\ref{nsersic-nsersicB-BT-3d-space} to be correlated with $B/T$, and that both quantities can be considered as proxies of the Hubble type of the galaxy. Median values and standard errors are computed in logarithmic and linear bins of $B/T$ and shown as black dots and error bars.

Figure \ref{mag-diff-BD-1p-vs-BT} shows that for small values of $B/T$, the bulge-disc decomposition leads to brighter magnitudes than the single-S\'ersic profile for $74.8\%$ of galaxies with $B/T \leq 0.1$, and even $78.7\%$ for $B/T \leq 0.05$, against a more balanced fraction for more prominent bulges, with $52.4\%$ of galaxies having brighter $m_\mathrm{B+D}$ for $B/T > 0.1$. Moreover, median magnitude differences are significantly negative and decrease for lower $B/T$, to reach a minimum value of $-0.08$ in the lowest $B/T$ interval considered ($[0.001,0.003]$) with an \rms dispersion of 0.08 mag (and $-0.07$, $-0.05$ and $-0.03$ mag offsets in the next three $B/T$ intervals of $[0.003,0.01]$, $[0.01,0.03]$, $[0.03,0.1]$, respectively). In this left part of the graph, the magnitude difference decreases for smaller S\'ersic indices (green to blue points), hence with brighter bulge-disc magnitudes than single-S\'ersic models. This can be explained by the fact that for these objects, the disc component, that accounts for $\geq 90\%$ of the flux, is constrained to be exponential (\ie $n=1$), whereas the single-S\'ersic profile index can vary and provide a better fit to the object, thus avoiding an over-estimation of the object's flux in its outer part, than given by the exponential disc. For $B/T\ge0.1$, the median magnitude difference increases, to reach values compatible with 0 (by less than twice the standard error). The dispersion in $B/T$ bins is in between $0.12$ and $0.13$ mag for all $B/T > 0.1$ intervals and there is a strong vertical colour gradient indicating that the sign and value of the magnitude difference depends on the value of the single-S\'ersic index measured for each $B/T$, with higher $n$ corresponding to brighter single-S\'ersic models compared to the bulge-disc decomposition, and inversely. 

To better understand the $m_\mathrm{1p}-m_\mathrm{auto}$ offsets in \fg\ref{mag-diff-BD-1p-vs-BT}, we plot in the left panel of \fg\ref{mag-diff-BD-1p-explain} the relation between $n$ and $B/T$ as in the top panel of \fg\ref{nsersic-nsersicB-BT-3d-space}, but here the points are colour-coded by the difference in \IE magnitude between the bulge-disc and the single-S\'ersic models. We observe that there appears to be a correlation between $n$ and $B/T$, which when linearly fitted in linear scales $n$ versus $B/T$ has a Pearson correlation coefficient of 0.57. We then perform a first-degree polynomial fit of $n$ as a function of $B/T$, formalised as $n=P(B/T) = a_1 B/T + a_0$, with best-fit parameters $a_1 = 3.96 \pm 0.012$ and $a_0 = 0.95 \pm 0.05$; this fit is plotted as the solid purple line in \fg\ref{mag-diff-BD-1p-explain}. This fitted function yields $P(0)=0.95 \approx 1$, so we retrieve the fact that bulge-less galaxies are pure exponential discs. We also obtain $P(1)=4.91$, and for the index of a de Vaucouleurs profile $n=4$, typical of elliptical galaxies, $P^{-1}(4) = 0.77$. Contrary to a preconceived idea that elliptical galaxies have $B/T=1$, both photometric analyses \citep{Quilley-2023-scaling-bulges-and-disks} and dynamical analyses \citep{Krajnovic-2013-ATLAS-3D-photo-kine-stellar-disks-in-ETG} of elliptical galaxies favour a small value of $B/T$ because these objects often require a weak disc component.

Furthermore, the $n$ versus $B/T$ linear fit interestingly appears as a demarcation line between galaxies with positive and negative magnitude differences: models are brighter for the single-S\'ersic case than in the bulge-disc decomposition when $n>P(B/T)$, and inversely. We then show in the right panel of \fg\ref{mag-diff-BD-1p-explain} the relation between the magnitude difference between the two models and the vertical logarithmic distance of each point $(B/T;n)$ to the fitted relation $n=P(B/T)$. After excluding the 625 outliers from the $n=P(B/T)$ relation satisfying $| \logten(\frac{n}{P(B/T)})| > 0.3$ (which enclose 395 of the 399 outliers identified in \fg\ref{all-radii-related-cmap-BT}), and galaxies with large magnitude difference set to $|m_\mathrm{B+D} - m_\mathrm{1p}| > 0.2$, we measure a Pearson correlation coefficient of $0.68$ between the two axes. A linear relation is then fitted to the right panel of \fg\ref{mag-diff-BD-1p-explain} using an orthogonal distance regression\footnote{Using the \texttt{Python scipy ODR} library. The `orthogonal distance regression' minimises the orthogonal distances to the fit rather than the vertical ones, therefore also taking into account the horizontal distances.} and this is plotted as a solid blue line with a $0.45$ coefficient of determination and $74\%$ of points being within $3\sigma$ of the relation, with $\sigma = 0.05$ the orthogonal \rms dispersion. We repeated this process starting from a first-degree polynomial fit of $\logten(n)$ (instead of $n$) as a function of $B/T$, since this may also appear as an acceptable fit (see \fg\ref{nsersic-nsersicB-BT-3d-space}). The correlation exhibited in \fg\ref{mag-diff-BD-1p-explain} validates our interpretation that the dispersion observed in the model photometry is the result of the distance from the relation between the single-S\'ersic profile index $n$ and the bulge-disc decomposition $B/T$ fitted in the left panel of \fg\ref{mag-diff-BD-1p-explain}.

Although the magnitude offset analysis is only reported for the \IE band, similar trends were obtained in the three NISP bands, when examining the magnitude differences as a function of single-S\'ersic $n$ and $B/T$ values in the corresponding bands (as in \fgs\ref{mag-diff-1p-auto-vs-nsersic} and \ref{mag-diff-BD-1p-vs-BT}). However, even though there is an offset compatible with 0  between the two magnitudes, both of them are about $0.2$ mag brighter than $m_{\rm auto}$, in all three \YE, \JE, and \HE bands, since the aperture radius is defined on the detection image, which is here that in the \IE band. 

To conclude, the above analysis allows us to answer the important issue of which of the magnitudes of the single-S\'ersic profile or bulge-disc model is likely to be closer to the true magnitude. For small $B/T$ hence negligible bulge component, it is $m_\mathrm{1p}$, whereas for galaxies with a significant bulge, it is $m_\mathrm{B+D}$. This is because in each case, the corresponding model offers more flexibility in adjusting the full light distribution of the objects, from the centre to the outskirts.

\section{Model colours and colour gradients \label{sct-color-gradients}}

To limit biases in measuring colours of galaxies, it is common to measure them as a difference between two magnitudes calculated within a common aperture \citep{De-Vaucouleurs-1977-B-photom-115-galaxies,De-Vaucouleurs-1977-UBV-colors-100-galaxies}. A substitute when using models can be to calculate colours from a fixed profile fitted in all considered bands, that is with identical parameters, except for the total flux, remaining free (for instance as implemented in \texttt{SExtractor} with the \texttt{DETMODEL} magnitudes). The colours of these models would then be fixed to the flux ratio, hence independent of any aperture size. However, we chose to calculate colours of the ERO-Perseus galaxy sample using the independent models in the various bands considered (approach 1 in \sct\ref{sct-model-fitting-config}). This gives the advantage of better fitting the 2D brightness distributions of each object in each band. It also allows us to study higher order effects such as band-to-band variations in the structural parameters of a galaxy model or one of its components, a tool for measuring colour gradients of a profile model (see \sct\ref{sct-bulge-disk-gradient}). The tight correlations between the structural parameters measured with a common single-S\'ersic profile in the three NISP bands compared to those of independent profiles in the three bands, also supports this choice (see Appendix \ref{appendix-common-profile}).

\begin{figure*}
    \centering
    \includegraphics[width=\columnwidth]{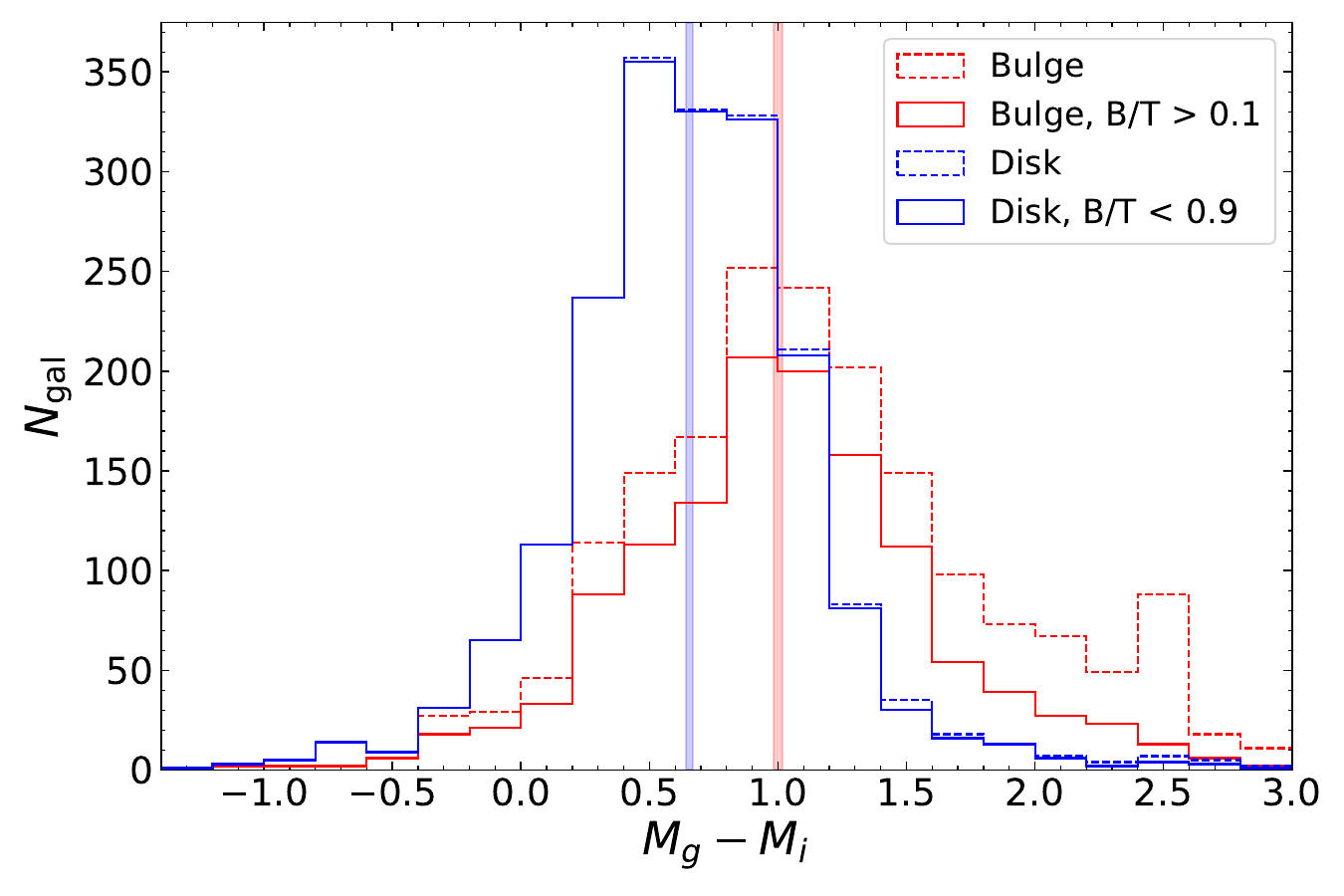}
    \includegraphics[width=\columnwidth]{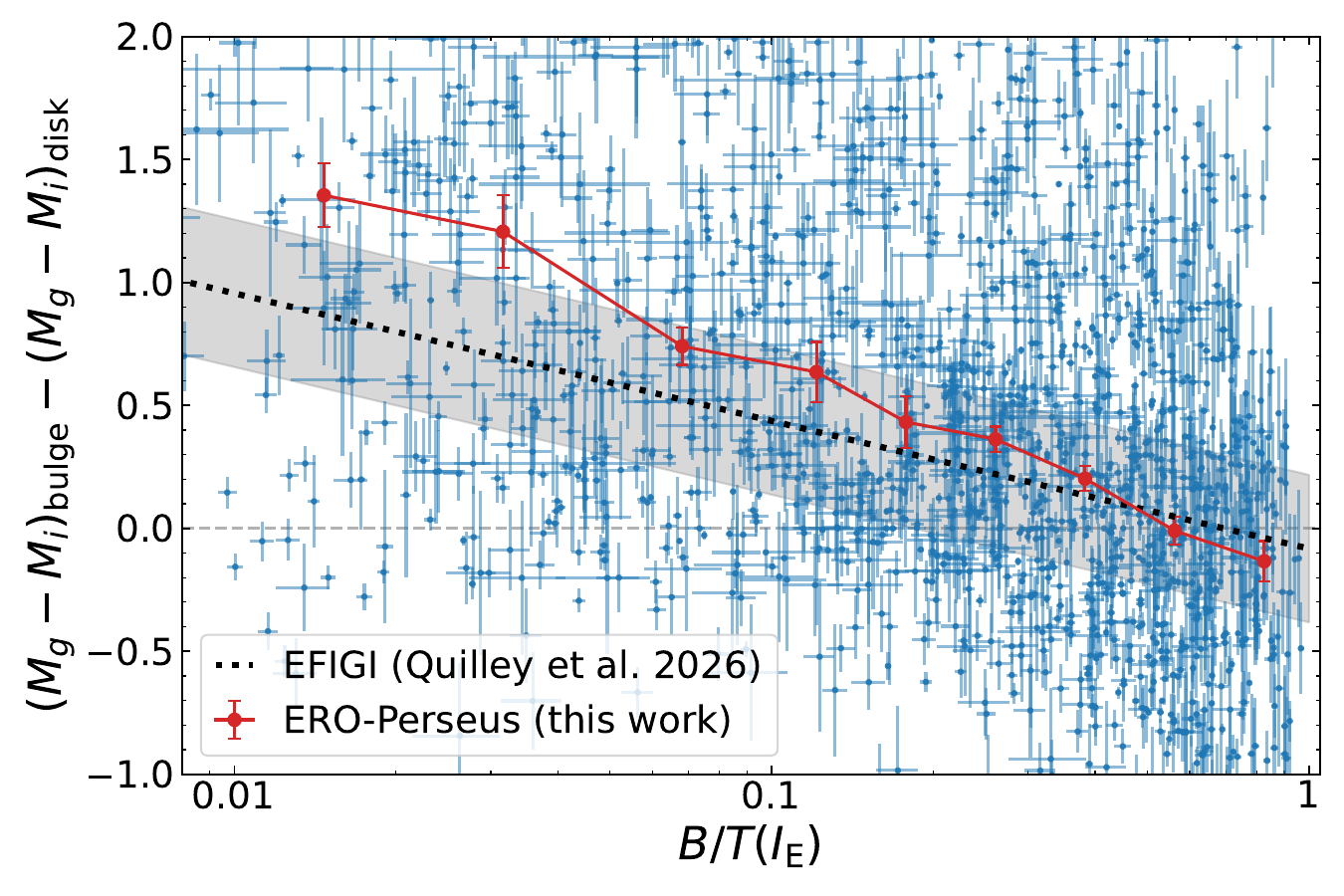}
    \caption{Left: Histograms of the rest-frame $M_g-M_i$ colours for the bulges and discs of the $1885$ galaxies to $\IE < 21$ with reliable models (see text) in red and blue, respectively. The dashed lines correspond to the full samples, whereas the solid lines correspond to the components enclosing more than $10\%$ of the galaxy total light. The vertical lines represent the median, and their width indicates the bootstrap errors for the restricted samples. Right: Bulge-disc rest-frame $M_g-M_i$ colour difference between the bulge and the disc of each galaxy as a function of $B/T(\IE)$. The red dots indicate the median colour difference and associated bootstrap error in bins of $B/T$. The dashed black line indicates the relation found for nearby galaxies with the EFIGI sample in \cite{Quilley-2025-colors-color-gradients} using $B/T$ in the $i$ band, and the grey shaded region indicates the $\pm1\sigma$ dispersion around it. On average, bulges appear redder than discs. Moreover, the behaviour of disc-dominated and bulge-dominated galaxies is distinct, indicating that the former host bulges redder than their discs, whereas the latter have their bulges and discs of similar colour, and this variation depends on the value of $B/T$.}
    \label{bulge-disk-colors}
\end{figure*}

\subsection{Bulge and disc colour dichotomy
\label{sct-bulge-disk-color}}

We show in \fg\ref{bulge-disk-colors} the distribution of the rest-frame $M_g - M_i$ colours separately for the bulges and discs of galaxies in the left panel, and the difference between the bulge and disc colour of each galaxy versus its $B/T$ in the right panel. Photometric redshifts are used to derive these rest-frame colours (see \sct\ref{sct-methodo-SED}). The choice of the $g-i$ colour is motivated by its wide use in ground-based surveys, and is well constrained with the bands available at the redshifts probed here. Indeed, at the median redshift $z_{\rm med}=0.24$ of our sample, the observed \IE and \YE bands are shifted to the $g+r$ and $i+z$ bands combined, respectively. Similarly, at the highest redshifts probed (95\% percentile) $z_{95}=0.54$, the \IE, \YE, and \JE bands are shifted to the $u+g$ bands combined, the reddest half of $r$ and bluest half of $i$, and the reddest half of $i$ along with the $z$ band, respectively. Starting from the $2047$ galaxies with $\IE < 21$ and meaningful bulge and disc models (see \sct\ref{sct-check-Re}), we further remove galaxies for which the SED fit has failed to produce $M_g - M_i$ colours for either their bulge or disc. This removes points accumulating at $M_g-M_i=-0.2146$ (produced by the SED 31 -- SB11\_A\_0.sed -- with $A_V=0.0$), or at $M_g-M_i=2.5423$ (produced by the SED 8 -- S0\_A\_0.sed -- and $A_V=3.0$), leaving 1885 galaxies whose bulge-disc colours are analysed.

In the left panel of \fg\ref{bulge-disk-colors}, we examine the colour distributions of bulges and discs for the whole sample, and for the components with more than $10\%$ of the total galaxy light (in \IE), so as to avoid drawing conclusions from low flux bulge or disc components which are affected by larger uncertainties, and may not actually model a physical bulge or disc. The $B/T>0.1$ added criterion discards a tail of 269 bulges with very red colours ($M_g - M_i\gtrsim2$) which are most likely non-physical, but also exclude 359 bulges with $B/T\le0.1$ and $M_g - M_i$ colours that cover the full 0.0--2.0 range, among which there are many apparently well modelled faint bulges. The modes of both distributions of $M_g - M_i$ bulge colours are $\sim1.0$ mag. The median (with associated bootstrap errors) and dispersions (estimated as half the 16--84th percentile range) decrease from $1.11\pm0.02$ to $1.00\pm0.02$ mag, and from 0.77 to 0.55, respectively, when restricting the samples to $B/T > 0.1$. For the distributions of disc colour, the $B/T<0.9$ selection leads to almost identical results as for the full sample: the latter and the restricted sample have median absolute colours of $0.66\pm0.01$ and $0.65\pm0.01$, and their dispersions are $0.40$ and $0.39$, respectively. The left panel of \fg\ref{bulge-disk-colors} therefore shows a difference in the bulge and disc colours of galaxies with a difference in the medians of the histograms of $0.34$ in $M_g - M_i$ for the $B/T$ restricted sample containing 1278 and 1858 galaxies, respectively, which is significant at the $17\sigma$ level using the bootstrap uncertainties.

The right panel of \fg\ref{bulge-disk-colors} shows the distribution of bulge-disc $M_g - M_i$ colour difference of all galaxies as a function of $B/T$ measured in the \IE band. The frequent positive differences (for 64\% of the full sample) confirm that bulges are redder than the discs they are embedded in, not only that the bulge population is overall redder than the disc population. Moreover, there is a clear morphological trend with the difference in bulge-disc colour, with the largest differences for small red bulges embedded in blue discs, and a progressive decrease with increasing $B/T$ to reach similar bulge and disc colours in bulge-dominated systems (i.e. $B/T\gtrsim0.5$). The median colour differences are for instance $0.74 \pm 0.08$ and $0.64\pm0.12$ for $B/T \in [0.05, 0.1]$ and $[0.1,0.15]$ respectively, that is, bulges are on average redder than their discs with a significance level of $4$ to $7$ $\sigma$, whereas in the three bins at $B/T > 0.3$, the median differences are within $2.9$, $0.1$ and $1.2$ $\sigma$ of 0, hence compatible with a null difference.

The bulge and disc colour differences shown in \fg\ref{bulge-disk-colors} are similar to those measured in previous studies based on bulge and disc decompositions of galaxies observed from the ground, which all agree on the redder colour of bulges compared to their discs \citep{Mollenhoff-2004-BD-spirals-UBVRI, Vika-2014-multiband-BD-decomposition-MegaMorph, Kennedy-2016-GAMA-color-gradients-vs-BT-color-BD, Casura-2022-bulge-disk-decomposition-GAMA, Quilley-2025-colors-color-gradients}.
\cite{Vika-2014-multiband-BD-decomposition-MegaMorph} also demonstrated on a sample of 163 galaxies of various Hubble types that spiral types have stronger bulge-disc colour differences than  lenticulars and ellipticals. By analysing the 4458 visually classified galaxies of the EFIGI catalogue \citep{Baillard-2011-EFIGI}, \citet{Quilley-2025-colors-color-gradients} further demonstrated that the bulge and disc rest-frame optical colours are dependent on the Hubble type for the considered galaxies, with rather stable red colours for all bulges, whereas disc colours are bluer than their bulges for late type spirals, and become redder backwards along the Hubble sequence to reach similarly red colours as the bulges in lenticulars.
\citet{Sachdeva-2017-bulge-growth}, \citet{Nedkova-2024-bulge-disc-decomposition-UVJ-diagrams-size-mass-relations}, and \citet{Dimauro-2022-bulge-growth} also measured a bulge-disc colour dichotomy from HST imaging out to redshifts of 1, 1.5 and 2, respectively.

The interpretation of the bulge-disc colour dichotomy detected here in terms of stellar age and metallicity gradients, as well as internal dust extinction are discussed in \sct\ref{sct-color-variation-interpretation} below.

\subsection{Variation of structural parameters across bands \label{sct-bulge-disk-gradient}}

\begin{figure*}
\includegraphics[width=\textwidth]{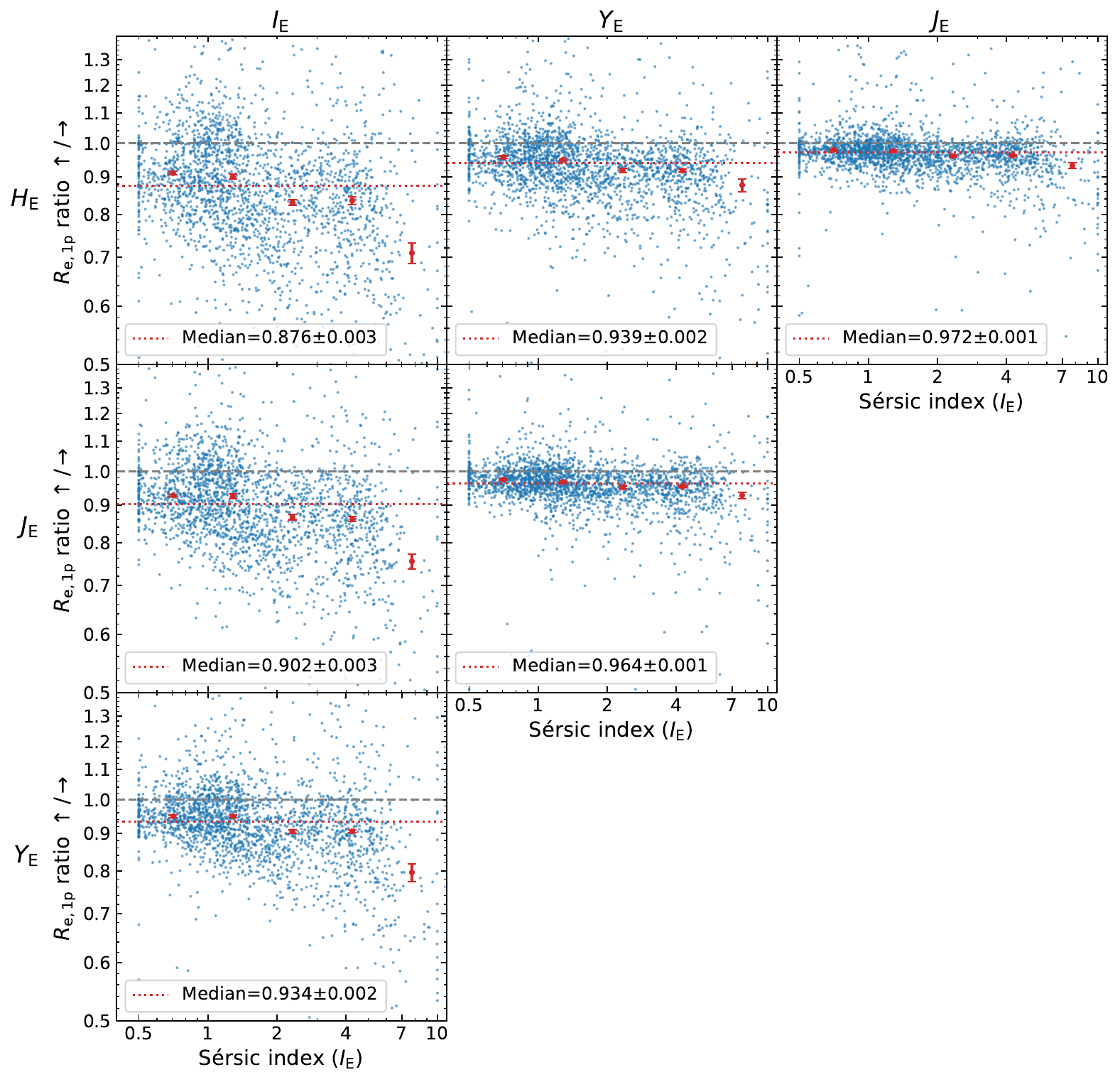}
\caption{Ratios of the single-S\'ersic effective radius $R_\mathrm{e,1p}$ between different bands as a function of the S\'ersic index measured in the \IE VIS band for the $2328$ galaxies to $\IE\le21$, with $R_\mathrm{e,disc} /R_\mathrm{e,bulge} \ge 1$ or $R_\mathrm{e,disc}/R_\mathrm{e,1p} > 0.9$ in the \IE band. All possible pairs of \Euclid bands are shown in the six panels, with the band whose $R_\mathrm{e,1p}$ is the numerator (denominator) appearing on the left (top). Overall median values of the $R_\mathrm{e,1p}$ ratio appear on each panel as a red dotted horizontal line, whereas the red points and error bars correspond to the median and standard error for bins of $n_{\IE}$ listed in \tab\ref{tab-Re-ratio-1p}. Redder bands have smaller $R_\mathrm{e,1p}$ than bluer bands for the majority of galaxies, and this effect is stronger for bands that are further apart in wavelength. Most galaxies have blue to red inward gradients, which are the largest between the VIS band and any NISP bands, and are interpreted as resulting mostly from the colour dichotomy of bulges and discs shown in \fg\ref{bulge-disk-colors}.}
\label{color-grad-Re-1p}
\end{figure*}

We now examine how the structural parameters describing our sample of galaxies vary from band to band, as a demonstration of the ability to measure colour variations within galaxies in \Euclid images. Figure \ref{color-grad-Re-1p} displays in its different panels the ratios of the effective radii $R_\mathrm{e,1p}$ measured for the single-S\'ersic profile for all the different pairs of \Euclid bands, as a function of the VIS \IE S\'ersic index $n_{\IE}$ (in logarithmic bins), that characterises galaxy morphology. In \fg\ref{color-grad-Re-1p}, the ratio of any pair of effective radii is computed as the value in the redder band divided by the value in the bluer band.

Figure \ref{color-grad-Re-1p} show that the median $R_\mathrm{e,1p}$ ratios per $n_{\IE}$ interval as well as over the full range of \IE S\'ersic indices (shown as dashed red horizontal lines) are systematically below 1, therefore indicating smaller $R_\mathrm{e,1p}$ for redder bands. The various median ratios are shown in \tab\ref{tab-Re-ratio-1p}, and are in the $0.87$ to $0.93$ range between the VIS and the three NISP bands. Additionally, the ratio between two bands is closer to 1 for bands that are closer to each other in terms of wavelength coverage, suggesting monotonically more compact single-S\'ersic profiles with increasing wavelength. In the following, we denote as colour gradients the fact that the effective radii vary with wavelength.

\begin{table*}[ht]
\caption{Median values of the ratios of single-S\'ersic effective radii between pairs of bands.}
\resizebox{\linewidth}{!}{
\begin{tabular}{ c c c c c c c c }
\hline\hline
$n_{\IE}$ & \multirow{2}{*}{${N_\mathrm{gal}}^{(b)}$} & \multicolumn{6}{c}{Median $R_\mathrm{e,1p}$ ratio$^{(c)}$ over pairs of bands} \\  
interval$^{(a)}$ & & \YE/\IE & \JE/\IE & \HE/\IE & \JE/\YE & \HE/\YE & \HE/\JE \\
\hline
$[0.5, 0.9]$ & 467 & $0.949\pm0.004$ & $0.928\pm0.004$ & $0.911\pm0.005$ & $0.975\pm0.002$ & $0.957\pm0.004$ & $0.979\pm0.002$  \\
$[0.9, 1.7]$ & 769 & $0.949\pm0.004$ & $0.926\pm0.006$ & $0.901\pm0.007$ & $0.968\pm0.002$ & $0.950\pm0.003$ & $0.977\pm0.002$ \\
$[1.7, 3.0]$ & 422 & $0.905\pm0.005$ & $0.867\pm0.007$ & $0.831\pm0.008$ & $0.951\pm0.004$ & $0.919\pm0.006$ & $0.961\pm0.003$  \\
$[3.0, 5.5]$ & 416 & $0.906\pm0.004$ & $0.862\pm0.007$ & $0.835\pm0.010$ & $0.954\pm0.003$ & $0.918\pm0.004$ & $0.962\pm0.002$  \\
$[5.5, 10]$ & 151 & $0.796\pm0.022$ & $0.754\pm0.018$ & $0.709\pm0.023$ & $0.928\pm0.009$ & $0.877\pm0.018$ & $0.932\pm0.008$  \\
$[0.5, 10]$  & 2328 & $0.934\pm0.002$ & $0.902\pm0.003$ & $0.876\pm0.003$ & $0.964\pm0.001$ & $0.939\pm0.002$ & $0.972\pm0.001$ \\
\hline
\vspace{1pt}
\end{tabular}}
\small\textbf{Notes.}
The median values listed here correspond to those plotted in \fg\ref{color-grad-Re-1p}, computed over intervals of the S\'ersic index, for the $2328$ galaxies with $\IE\le21$ (excluding the spurious single-S\'ersic models, see text).
$^{(a)}$ The bins are defined to be of equal logarithmic width over the $[0.5, 10]$ full available interval for the \IE S\'ersic indices (see \sct\ref{sct-model-fitting-config}).
$^{(b)}$ Number of galaxies in the considered sample.
$^{(c)}$ Median values with bootstrap errors. 
\label{tab-Re-ratio-1p}
\end{table*}

\begin{table}[ht]
\caption{ 
Statistics of the difference of the single-S\'ersic indices $n$ between different bands.}
\resizebox{\linewidth}{!}{
\begin{tabular}{ l r r r r r r }
\hline\hline
& \multicolumn{6}{c}{Single-S\'ersic index difference between pairs of bands}\\ 
& \YE$-$\IE & \JE$-$\IE & \HE $-$\IE & \JE$-$\YE & \HE$-$\YE & \HE$-$\JE \\
\hline
Mean & 0.09 & 0.15 & 0.23 & 0.05 & 0.13 & 0.08 \\
Median & 0.08 & 0.12 & 0.15 & 0.03 & 0.06 & 0.03 \\
RMS & 1.06 & 1.14 & 1.37 & 0.58 & 0.76 & 0.61 \\
Boot. err.$^{(a)}$ & 0.005 & 0.006 & 0.007 & 0.002 & 0.003 & 0.002 \\
${q_{10}}^{(b)}$ & $-0.58$ & $-0.59$ & $-0.67$ & $-0.20$ & $-0.22$ & $-0.19$ \\
${q_{25}}^{(b)}$ & $-0.08$ & $-0.06$ & $-0.05$ & $-0.03$ & $-0.03$ & $-0.03$ \\
${q_{75}}^{(b)}$ & 0.25 & 0.32 & 0.42 & 0.13 & 0.20 & 0.12 \\
${q_{90}}^{(b)}$ & 0.68 & 0.83 & 1.15 & 0.35 & 0.60 & 0.36 \\ \hline
\vspace{1pt}
\end{tabular}}
\small\textbf{Notes.}
The statistics are computed for the $2328$ galaxies to $\IE\le21$. All pairs of bands are shown in the different columns.
$^{(a)}$ Bootstrap error. {$^{(b)}$} $q_n$ quantiles corresponding to the $n$th percentile.
\label{tab-n-sersic-color-grad}
\end{table}

All panels of \fg\ref{color-grad-Re-1p} show similar variations of the effective radii ratios with increasing $n_{\IE}$: the first two points (with $n_{\IE} \lesssim1.7$) have similar colour gradients (see \tab\ref{tab-Re-ratio-1p}). There is a systematic shift to a lower and similar ratio (hence stronger gradient) for the next two points with $n_{\IE} \in [1.7,5.5]$ (see \tab\ref{tab-Re-ratio-1p}). The decrease between the second and third $n_{\IE}$ interval are significant at the $4.8\sigma$, $3.4\sigma$, $2.5\sigma$, $5.0\sigma$, $2.6\sigma$, and $3.9\sigma$ confidence levels in the six panels ordered from left to right and from top to bottom. There appears to be a further decrease in the $R_\mathrm{e}$ ratio for $n_{\IE} \ge 4$, but the numbers are small for all pairs of bands at such steep profile indices. We emphasise that because of the dominant continuum in the stellar spectra, $R_\mathrm{e}$ ratios measured in various pairs of bands are correlated, as illustrated in \fg\ref{color-NISP} (see Appendix \ref{appendix-common-profile}). Therefore, galaxies with a ratio having a large (small) deviation from unity in one panel of \fg\ref{color-grad-Re-1p} also have a large (small), deviation from unity in all the other panels. 

Given that bulges are on average redder than discs in spiral galaxies, this systematic increase of the colour gradient for $n_{\IE} \gtrsim 1.7$ compared to smaller values of $n_{\IE}$ is likely to result from the transition between late-type galaxies with a negligible or absent bulge ($n_{\IE} \in [0.5,1.5]$) towards earlier spiral types, hosting a bulge that encloses a larger fraction of the galaxy light: the colour gradient in the galaxies modelled as a single-S\'ersic profile therefore results in part from the bulge-disc colour difference shown in \sct\ref{sct-bulge-disk-color}. The fact that the transition in \fg\ref{color-grad-Re-1p} occurs at $n_{\IE} \approx 1.7$ may also be linked to the transformation of the pseudo-bulges of late spiral galaxies into classical bulges (with steeper profiles) hosted by earlier spiral and lenticular types, which was measured by \cite{Quilley-2023-scaling-bulges-and-disks} to take place in the $0.08\le B/T\le0.15$ interval for nearby SDSS galaxies. 

In \tab\ref{tab-n-sersic-color-grad} we also examine the variations in the single-S\'ersic profiles across the $\IE\le21$ sample by computing the mean, median and \rms dispersion, standard error and various quantiles of the difference between the S\'ersic indices of all pair of \Euclid bands, in the order of the index of the redder band minus that of the bluer one. The median difference takes values between $0.03$ to $0.15$ depending on the pair of bands, which indicates that redder bands not only have more concentrated light distribution (with smaller effective radii as shown in \fg\ref{color-grad-Re-1p}), but this light is also distributed in steeper profiles than in bluer bands. The effects are significant at the $3.8\sigma$, $5.2\sigma$, $5.3\sigma$, $2.5\sigma$, $4.0\sigma$, and $2.2\sigma$ confidence levels for the six listed pairs of bands. This result is again consistent with the bulges being on average redder than discs and having steeper profiles. The differences in S\'ersic indices between bands also increase with the distance between them in their central wavelength.

\begin{figure*}
\includegraphics[width=\textwidth]{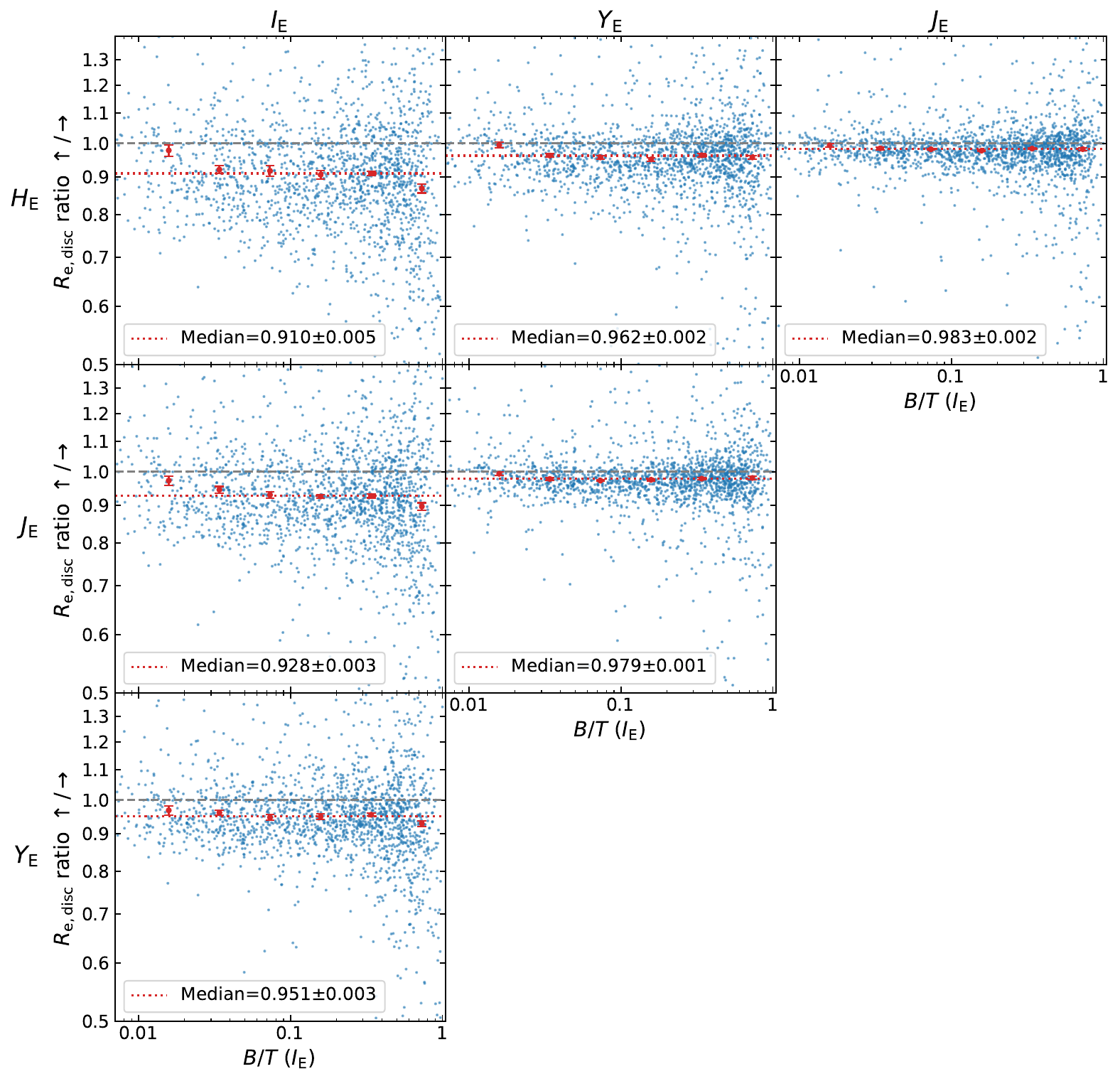}
\caption{Ratios of the effective radii of galaxy discs between different VIS and NISP bands as a function of the bulge-to-total ratio $B/T$ measured in the \IE band, from the bulge-disc decompositions of the $2047$ galaxies to $\IE\le21$, with $R_\mathrm{e,disc} / R_\mathrm{e,bulge} \ge 1$ and $R_\mathrm{e,disc}/R_\mathrm{e,1p}>0.9$ in the \IE band. All pairs of \Euclid bands are shown in the six panels, with the band whose $R_\mathrm{e,disc}$ is the numerator (denominator) appears on the left (top). Overall median values of the $R_\mathrm{e}$ ratio appear on each panel as a red dotted horizontal line, whereas the red points and error bars correspond to the median and standard errors for bins of $B/T$. Redder bands have smaller $R_\mathrm{e,disc}$ than bluer bands for the majority of galaxies, and this effect is stronger for bands that are further apart in wavelength. Most galaxies have blue to red inward gradients in their discs, which are the largest between the \IE band and any NISP band, and are interpreted as spatial variations in the disc's stellar populations at different distances from the disc centres.}
\label{color-grad-Re-disk}
\end{figure*}

\begin{table*}[ht]
\caption{Median values for the ratios of the disc effective radii between pairs of bands.}
\resizebox{\linewidth}{!}{
\begin{tabular}{ c c c c c c c c }
\hline\hline
$B/T$ & \multirow{2}{*}{${N_\mathrm{gal}}^{(b)}$} & \multicolumn{6}{c}{Median $R_\mathrm{e,disc}$ ratio$^{(c)}$ over pairs of bands} \\  
interval$^{(a)}$ & & \YE/\IE & \JE/\IE & \HE/\IE & \JE/\YE & \HE/\YE & \HE/\JE \\
\hline
$[0.01,0.02]$ & 131 & $0.968\pm0.014$ & $0.973\pm0.014$ & $0.978\pm0.018$ & $0.993\pm0.005$ & $0.995\pm0.008$ & $0.994\pm0.005$ \\
$[0.02,0.05]$ & 200 & $0.961\pm0.007$ & $0.945\pm0.010$ & $0.921\pm0.012$ & $0.977\pm0.004$ & $0.963\pm0.005$ & $0.985\pm0.005$ \\
$[0.05,0.1]$ & 256 & $0.947\pm0.009$ & $0.930\pm0.010$ & $0.917\pm0.015$ & $0.972\pm0.003$ & $0.957\pm0.005$ & $0.982\pm0.003$ \\
$[0.1,0.2]$ & 300 & $0.950\pm0.008$ & $0.925\pm0.005$ & $0.906\pm0.012$ & $0.975\pm0.003$ & $0.951\pm0.005$ & $0.977\pm0.003$ \\
$[0.2,0.5]$ & 507 & $0.955\pm0.004$ & $0.927\pm0.006$ & $0.910\pm0.006$ & $0.978\pm0.002$ & $0.963\pm0.005$ & $0.984\pm0.003$ \\
$[0.5,1]$ & 547 & $0.929\pm0.008$ & $0.897\pm0.010$ & $0.868\pm0.012$ & $0.981\pm0.005$ & $0.957\pm0.005$ & $0.982\pm0.004$ \\
$[0.001,1]$ & 2047 & $0.951\pm0.003$ & $0.928\pm0.003$ & $0.910\pm0.005$ & $0.979\pm0.001$ & $0.962\pm0.002$ & $0.983\pm0.002$ \\
\hline
\vspace{1pt}
\end{tabular}}
\small\textbf{Notes.}
The median values listed here correspond to those plotted in \fg\ref{color-grad-Re-disk} and are computed over logarithmic intervals of $B/T$.
$^{(a)}$ The bins are defined to be of equal logarithmic width over the $[0.01, 1]$ interval for the VIS $B/T$, as smaller values are most likely the result of the S\'ersic component not modelling the bulge. 
$^{(b)}$ Number of galaxies in the considered sub-sample of the $2047$ galaxies with $\IE\le21$, excluding the spurious bulge and disc decompositions (see text).
$^{(c)}$ Median values with bootstrap errors. 
\label{tab-Re-ratio-disk}
\end{table*}

\begin{table*}[ht]
\caption{Median values for the ratios of the bulge effective radii between pairs of bands.}
\resizebox{\linewidth}{!}{
\begin{tabular}{ c c c c c c c c }
\hline\hline
$B/T$ & \multirow{2}{*}{${N_\mathrm{gal}}^{(a)}$} & \multicolumn{6}{c}{Median $R_\mathrm{e,bulge}$ ratio$^{(b)}$ over pairs of bands} \\  
interval & & \YE/\IE & \JE/\IE & \HE/\IE & \JE/\YE & \HE/\YE & \HE/\JE \\
\hline
$[0.0,0.2]$ & 964 & $1.192\pm0.058$ & $1.300\pm0.076$ & $1.375\pm0.075$ & $1.000\pm0.000$ & $1.000\pm0.001$ & $1.000\pm0.000$ \\
$[0.2,0.4]$ & 421 & $0.999\pm0.003$ & $1.000\pm0.003$ & $1.000\pm0.004$ & $1.000\pm0.002$ & $0.998\pm0.005$ & $0.998\pm0.004$ \\
$[0.4,0.6]$ & 350 & $0.986\pm0.011$ & $0.981\pm0.014$ & $0.994\pm0.014$ & $1.000\pm0.002$ & $0.991\pm0.007$ & $0.990\pm0.007$ \\
$[0.6,0.8]$ & 231 & $0.965\pm0.022$ & $0.968\pm0.017$ & $0.968\pm0.026$ & $1.000\pm0.003$ & $0.999\pm0.010$ & $0.999\pm0.003$ \\
$[0.8,1.0]$ & 81 & $0.988\pm0.022$ & $0.979\pm0.033$ & $0.940\pm0.031$ & $1.000\pm0.006$ & $0.999\pm0.008$ & $0.999\pm0.006$ \\
$[0,1]$ & 2047 & $1.000\pm0.002$ & $1.002\pm0.004$ & $1.012\pm0.009$ & $1.000\pm0.000$ & $1.000\pm0.000$ & $1.000\pm0.000$ \\
\hline
\vspace{1pt}
\end{tabular}}
\small\textbf{Notes.}
Some of the median values listed here are plotted in \fg\ref{color-grad-Re-bulge}, and all are similarly computed over intervals of $B/T$.
$^{(a)}$ Number of galaxies in the considered sub-sample of the $2047$ galaxies with $\IE\le21$, excluding the spurious bulge and disc decompositions (see text).
$^{(b)}$ Median values with bootstrap errors. 
\label{tab-Re-ratio-bulge}
\end{table*}

We now examine how the gradients in the effective radii of the single-S\'ersic model may result from analogous gradients within the bulge, the disc, or both. The various panels of \fg\ref{color-grad-Re-disk} show, similarly to \fg\ref{color-grad-Re-1p}, the ratios in disc effective radii between all pairs of \Euclid bands, for galaxies with $\IE\le21$, as a function of the $B/T$ ratio, which characterises the Hubble type. Figure \ref{color-grad-Re-disk} shows that colour gradients are also visible within the discs, with median $R_\mathrm{e,disc}$ ratios of 0.95, 0.93, and 0.91 for the three NISP bands \YE, \JE, and \HE, over the \IE band, respectively (see \tab\ref{tab-Re-ratio-disk}), but are weaker than those in the single-S\'ersic profiles, since the bulge-disc colour difference is not contributing to the radii ratio. Our careful verification of the validity of our bulge and disc models (see \sct\ref{sct-quality-check}), as well as the persistence of disc colour gradients for low $B/T$ values and their lack of significant variation with $B/T$ (see \fg\ref{color-grad-Re-disk}), all suggest that the disc gradients are unlikely to result from the covariances between the bulge and disc components of the bulge-disc decomposition (this would cause bulge contamination into the disc).

\begin{figure}
\includegraphics[width=\columnwidth]{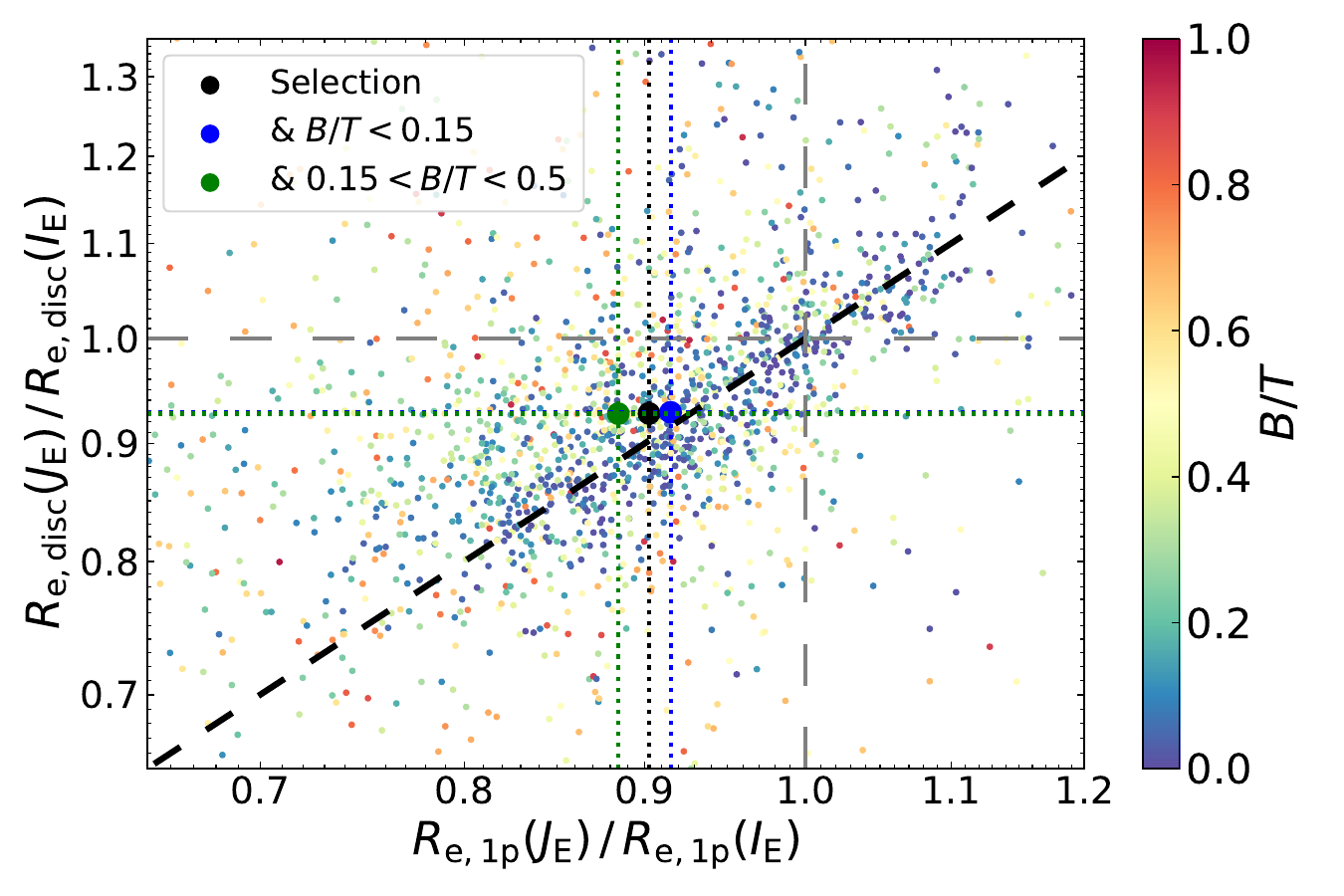}
\caption{Comparison between the ratio of effective radii measured in the \JE and \IE bands for the single-S\'ersic profile ($x$ axis) and for the disc component ($y$ axis) of the $2047$ galaxies to $\IE\le21$, with $R_\mathrm{e,disc} / R_\mathrm{e,bulge} \ge 1$ and $R_\mathrm{e,disc}/R_\mathrm{e,1p}>0.9$ in the \IE band. The colour of the points represent the bulge-total ratio $B/T$ in the bulge-disc decompositions. The black, blue, and green crosses correspond to the medians for the $[0,1]$, $[0,0.15]$, and $[0.15,0.5]$ $B/T$ subsamples, after selecting galaxies with both $R_\mathrm{e}$ ratios in the $[0.7,1.2]$ interval. Both ratios, hence colour gradients, show a similar behaviour, as seen in \fgs\ref{color-grad-Re-1p} and \ref{color-grad-Re-disk}, with ratios below unity indicating that redder stellar populations are more concentrated in galaxy centres. The disc and single-S\'ersic ratios are similar for $B/T \approx 0$ but for larger bulges, the single-S\'ersic ratio decreases further away from 1, whereas the disc ratio remains stable (bootstrap uncertainties are provided in the text). Because single-S\'ersic profile models both the bulge and the disc, and the latter is made up of bluer stars than the former in spiral galaxies, it displays a stronger overall colour gradient in galaxies with more prominent bulges.}
\label{color-grad-Re-disk-vs-1p}
\end{figure}

For direct comparison of the colour gradients of the single-S\'ersic profile and disc component, we compare in \fg\ref{color-grad-Re-disk-vs-1p} the $R_\mathrm{e,disc}$ and $R_\mathrm{e,1p}$ ratios between the \IE and \JE bands for the intersection of the samples after exclusion of the outliers in effective radii for both types of fits, the colour of the points representing their $B/T$ value. The bottom-left quadrant that corresponds to `redder inside - bluer outside' for both the disc and the single-S\'ersic models is the most populated, with $62.3\%$ of the sample against $18.4\%$, $6.1\%$, and $13.3\%$ in the top-left, bottom-right, and top-right quadrants, respectively. We further notice in \fg\ref{color-grad-Re-disk-vs-1p} that blue points representing bulge-less galaxies are aligned along this diagonal (the 575 galaxies with $B/T < 0.1$ have a Pearson correlation coefficient of 0.65), due to their similar values of effective radii of discs and single-S\'ersic fits, whereas green to orange points corresponding to discs hosting non-negligible bulges are preferentially spread out above the diagonal. This is illustrated quantitatively by the crosses plotted in \fg\ref{color-grad-Re-disk-vs-1p} indicating the median values and bootstrap errors for the band-to-band $R_\mathrm{e}$ ratios of galaxies limited to the $[0.6,1.2]$ interval for both \JE-to-\IE ratio, and for all values of $B/T$ (black circle and vertical line), for $B/T < 0.15$ (blue circle), and for $B/T \in [0.15,0.5]$ (green circle). Galaxies with non-existent to small bulges ($B/T < 0.15$) thus have consistent median values of their single-S\'ersic and disc \JE-to-\IE $R_\mathrm{e}$ ratio of $0.915$ $\pm0.004$ and $0.929$ $\pm0.005$, respectively (bootstrap errors were calculated for all median values in this graph). For galaxies with a more prominent bulge ($B/T \in [0.15,0.5]$), its presence strengthens the single-S\'ersic colour gradient but not that of the disc, therefore decreasing the $R_\mathrm{e,1p}$ ratio down to a median of $0.885$ $\pm0.006$, hence a $5.8 \sigma$ decrease below that of the $R_\mathrm{e,disc}$ ratio, whose median remains stable at  $0.927$ $\pm0.004$. The robustness of the \JE to \IE disc gradients to variations in $B/T$, contrary to the gradients in the single-S\'ersic models, suggests that the disc measurements are weakly affected by possible degeneracies between the bulge and disc components, hence by the bulge-disc red-blue dichotomy. Similar results are obtained for the gradients between \IE and the other two NISP bands.

\begin{figure}
\includegraphics[width=\columnwidth]{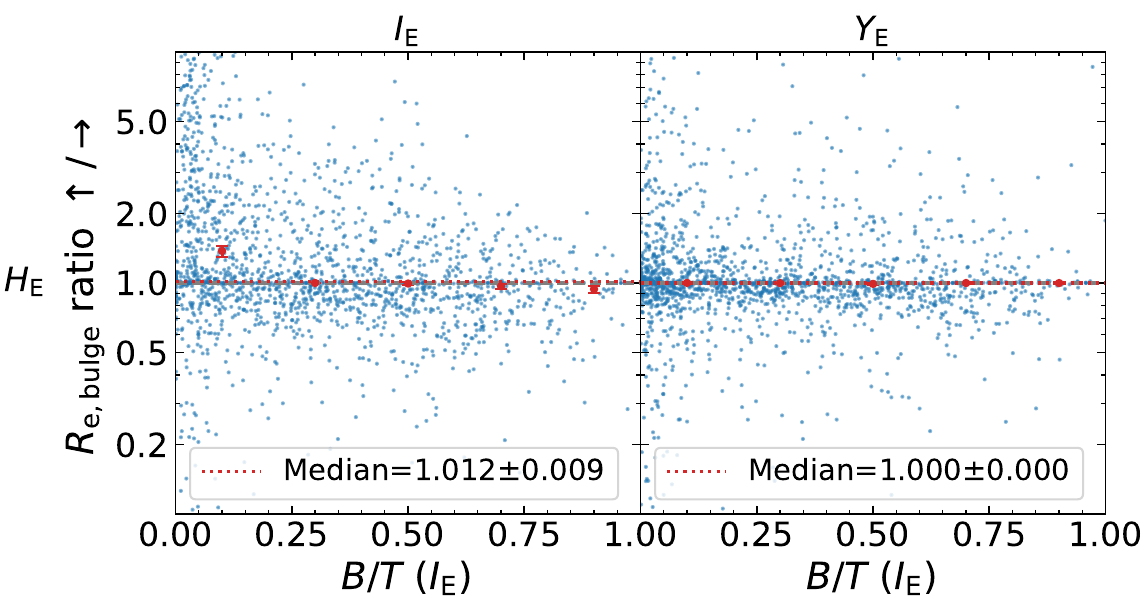}
\caption{Ratios of the effective radii of galaxy bulges between different VIS and NISP bands as a function of $B/T$ measured in the \IE band from the $2047$ galaxies to $\IE\le21$, with $R_\mathrm{e,disc} / R_\mathrm{e,bulge} \ge 1$ and $R_\mathrm{e,disc}/R_\mathrm{e,1p}>0.9$ in the \IE band. Only two pairs of \Euclid bands are shown as examples of an overall similar behaviour, with the band whose $R_\mathrm{e,disc}$ is the numerator (denominator) appearing on the left (top). Overall median values of the bulge $R_\mathrm{e}$ ratio appear on each panel as a red dotted horizontal line, whereas the red points and error bars correspond to the median and standard errors for bins of $B/T$. The median ratios are all consistent with a value of 1, except for the first interval in the left panel, affected by a larger dispersion and undersampling of the PSF, indicating that the present approach and the depth and resolution of the ERO-Perseus images do not allow one to detect colour gradients within the bulges of galaxies.}
\label{color-grad-Re-bulge}
\end{figure}

Finally, in order to examine whether colour gradients are detected in bulges, we calculate the ratios in bulge effective radii between pairs of \Euclid bands as a function of \IE $B/T$, for the same sample as in \fg\ref{color-grad-Re-disk}. We only display in \fg\ref{color-grad-Re-bulge} the \HE over \IE, and \HE over \YE ratios corresponding to the farthest apart infrared-optical and infrared-infrared pairs of bands (other pairs of bands show a similar behaviour at all values of $B/T$, see \tab\ref{tab-Re-ratio-bulge}). Unlike the median ratios for the single-S\'ersic profiles and disc components, the median ratios for the overall bulge sample have a value consistent with unity for all pairs of bands, hence no colour gradient is detected within bulges. The PSF and its undersampling (see \sct\ref{sct-psf}) having potentially more impact on bulges with small effective radii (hence with small $B/T$ values) they may contribute to the increased scatter in the lowest $B/T$ bin, particularly in the \IE band.

Variations in effective radii of galaxies have been measured from various ground-based surveys using only single-S\'ersic fits in optical in near-infrared bands and provide comparable results. \citet{Ko-Im-2005-color-gradients-ETG} first measured a lower $R_\mathrm{e}$ in the $K$ than in the $V$ band for 273 nearby early-type galaxies (using de Vaucouleurs profile fits), with a weaker variation in dense galaxy environments, interpreted as further mixing of different stellar populations due to a history with more frequent mergers. Colour gradients were then also observed in 5080 early-type galaxies to $z<0.1$ by \cite{La-Barbera-2010-color-gradients}, who quantified the decrease in the single-S\'ersic effective radii with wavelength across optical and near-infrared bands ($grizYJHK$). Studies of even larger samples from the Galaxy And Mass Assembly (GAMA) survey out to $z<0.3$, confirmed the prevalence of optical and near-infrared ($ugrizYJHK$) $R_\mathrm{e}$ colour gradients within all galaxy types (also based on single-S\'ersic fits). The derived variations in $R_\mathrm{e}$ range from a 15\% to 50\% decrease with wavelength, depending on the considered pairs of bands and the galaxy types, with stronger effects for early-type galaxies \citep{Kelvin-2012-GAMA-sersic-fits, Hausler-2013-multiband-sersic-profile-MegaMorph, Vulcani-2014-GAMA-color-gradients, Kennedy-2015-GAMA-color-gradients-vs-z-lum}. In addition, these analyses highlighted an increase in the S\'ersic index $n$ with wavelength by sometimes as much as 50\%, and studied these dependencies in galaxy populations of various colours and morphologies. The stronger gradient for larger values of $n$ measured here (shown in \fg\ref{color-grad-Re-1p} and \tab\ref{tab-Re-ratio-1p}) is for example consistent with the results of \cite{Vulcani-2014-GAMA-color-gradients}, who found a steeper decrease of $R_\mathrm{e}$ with wavelength for galaxies with $n>2.5$ compared to those with $n<2.5$. Lastly, non-parametric morphology of several hundreds of galaxies using integral field spectroscopy yield an analogous galaxy size increase with wavelength in the optical range \citep{Nersesian-2023-non-param-color-gradients}.

The present analysis provides the first measure of disc gradients out to redshifts of $0.5$, and is made possible by the exquisite \Euclid angular resolution. Disc colour gradients that vary with $B/T$ are also observed in the EFIGI nearby galaxies by \cite{Quilley-2025-colors-color-gradients}, using images from the Sloan Digital Sky Survey (SDSS) and comparing the disc effective radii in the $g$, $r$, and $i$ bands. This EFIGI analysis also detects colour gradients in the bulges of spiral galaxies with a large dispersion that is interpreted as the signature of bulge growth.

Direct comparison between \Euclid bands and the SDSS filter system, on which are based many previous analyses, is, however, not straightforward due to the large width of the \IE band that encompasses the SDSS $r$, $i$, and $z$ bands. For galaxies at $z=0.24$, which is the median redshift of galaxies with $\IE\le21$ in this study \citep{EROPerseusOverview}, the observed \IE band includes in rest-frame the red side (above about $4300\,\AA$) of the $g$ band, and the blue side (below about $7400\,\AA$) of the $r$ band, whereas the \YE, \JE, and \HE bands correspond to rest-frame wavelengths of approximately $7600$--$10\,000\,\AA$ (matching the union of the SDSS $r$ and $z$ bands), $9300$--$13\,000\,\AA$ (the \YE band and the blue part of the \JE band), and $12\,100$--$16\,600\,\AA$ (the \JE band). 

At higher redshifts in the $z=0.5$--$2.0$ range for which \Euclid was designed, the \Euclid bands become narrower and closer to the common optical bands when blue-shifted to the rest-frame. For instance, at $z=0.5$, \IE detects the 3300--6200$\,\AA$ interval in the rest-frame, which corresponds to the $u$ and $g$ bands plus the blue part of the $r$ band, and at $z=2$ the three NISP bands \YE, \JE, and \HE correspond roughly to the SDSS $u$, $g$, and $r$ bands. However, only a few such high-redshift galaxies are observed in the ERO-Perseus images at $\IE\le21$, the limiting magnitude used here for bulge-disc decomposition (see \sct\ref{sct-quality-check}), since only 1\% of galaxies have $z>0.58$ at this limiting magnitude. 
To be able to resolve the inner structure of high-redshift galaxies at $\IE=21$ and fainter, the enhanced depth of the Euclid Deep Survey will be needed (see predictions from \citealt{Bretonniere-EP13}).

\subsection{Interpretation of colour variations in bulges and discs\label{sct-color-variation-interpretation}}

The median $\IE - \JE$ bulge-disc colour differences presented in \sct\ref{sct-bulge-disk-color} as well as the colour gradients described in \sct\ref{sct-bulge-disk-gradient} could be interpreted at face value as gradients in the dominant stellar populations, which could be distributed differently within bulges and the internal part of galaxies compared to within the disc and its external parts, respectively. Stellar metallicity variations and their degeneracies with stellar age complicate their disentanglement within the stellar gradients \citep{Worthey-1994-disentanglement-age-metallicity}. Nevertheless, direct evidence of age and metallicity gradients across Hubble types and their bulge and disc components was provided by the integral field spectroscopy analysis of the CALIFA sample by \citet{Gonzales-Deldago-2015-CALIFA-Hubble-sequence-age-metallicity-gradients}. More recently, \citet{Jegatheesan-2024-MANGAIII-assembly-histories-bulges-disks} showed using the MANGA III sample that bulges are in general older and more metal-rich than discs across the Hubble sequence.

Numerous analyses of observed galaxy samples are quoted at the end of \scts\ref{sct-bulge-disk-color} and \ref{sct-bulge-disk-gradient}, regarding the colour and colour gradients of galaxies and showed that attenuation by internal dust can affect them as a function of disc opacities and inclinations. Radiative transfer modelling of the stellar light attenuation by dust within the bulges and discs of galaxies have provided quantitative measures of the impact of dust extinction on their colours and effective radii. For a giant late-type galaxy like the Milky Way, \citet{Pierini-2004-dust-extinction-bulge-disk} derived a $B-V$ reddening of $\le0.075$ for the bulge (for all inclinations), and $0.03$ for face-on discs (steeply increasing with inclination, up to $0.2$ when edge-on). Still with radiative transfer calculations, \citet{Gadotti-2010-dust-extinction-bulge-disk} calculated that with dust extinction, disc effective radii are overestimated by up to $30$\% depending on the face-on optical depth and inclination ($\le60^\circ$), whereas bulge effective radii and S\'ersic indices are underestimated by up to a factor of 2. \citet{Popescu-2011-SED-disk-dust-PAH} further explored the ultraviolet-to-far infrared modelling of galaxy discs as a function of the $B/T$ morphological parameter and disc inclination, by adding a clumpy dust component to the diffuse one in the old and young disc stellar populations. Using bulge-disc decomposition of simulated images of discs and bulges produced with radiative transfer calculations, \citet{Pastrav-2013-dust-photometry-disk-bulge-spirals} showed that dust extinction increases the effective radii of discs (and reduces their S\'ersic indices), with a larger effect at shorter wavelength and with increasing disc inclination, but with a small impact on the effective radii of bulges.

Interestingly, \citet{Mosenkov-2019-dust-emission-FIR} directly mapped the bulk of dust emission in spiral galaxies by modelling the far infrared emission of disc galaxies with single-S\'ersic profiles. For galaxies with no star formation in their centre, they measured a dust depletion in the inner regions, at 30 to 40\% of the $R_{25}$ radius to the 25th mag isophote in the optical. Using $R_{25}/R_\mathrm{e,\rm disc}$ estimates from nearby galaxies \citep{Quilley-2023-PhD-thesis}, the radius of this depletion region is a factor of $0.6$ times the bulge effective radius for an average S0 type, which increases up to factors of 1.3 for an Sab, and 2 for an Sm, therefore limiting the effects of extinction in most bulges.

These various analyses illustrate the large variability in the impact of dust extinction on galaxy structural parameters. Disentangling dust extinction from the relative effects of stellar age and metallicity has been less explored. Based on gradients calculated from the 1D profiles of nearby SDSS galaxies ($z\lesssim0.05$), \citet{Tortora-2010-color-gradients-correlation-mass} studied the relative roles of metallicity and stellar age gradients onto the colour gradients from single-S\'ersic profiles, with a dominant role of the metallicity in late-type galaxies. Using a similar 1D approach on an order of magnitude larger sample, \citet{Liao-2023-color-gradients-lowz-DESI} measured optical colour gradients of the inner discs of spiral and irregular galaxies to $z\le0.1$, and evidence of a dust contribution, beyond age and metallicity gradients, derived from integral field spectroscopy.
 
The analysis of \citet{Baes-2024-Re-mass-light-wavelength} based on a hydrodynamical cosmological simulation proposes an interesting way to lift the degeneracies in the origin of the colour gradients by showing for galaxies of all types from a cosmological simulation, that the colour gradients in effective radii (measured by elliptical growth curves) actually originate from gradients in both the ages and metallicities of the stellar populations, as well as from the dust attenuation effects, with a relative contribution of around $80\%$ and $20\%$, respectively. However, other studies have found a more significant impact of dust on the observed colour gradients \citep{Marshall-2022-impact-dust-sizes-galaxies-EoR, Miller-2023-color-gradients-mass-vs-light-sizes, Nedkova-2024-mass-size-z-3}, mainly at higher redshifts, so this issue remains an active area of research. Still, assuming that the result from \citet{Baes-2024-Re-mass-light-wavelength} is valid, we could conclude that the bulge-disc colour difference and colour gradients detected in \scts\ref{sct-bulge-disk-color} and \ref{sct-bulge-disk-gradient} could be mostly the result of gradients in the stellar population, with older (or more metal rich) stars in the bulges and internal discs, and younger (or less metal rich) stars in the external parts of discs.

\section{Implications of bulge-disc morphology on cosmic shear and spin alignments \label{sct-check-shear-spin}}

Both the biases in the structural parameters of galaxies detected by \Euclid presented in \sct\ref{sct-quality-check}, and the colour variations with morphology described in \sct\ref{sct-color-gradients} might affect the cosmological analyses of the mission, since they are based on galaxy shapes. Measurements of the cosmic shear from the Euclid Wide Survey \citep{Scaramella-EP1}, one of the primary probe of \Euclid, is aimed at constraining the dark matter and the dark energy of the Universe. The shear is defined as the weak lensing of background galaxies by the large-scale mass distribution intervening along the line-of-sight, which causes coherent stretching and alignment of neighbouring galaxies on the sky. This therefore requires the measurement of the apparent axis ratio of galaxies and their relative PA (see \sct\ref{sct-check-angle}). Figure \ref{aspect-ratio-bias} shows that deriving each galaxy's axis ratio from the single-S\'ersic fits will be susceptible to significant dispersion as well as systematic biases, due to the presence of bulges with varying $B/T$, hence depending on their intrinsic morphologies. The colour variations across galaxy surface brightness distributions shown in \sct\ref{sct-color-gradients} will also potentially have a double impact on the biases in the shear, due to the varying PSF with colour within the wide \IE\ band \citep{Voigt-2012-galaxy-color-gradients-shear, Semboloni-2013-shear-bias-galaxy-profile-color,Er-2018-color-gradient-bias-shear}, and the choice of the PSF for modelling galaxies with spatially variable colours. Therefore, any forward modelling approach to control the systematics should include detailed knowledge of the morphology of the sources, in particular the distribution functions of their bulge and disc parameters with wavelength at the various redshifts at which they are observed. 

Another analysis in which \Euclid could have a significant impact is the measurement of the spin alignment of discs with cosmic filaments \citep{Dubois-2014-cosmic-web-spins}. The spins of discs are defined as their rotation axes, and their angle with the filaments can be derived from the combination of their axis ratio and their PA, compared to the orientation of the filaments. If the graphs in \fg\ref{angle-diff} indicate that the PA of the single-S\'ersic profile is a good approximation of the disc PA from the bulge-disc decomposition (see \sct\ref{sct-check-angle}), measuring the disc inclinations will be affected by the biases in the axis ratio if one uses the single-S\'ersic profiles (see \sct\ref{sct-check-elong}). One should instead use the disc axis ratios measured from the bulge-disc decompositions. 

\section{Summary\label{sct-summary}}

In this study, we use the first \Euclid data obtained within the EROs of the Perseus cluster \citep{EROData,EROPerseusOverview}, to test the ability of \Euclid images to characterise the morphology of $2445$ and $12\,786$ galaxies at $\IE\le21$ and $\IE\le23$, respectively. After calculating the point-spread functions in the VIS \IE band, and the NISP \YE, \JE, and \HE bands, with $0\arcsecf16$, $0\arcsecf47$, $0\arcsecf49$, $0\arcsecf50$ FWHM, respectively, we perform luminosity model fitting of the VIS and NISP images with the \texttt{SourceXtractor++} software \citep{Bertin-2020-SourceXtractor-plus-plus, Kummel-2022-SE-use}, whose efficiency was demonstrated in the \Euclid Morphology Challenge \citep{Bretonniere-EP26}. We also perform SED-fitting of the model photometry obtained with \texttt{HyperZSpec} and derive absolute magnitudes for whole galaxies, bulges and discs with \texttt{Hyp\_Magabs}.

Galaxy brightness distribution models have elliptical symmetry and are either a single-S\'ersic profile or the sum of a S\'ersic bulge and an exponential disc. We analyse the results of several fitting configurations, given that the single-S\'ersic and bulge-disc modelling used here have minimal priors: there is no condition relating a given parameter between the bulge and disc components or between different bands, except in a specific test performed on the three NISP bands (see Appendix \ref{appendix-common-profile}). Visual examination of numerous fitted galaxies in the \IE images shows nearly ubiquitous bulge and disc components in galaxies that are not irregulars, with all values from 0 to 1 in their bulge-to-total light ratio $B/T$. Moreover, disc galaxies frequently show well defined spirals arms,  flocculence due to \ion{H}{ii} regions, patchy dust, bars and rings, hence resembling altogether the morphological types of the Hubble sequence built on nearby galaxies \citep{Hubble-1926-extragalactic-nebulae, De-Vaucouleurs-1959-class-morph}, and  further explored by \citet{Baillard-2011-EFIGI}. 

To assess the consistencies and discrepancies between the single-S\'ersic profile model and the bulge-disc decomposition, we analyse the multi-variate distributions of the derived fluxes and structural parameters, \ie their major axis effective radii, centres, S\'ersic indices, axis ratios of elliptical models, position angles of ellipse major axes, and total magnitudes of the $2445$ galaxies with $\IE\le21$, the magnitude limit at which the \Euclid morphology challenge showed that bulge-disc decomposition can reliably recover the object parameters \citep{Bretonniere-EP26}.

By comparing the effective radii of the single-S\'ersic model with those for the bulge and disc model components, in \sct\ref{sct-check-Re}, we show consistency between the former and the dominant disc or bulge component for extreme bulge-to-total light ratios of $B/T\approx0$ or $B/T\approx1$, respectively. For 91\% of galaxies, the disc effective radius is larger than the bulge radius, and the single-S\'ersic effective radius is intermediate between them depending on the $B/T$ value, which describes the full interval from 0 to 1. The single-S\'ersic effective radius should therefore be used with caution. We also derive a relation between the effective radii of the bulge, disc, single-S\'ersic models, and $B/T$. Visual examination shows that the 9\% of outlier fits in the plane of ratios of effective radii include non-physical models of the bulge and disc, which are inverted or their centres are kept apart within the galaxy, or the bulge is contaminated by a bar which may strongly bias its parameters. The outliers also include galaxies for which the single-S\'ersic model cannot fit both the peaked bulge and the extended disc, which a bulge-disc decomposition succeeds to do. Examples of these various fits are shown in Appendix A. 

In \sct\ref{sct-check-position} we report on measured offsets between the centre of the single-S\'ersic model and either the bulge or disc component centres by the single-S\'ersic effective radius or more for a total of $3.3\%$ of galaxies, with separations for the rest of the objects peaking at less than one-tenth of the single-S\'ersic effective radius for both components. Exploring these offsets in the profile centres are useful as they allow us to detect outliers such as asymmetric galaxies due to fly-bys or mergers, as well as non-physical bulge-disc fits in which the galaxies cannot be modelled by a central light excess and a surrounding disc (for example, edge-on galaxies with strong dust lanes masking any bulge), or because of a contaminating star or strong \ion{H}{ii} region (examples of such fits are shown in Appendix A). 

In \sct\ref{sct-check-n} we examine how that the indices of the single-S\'ersic model $n$ and that of the bulge component $n_\mathrm{B}$ from the bulge-disc decomposition vary with $B/T$. If $n$ is commonly used to roughly split early and late galaxy types, there are large degeneracies in its measurement, and it cannot be used to separate Hubble morphological types. This requires bulge-disc decomposition and the use of reliable measures of $B/T$ as a proxy for type (which could be complemented by bulge and disc parameters).

Comparison of the axis ratio of the disc model component with the single-S\'ersic profile model in \sct\ref{sct-check-elong} shows that they agree on average only for face-on discs with low values of $B/T$. For disc galaxies with some inclination and $B/T\simeq0.1$--$0.9$, that is intermediate and early spirals, as well as lenticulars, the single-S\'ersic modelling overestimates their disc axis ratio, hence underestimating the disc inclination, by an amount that increases with both $B/T$ and the disc inclination itself. The single-S\'ersic axis ratio is also an  overestimate of the bulge axis ratio for weakly inclined discs, or an underestimate of the bulge axis ratio for highly inclined galaxies, with a decreasing effect for more prominent bulges. Contrary to the other parameters, the position angles of the single-S\'ersic profile major axes can be used as reliable indicators of the galaxy disc major axis position angles (\sct\ref{sct-check-angle}).

We measure in \sct\ref{sct-model-photo} the systematic differences between adaptive aperture, single-S\'ersic, and bulge-disc magnitudes. We calculate median VIS offsets between the models and aperture magnitudes of 0.04 mag for the single-S\'ersic profile and 0.07 mag for the bulge-disc decomposition, because the bulge-disc decomposition better models simultaneously the central bulge peak and the disc extension when there is a significant bulge. In contrast, model magnitudes of galaxies with negligible bulge components benefit from the free index of the single-S\'ersic parameter. We also show that these model-aperture magnitude offsets depend on the morphology of galaxies and steadily increase with the index of the single-S\'ersic model and $B/T$ of the bulge-disc decomposition, from $-0.08$ to 0.01 mag for $B/T$ values from 0.001 to 1 (and with \rms dispersions increasing from 0.05 to 0.15 mag). Moreover, we show that the offsets between the two types of model magnitudes can be parametrised linearly by the deviation from the relation between the single-S\'ersic index and $B/T$.

One important result of the bulge-disc decompositions performed on the ERO-Perseus field galaxies (see \sct\ref{sct-bulge-disk-color}) is that the median $M_g - M_i$ rest-frame colour of the significant disc components is $0.64\pm0.01$, whereas the significant bulges have a redder median colour of $0.98\pm0.02$ (significant meaning a component enclosing more than 10\% of the total galaxy light). Using the standard errors, this colour difference is significant at the $17\sigma$ level. The bulge-disc $M_g - M_i$ colour difference is the strongest for small bulges ($B/T < 0.05$), and decreases progressively with increasing $B/T$ to reach similar colours of bulges and discs for bulge-dominated galaxies ($B/T > 0.5$).

Finally, by comparing the effective radii of the single-S\'ersic fits in the VIS and NISP bands, we report in \sct\ref{sct-bulge-disk-gradient} on systematic offsets from unity in their ratios at the 7 to 13\% level between VIS and each of the NISP bands (with percent level uncertainties on these values), and at the 3 to 6\% level among the NISP bands. These effective radii offsets are interpreted as colour gradients in the modelled galaxies, and they increase with the index of the single-S\'ersic profile, which is likely due in part to the bulge-disc colour difference, and to the increase in the $B/T$ ratio with S\'ersic index. Larger S\'ersic indices, hence steeper profiles, are also detected in redder bands. Remarkably, when separating the bulge and disc components, these band-to-band gradients remain in the disc models at the 5 to 10\% level. They are undetectable in the bulge models. 

\section{Conclusions and perspectives\label{sct-conclusions}}

While it has been previously shown from space-based imaging that the Hubble sequence of ellipticals, lenticulars, spirals, and irregular galaxies was in place at redshifts $z\approx1$--3, with the current analysis we took one step further by performing a quantitative morphological analysis of the properties of galaxies detected with \Euclid imaging. Through bulge-disc decomposition to $\IE\le21$ of the background galaxies in the EROs of the Perseus cluster, we have shown that galaxies at $z\le0.54$ that are not of irregular type show ubiquitous bulge and disc components, with a $B/T$ taking all values from 0 to 1. We noticed by visual examination of hundreds of modelled galaxies (a majority with $\IE\lesssim21$) that disc galaxies mainly exhibit dynamical structures such as spiral arms, bars, and rings and textures such as flocculence and dust. The amplitude and frequency of these features, however, may be different from those in the nearby Universe \citep{de-Lapparent-2011-EFIGI-stats}, and other features unseen in the nearby Universe may have been missed in the limited number of visually examined galaxies. These galaxies nevertheless resemble in their morphologies the present-time morphological types in the Hubble sequence \citep{Hubble-1926-extragalactic-nebulae} seen from the ground and at redshifts $\leq0.05$ \citep{Baillard-2011-EFIGI}, even though we are observing them up to 5 billion years ago.
The fainter galaxies with $21\lesssim\IE\lesssim23$ are too poorly resolved to draw detailed conclusions about them, but they show consistent single-S\'ersic fit parameters with galaxies at $\IE\lesssim21$. 

In this present analysis, we emphasise the use of bulge-disc decomposition for reliable characterisation of galaxy morphologies and sizes for galaxies hosting both a bulge and a disc component. The single-S\'ersic parameters were shown to be intermediate combinations of bulge and disc parameters, depending on the value of $B/T$. These biases measured in characterising galaxy morphology by a single-S\'ersic profile apply to all lenticulars and spiral galaxies and hence to the vast majority of galaxies per unit volume at a given absolute magnitude limit. Bulge-disc decomposition should be performed to the reliable detection limits of these components. If based on single-S\'ersic modelling, morphological studies beyond these limits are difficult to compare with bulge-disc measurements.

A significant population of outliers are the barred galaxies in which, without priors constraining the bulge to fit the central peak, a bulge-disc decomposition tends to fit an elongated S\'ersic profile onto the bar, which biases the bulge effective radius, axis ratio, S\'ersic index, and flux (hence $B/T$). This major degeneracy should be circumvented either by adding a bar component to the model (and potentially increasing the degeneracies) or using priors on the bulge and disc parameters, as allowed by \texttt{SourceXtractor++}. The latter option was successfully applied to the EFIGI nearby galaxies using the constraint that a bulge is a weakly elongated peak of light near the centre of the disc \citep{Quilley-2022-bimodality, Quilley-2023-PhD-thesis}. More complex models including bars, rings, and even spiral arms would only make small differences in measuring fluxes and characterising morphologies (via $B/T$ and the parameters of the bulge and disc components), whereas the gain in precision for sizes and fluxes from single-S\'ersic modelling to bulge-disc decomposition is significant.

One remarkable result of the present analysis is the bulge-disc colour difference and single-S\'ersic colour gradients detected at \IE$\le21$, which are similar to those obtained in other analyses from ground-based data and at lower redshifts. The measured disc colour gradients are, however, unprecedented at redshifts up to $z\le0.5$ and are in agreement with similar evidence of disc galaxies at $z\le0.05$ \citep{Quilley-2025-colors-color-gradients}. 
Altogether, the characteristics of bulges and discs obtained from these \Euclid data reinforces the derived bulge-disc decompositions as realistic approximations of the physical components of the modelled galaxies. These results are also consistent with the picture of galaxy evolution from later to earlier Hubble types by bulge growth and disc quenching proposed by \citet{Quilley-2022-bimodality}. These colour effects will be key to providing renewed insight into the stellar assembly of bulges and discs, when better measured and analysed with the larger statistics (and depth) of the forthcoming Euclid Wide and Deep Surveys \citep{Scaramella-EP1}.

In the meantime, the Q1 data (covering the three Euclid Deep Survey fields, that is $53$ deg$^2$, to the depth of the Euclid Wide Survey) have become available, and the data increase the available statistics by a factor of approximately $100$ compared to the single ERO-Perseus field. We emphasise that the non-negligible redshift interval spanned by \Euclid's forthcoming data releases will enable studies with sufficient statistics in order to define sub-samples per redshift interval. The photometric redshift estimates derived from the external photometry will be critical for such analyses \citep{Laigle-2019-Horizon-SED-fitting-performance-surveys, EC-Desprez-2020-EP-X-photo-z-challenge, EP-Paltani}. The stellar masses, star-formation rates, and absolute sizes of the galaxies provided by the photometric redshift analyses will also allow us to further study the variations in the colours and colour gradients of the Hubble types in terms of  galaxy formation and evolution.

We caution that the chromatic surface brightness modulation (sometimes summarised as the differential k-correction) that affects all galaxies  \citep{Papaderos-2023-chromatic-SB-modulation} impacts the bulge-disc colour difference as well as the colour gradients of galaxies for increasing redshifts. For example, at redshifts of 1 and beyond, the \IE band detects mostly UV emission, in which bulges become dramatically attenuated compared to discs.

Intrinsic effects due to possible variations with redshift in the spatial distribution of galaxy stellar populations across the morphological components will only be reliably studied using full forward modelling, which should include realistic SED modelling of the galaxy bulges and discs, including their specific internal extinction by dust. Such an approach was proposed by \citet{Livet-2021-dimensional-reduction-deep-images} based on massive dimensional reduction of deep survey images with neural networks, allowing catalogue-free modelling and inference, which circumvents the various observational biases \citep{Carassou-2017-inferring-photometric-size-evolution} that differentially affect the bulges and discs of galaxies at increasing redshifts \citep{Papaderos-2023-chromatic-SB-modulation}.

The billions of galaxies that will be detected in the \Euclid surveys thanks to a high angular resolution and wide field of view will require reliable automated morphological classification in order to create an inventory of the various galaxy populations and characterise their stellar contents. Profile modelling is a deterministic avenue towards dimensionality reduction, and it does not require a labelled sample. The current analysis illustrates the potential of bulge-disc decomposition for morphological analyses, and the limitations of the faster single-S\'ersic fits.

As a result, any forward modelling for controlling the systematics in the \Euclid cosmic shear measurements must include realistic distribution functions of the bulge and disc parameters. Measurements of the galaxy spin alignments with the cosmic web, which also require precise measurements of both the galaxy disc inclinations and their position angles, should gain in precision and reduced systematics by using the disc parameters derived from bulge-disc decompositions (see \sct\ref{sct-check-shear-spin}).

These cosmological analyses will in turn require that reliable morphological galaxy catalogues be obtained by the \Euclid surveys. As part of the \Euclid legacy science, these catalogues will also allow us to study the evolution of the Hubble sequence with large statistical samples since the cosmic noon epoch at $z\sim2$ \citep{Madau-Dickinson-2014-cosmic-SFH}. By allowing us to measure the frequency variations of the Hubble types with redshift and across large-scale structures, the \Euclid surveys are expected to provide unprecedented insight into the morphological transformations of galaxies in relation to their star-formation and merger histories from cosmic noon to the present epoch. 

\medskip

\section*{Data availability}Tables A.1, A.2, and A.3 are only available in electronic form at the CDS via anonymous ftp to \url{cdsarc.u-strasbg.fr} (130.79.128.5) or via \url{http://cdsweb.u-strasbg.fr/cgi-bin/qcat?J/A+A/}.}

\begin{acknowledgements}
Louis Quilley acknowledges funding from the CNES postdoctoral fellowship program.
\AckERO  
\AckEC
\end{acknowledgements}

\bibliographystyle{aa}
\bibliography{biblio}

\begin{appendix}

\onecolumn

\section{Data tables}

The data used in the present analysis are made available in FITS format at the \textit{Centre de Donn\'ees astronomiques de Strasbourg} (CDS)  for all galaxies with \IE$\le23$ for single-S\'ersic profiles and \IE$\le21$ for bulge and disc profiles. Tables \ref{tab:single_sersic} and \ref{tab:bulge_disk} describe the derived parameters from the single-S\'ersic models and the bulge-disc decompositions of the projected brightness distribution of galaxies, respectively. The parameters obtained by the SED fits to the model photometry are listed in Table \ref{tab:sed_fitting}.

\begin{table*}[htbp]
\centering
\caption{Single S\'ersic profile fit parameters.}
\label{tab:single_sersic}
\begin{threeparttable}
\begin{tabular}{lll}
\toprule
\textbf{Parameter} & \textbf{Description} & \textbf{Units/Notes} \\
\midrule
\verb|ID| & Unique identifier for the source  & \\
 & from \cite{EROPerseusOverview} & \\
\verb|Separation| & Separation between modelled sources and  & arcsec \\
 & detection from \cite{EROPerseusOverview} & \\
\multicolumn{3}{l}{\textit{Source identification and geometry:}\tnote{1}} \\
\verb|source_flags| & Flags indicating source properties or issues & \\
\verb|pixel_centroid_x,pixel_centroid_y| & X-, Y-coordinates of the centroid in pixels & pixels\tnote{2} \\
\verb|world_centroid_alpha| & Right ascension (J2000) of the centroid & degrees \\
\verb|world_centroid_delta| & Declination (J2000) of the centroid & degrees \\
\verb|ellipse_a| & Semi-major axis of the ellipse & pixels \\
\verb|ellipse_b| & Semi-minor axis of the ellipse & pixels \\
\verb|ellipse_theta| & Position angle of the ellipse major axis\footnote{This is the position angle between the major axis of the ellipse and the CCD $x$-axis, measured counter-clockwise between $-\pi/2$ and $\pi/2$ radians.} & degrees \\
\verb|ellipse_c_xx,ellipse_c_yy,ellipse_c_xy| & Second moments of the ellipse & pixels$^2$\\
\verb|area| & Area of the source & pixels$^2$ \\
\verb|elongation| & Elongation of the source & \\
\verb|ellipticity| & Ellipticity of the source & \\
\addlinespace[0.1cm]
\multicolumn{3}{l}{\textit{Photometry:}\tnote{1}} \\
\verb|isophotal_flux|, \verb|isophotal_flux_err| & Isophotal flux and error & counts \\
\verb|isophotal_mag|, \verb|isophotal_mag_err| & Isophotal magnitude and error & mag \\
\verb|auto_flux_[BAND]|, \verb|auto_flux_err_[BAND]| & Auto flux and error in bands\tnote{3} & counts \\
\verb|auto_mag_[BAND]|, \verb|auto_mag_err_[BAND]| & Auto magnitude and error in bands\tnote{3} & mag \\
\verb|flux_radius_1_[BAND]| & Radius containing half the flux in bands\tnote{3} & pixels \\
\addlinespace[0.1cm]
\multicolumn{3}{l}{\textit{Association and fitting:}\tnote{1}} \\
\verb|assoc_match| & Association match flag & \\
\verb|fmf_reduced_chi_2| & Reduced chi-squared of the fit & \\
\verb|fmf_iterations| & Number of iterations for the fit & \\
\verb|fmf_stop_reason| & Reason for stopping the fit & \\
\verb|fmf_duration| & Duration of the fit & seconds \\
\verb|fmf_flags| & Flags for the fit & \\
\verb|fmf_chi2_per_meta_[1-5]| & Chi-squared per meta-iteration & \\
\verb|fmf_iterations_per_meta_[1-5]| & Iterations per meta-iteration & \\
\verb|fmf_meta_iterations| & Number of meta-iterations & \\
\addlinespace[0.1cm]
\multicolumn{3}{l}{\textit{Single S\'ersic profile parameters:\tnote{3}}} \\
\verb|x_[BAND]|, \verb|x_[BAND]_err| & X-coordinate of centroid and error& pixels \\
\verb|y_[BAND]|, \verb|y_[BAND]_err| & Y-coordinate of centroid and error & pixels \\
\verb|flux_[BAND]|, \verb|flux_[BAND]_err| & Total model flux and error & counts \\
\verb|mag_[BAND]|, \verb|mag_[BAND]_err| & Total model magnitude and error & mag \\
\verb|nsersic_[BAND]|, \verb|nsersic_[BAND]_err| & S\'ersic index and error & \\
\verb|rad_[BAND]|, \verb|rad_[BAND]_err| & Effective radius along the semi-major axis, and error & pixels \\
\verb|rad_arc_[BAND]|, \verb|rad_arc_[BAND]_err| & Effective radius along the semi-major axis, and error & arcsec \\
\verb|aspect_[BAND]|, \verb|aspect_[BAND]_err| & Axis ratio and error & \\
\verb|angle_[BAND]|, \verb|angle_[BAND]_err| & Position angle and error & radians \\
\verb|angle_deg_[BAND]|, \verb|angle_deg_[BAND]_err| & Position angle and error & degrees \\
\bottomrule
\end{tabular}
\begin{tablenotes}
\item[1] More information about the computation of these parameters is available in the documentation of \texttt{SourceXctractor++} at \url{https://sourcextractorplusplus.readthedocs.io/en/latest/}
\item[2] The VIS image is used as a detection image so all sizes given in pixels use the VIS pixel scale of \ang{;;0.1}
\item[3] \verb|[BAND]| is one of: \verb|IE|, \verb|YE|, \verb|JE|, \verb|HE|.
\end{tablenotes}
\end{threeparttable}
\end{table*}

\begin{table}[htbp]
\centering
\caption{Parameters of the bulge (S\'ersic) and disc (exponential) components of the fits.}
\label{tab:bulge_disk}
\begin{threeparttable}
\begin{tabular}{lll}
\toprule
\textbf{Parameter}\tnote{1} & \textbf{Description} & \textbf{Units/Notes} \\
\midrule
\verb|x_[B/D]_[BAND]|\tnote{2} & X-coordinate of the bulge/disc centroid & pixels\tnote{3} \\
\verb|x_[B/D]_[BAND]_err| & Error on the X-coordinate of the bulge/disc centroid & pixels \\
\verb|y_[B/D]_[BAND]| & Y-coordinate of the bulge/disc centroid & pixels \\
\verb|y_[B/D]_[BAND]_err| & Error on the Y-coordinate of the bulge/disc centroid & pixels \\
\verb|flux_[BAND]| & Total flux in the band & counts \\
\verb|flux_[BAND]_err| & Error on the total flux in the band & counts \\
\verb|flux_[B/D]_[BAND]| & Bulge/disc flux in the band & counts \\
\verb|flux_[B/D]_[BAND]_err| & Error on the bulge/disc flux in the band & counts \\
\verb|mag_[BAND]| & Total magnitude in the band & mag \\
\verb|mag_[BAND]_err| & Error on the total magnitude in the band & mag \\
\verb|mag_[B/D]_[BAND]| & Bulge/disc magnitude in the band & mag \\
\verb|mag_[B/D]_[BAND]_err| & Error on the bulge/disc magnitude in the band & mag \\
\verb|BT_[BAND]| & Bulge-to-total flux ratio in the band& \\
\verb|BT_[BAND]_err| & Error on the bulge-to-total flux ratio in the band & \\
\verb|aspect_[B/D]_[BAND]| & Aspect ratio of the bulge/disc & \\
\verb|aspect_[B/D]_[BAND]_err| & Error on the aspect ratio of the bulge/disc & \\
\verb|angle_[B/D]_deg_[BAND]| & Position angle of the bulge/disc & degrees \\
\verb|angle_[B/D]_deg_[BAND]_err| & Error on the position angle of the bulge/disc & degrees \\
\verb|angle_[B/D]_[BAND]| & Position angle of the bulge/disc & radians \\
\verb|angle_[B/D]_[BAND]_err| & Error on the position angle of the bulge/disc & radians \\
\verb|rad_[B/D]_arc_[BAND]| & Effective radius along the semi-major axis of the bulge/disc & arcsec \\
\verb|rad_[B/D]_arc_[BAND]_err| & Error on the effective radius along the semi-major axis of the bulge/disc & arcsec \\
\verb|rad_[B/D]_[BAND]| & Effective radius of the bulge/disc & pixels \\
\verb|rad_[B/D]_[BAND]_err| & Error on the effective radius of the bulge/disc & pixels \\
\verb|nsersic_B_[BAND]| & S\'ersic index of the bulge & \\
\verb|nsersic_B_[BAND]_err| & Error on the S\'ersic index of the bulge & \\
\bottomrule
\end{tabular}
\begin{tablenotes}
\item[1] For the sake of conciseness, only the columns corresponding to the S\'ersic and exponential profile parameters are listed. All columns regarding source identification and geometry, photometry, association and fitting are the same columns as in \tab\ref{tab:single_sersic}.
\item[2] \verb|B| and \verb|D| refer to bulge and disc, respectively. \verb|BAND| is one of the \verb|IE|, \verb|YE|, \verb|JE|, and \verb|HE| bands.
\item[3] The VIS image is used as a detection image so all sizes given in pixels use the VIS pixel scale of \ang{;;0.1}
\end{tablenotes}
\end{threeparttable}
\end{table}

\begin{table*}[htbp]
\centering
\caption{SED-fitting results from the model photometry.}
\label{tab:sed_fitting}
\begin{threeparttable}
\begin{tabular}{lll}
\toprule
\textbf{Parameter} & \textbf{Description} & \textbf{Units/Notes} \\
\midrule
\verb|ID| & Unique identifier for the source from \cite{EROPerseusOverview} & \\
\verb|zphot| & Photometric redshift of the source from \cite{EROPerseusOverview} & \\
\verb|iSED| & Index of the best-fit SED model & \\
\verb|age| & Age of the stellar population & Gyr \\
\verb|A_V| & V-band extinction & mag \\
\verb|M_u| & Absolute magnitude in the CHFT-MegaCam u-band & mag \\
\verb|M_g| & Absolute magnitude in the CHFT-MegaCam g-band & mag \\
\verb|M_r| & Absolute magnitude in the CHFT-MegaCam r-band & mag \\
\verb|M_i| & Absolute magnitude in the CHFT-MegaCam i-band & mag \\
\verb|M_z| & Absolute magnitude in the CHFT-MegaCam z-band & mag \\
\verb|M_IE| & Absolute magnitude in the Euclid I-band & mag \\
\verb|M_YE| & Absolute magnitude in the Euclid Y-band & mag \\
\verb|M_JE| & Absolute magnitude in the Euclid J-band & mag \\
\verb|M_HE| & Absolute magnitude in the Euclid H-band & mag \\
\bottomrule
\end{tabular}
{\small\textbf{Note.} As four different types of model photometry were computed (single-S\'ersic, bulge-disc, bulge and disc), four sets of SED-fitting parameters were derived, which are listed in four separate tables, all with the same listed parameters as described here.}
\end{threeparttable}
\end{table*}

\newpage

\section{Examples of model fitting images}

\begin{figure*}
    \includegraphics[width=0.163\textwidth]{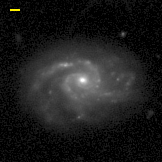}
    \includegraphics[width=0.163\textwidth]{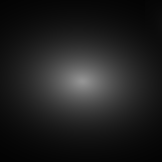}
    \includegraphics[width=0.163\textwidth]{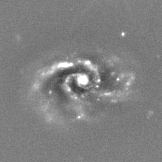}
    \includegraphics[width=0.163\textwidth]{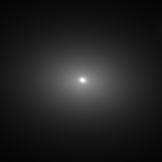}
    \includegraphics[width=0.163\textwidth]{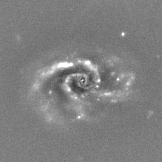}
    \includegraphics[width=0.163\textwidth]{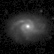}
    \caption{Object 1 (from left to right VIS \IE data, and residual images, bulge-disc model and residual images, NISP \YE data - the yellow dash in the \IE data image indicates 1 arcsec). This is a late spiral galaxy whose single-S\'ersic model underestimates the bulge flux, since a single component fit is not complex enough to properly model both the bulge and disc surface brightness distributions, whereas the bulge and disc decomposition is successful in modelling both components, although the exponential profile of the disc might be a little too steep. The bulge, disc, spiral arms, and flocculence are visible in the NISP \YE band image, but with lower angular resolution.}
    \label{fig:exple-1}
\end{figure*}

\begin{figure*}
    \includegraphics[width=0.163\textwidth]{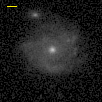}
    \includegraphics[width=0.163\textwidth]{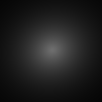}
    \includegraphics[width=0.163\textwidth]{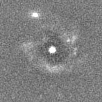}
    \includegraphics[width=0.163\textwidth]{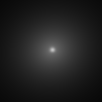}
    \includegraphics[width=0.163\textwidth]{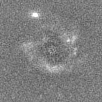}
    \includegraphics[width=0.163\textwidth]{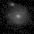}
    \caption{Object 2: same as for Object 1.}
    \label{fig:exple-2}
\end{figure*}

\begin{figure*}
    \includegraphics[width=0.163\textwidth]{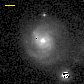}
    \includegraphics[width=0.163\textwidth]{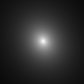}
    \includegraphics[width=0.163\textwidth]{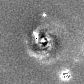}
    \includegraphics[width=0.163\textwidth]{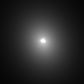}
    \includegraphics[width=0.163\textwidth]{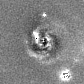}
    \includegraphics[width=0.163\textwidth]{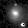}
    \caption{Object 3: same as for Object 1. Only outer spiral arms are visible in the \YE  image.}
    \label{fig:exple-3}
\end{figure*}

\begin{figure*}
    \includegraphics[width=0.163\textwidth]{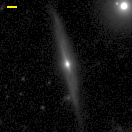}
    \includegraphics[width=0.163\textwidth]{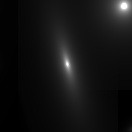}
    \includegraphics[width=0.163\textwidth]{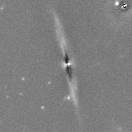}
    \includegraphics[width=0.163\textwidth]{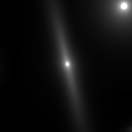}
    \includegraphics[width=0.163\textwidth]{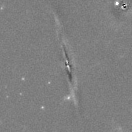}
    \includegraphics[width=0.163\textwidth]{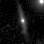}
    \caption{Object 4: an edge-on disc and barred galaxy with a warped disc and dust lanes (the former might be due to the asymmetry in the latter), and a significant bulge ($B/T=0.34$) which is well modelled by the bulge and disc decomposition. The single-S\'ersic model focuses on the bulge, bar, and inner disc, but leaves unmodelled some flux in the outer disc. This galaxy is an example of how intermediate $B/T$ galaxies have a single-S\'ersic effective radius intermediate between those of their bulge and disc (in the top-right-quadrant of \fg\ref{all-radii-related-cmap-BT}, see also \fg\ref{radii-1p-from-BD-values}), with here $R_\mathrm{e,1p}=21.20$, $R_\mathrm{e,bulge} = 3.57$, and  $R_\mathrm{e,disc} = 31.56$. It is also an example of the single-S\'ersic axis ratio being biased by the bulge component, with $(b/a)_{\rm 1p} = 0.27$ overestimating the disc value $(b/a)_{\rm disc} = 0.13$, which should rather be used to infer the galaxy's 3D orientation. The asymmetry in the disc is visible in the \YE image, but the bar and dust lanes are not detectable due to the lower angular resolution.}
    \label{fig:exple-new-1}
\end{figure*}

\begin{figure*}
    \includegraphics[width=0.163\textwidth]{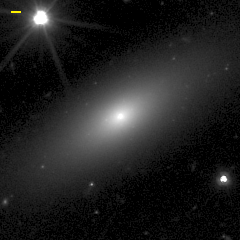}
    \includegraphics[width=0.163\textwidth]{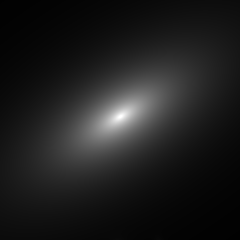}
    \includegraphics[width=0.163\textwidth]{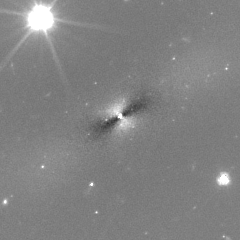}
    \includegraphics[width=0.163\textwidth]{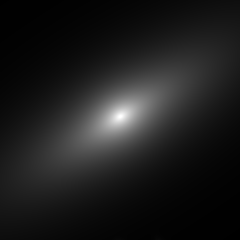}
    \includegraphics[width=0.163\textwidth]{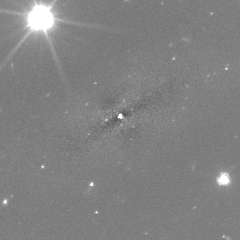}
    \includegraphics[width=0.163\textwidth]{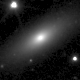}
    \caption{Object 5: an inclined and barred lenticular galaxy with a prominent bulge ($B/T = 0.31$) well fitted by both models, but with more accuracy using the bulge-disc decomposition in and around the bulge. The effective radius and axis ratio of the single-S\'ersic fit present the same limits in describing the galaxy as those described for Object 4 (see values in \tab\ref{tab-values-model-exples}). This featureless lenticular galaxy is not degraded in its visible morphology in the lower resolution \YE image.}
    \label{fig:exple-new-2}
\end{figure*}

\begin{figure*}
    \includegraphics[width=0.163\textwidth]{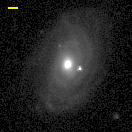}
    \includegraphics[width=0.163\textwidth]{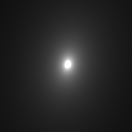}
    \includegraphics[width=0.163\textwidth]{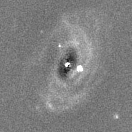}
    \includegraphics[width=0.163\textwidth]{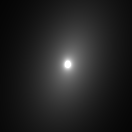}
    \includegraphics[width=0.163\textwidth]{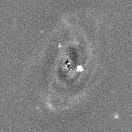}
    \includegraphics[width=0.163\textwidth]{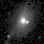}
    \caption{Object 6: same as for Object 1, with the single-S\'ersic model overestimating the flux in the inner part of the disc due to an anomalously high $n=6.90$ (whereas the bulge-disc decomposition has $B/T=0.23$ and $n_\mathrm{B}=2.41$), as well as an overestimated single-S\'ersic effective radius, about twice that for the disc (this galaxy is in the outlier upper-left region of \fg\ref{all-radii-related-cmap-BT}). Only external spiral arms are barely visible in the \YE image.}
    \label{fig:exple-4}
\end{figure*}

\begin{figure*}
    \includegraphics[width=0.163\textwidth]{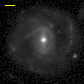}
    \includegraphics[width=0.163\textwidth]{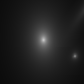}
    \includegraphics[width=0.163\textwidth]{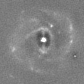}
    \includegraphics[width=0.163\textwidth]{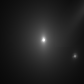}
    \includegraphics[width=0.163\textwidth]{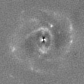}
    \includegraphics[width=0.163\textwidth]{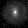}
    \caption{Object 7: same as for Object 1. This galaxy displays a bar and an inner ring at the end of the bar. The bar biases both the single-S\'ersic and bulge models, with respective axis ratios of 0.42 and 0.65, despite a nearly circular bulge. This also leads to an overestimation of $R_\mathrm{e,1p} =1.12 R_\mathrm{e,disc}$ (hence this object lies in the outlier upper-left region of \fg\ref{all-radii-related-cmap-BT}). Because of the lower resolution of the \YE image, the inner ring and bar appear as a lens structure inside the external spiral arms.}
    \label{fig:exple-5}
\end{figure*}

\begin{figure*}
    \includegraphics[width=0.163\textwidth]{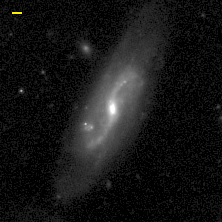}
    \includegraphics[width=0.163\textwidth]{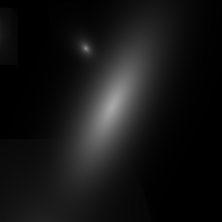}
    \includegraphics[width=0.163\textwidth]{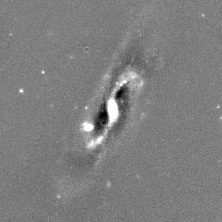}
    \includegraphics[width=0.163\textwidth]{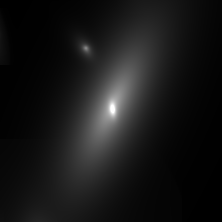}
    \includegraphics[width=0.163\textwidth]{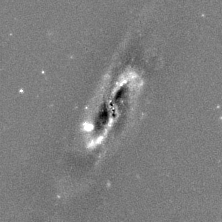}
    \includegraphics[width=0.163\textwidth]{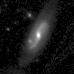}
    \caption{Object 8: an inclined late spiral galaxy with a bar. Both the single-S\'ersic and the bulge-disc decomposition fail in modelling a physical bulge, whereas the modelling of the disc component are similar and acceptable. Moreover, the weak bulge and the strong bar are fitted by an elongated bulge component (hence the small axis ratios of $0.39$), but with different position angles of the bulge and disc major axes (with an offset of $31.3^\circ$). The bulge, disc, spiral arms, and part of the flocculence are visible in the \YE lower resolution image.}
    \label{fig:exple-6}
\end{figure*}

\begin{figure*}
    \includegraphics[width=0.163\textwidth]{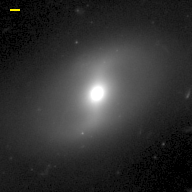}
    \includegraphics[width=0.163\textwidth]{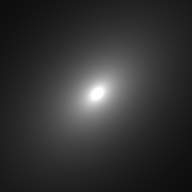}
    \includegraphics[width=0.163\textwidth]{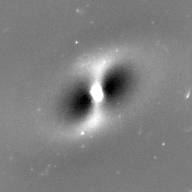}
    \includegraphics[width=0.163\textwidth]{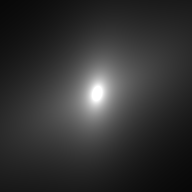}
    \includegraphics[width=0.163\textwidth]{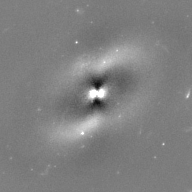}
    \includegraphics[width=0.163\textwidth]{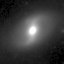}
    \caption{Object 9: an inclined and barred lenticular galaxy, with the same comment as for Object 8 regarding the bulge, disc and single S\'ersic fits, and impact of the bar, but with a more prominent bulge ($B/T=0.48$), whose value is certainly overestimated due to the bulge model including some of the bar flux. The effective radius and axis ratio of the single-S\'ersic fit present similar characteristics and limitations in describing the galaxy as those mentioned for Object 4 (see values in \tab\ref{tab-values-model-exples}). Due to its coarse morphological features, this lenticular galaxy is not degraded in the \YE image.}
    \label{fig:exple-new-3}
\end{figure*}

\begin{figure*}
    \includegraphics[width=0.163\textwidth]{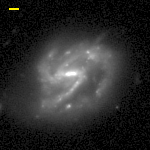}
    \includegraphics[width=0.163\textwidth]{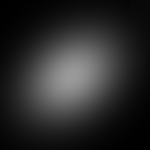}
    \includegraphics[width=0.163\textwidth]{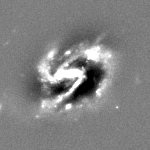}
    \includegraphics[width=0.163\textwidth]{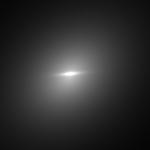}
    \includegraphics[width=0.163\textwidth]{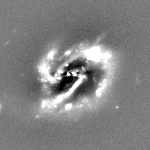}
    \includegraphics[width=0.163\textwidth]{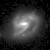}
    \caption{Object 10: same as for Object 8 (with a weak bulge and bar). The bulge, disc, spiral arms and flocculence are visible in the \YE  image, but with a lower angular resolution.}
    \label{fig:exple-7}
\end{figure*}

In this appendix we provide data, model, and residual images, for both the single-S\'ersic fits and the bulge-disc decompositions, so as to show typical and interesting examples of the fits performed to the galaxy projected  brightness distributions. Figures \ref{fig:exple-1} to \ref{fig:exple-15} show the VIS data, model, and residual images for the single-S\'ersic fit and the bulge-disc decomposition, along with the NISP \YE image, in order to compare both models for a set of galaxies illustrating the various situations discussed in \sct\ref{sct-quality-check}. Comparison of the NISP \YE and VIS \IE data images illustrates the three times lower angular resolution of the NISP instrument, hence the loss of information in the model fitting. Only galaxies with $\IE\lesssim18.5$ appear to have their \IE morphologies recognisable in the \JE image.

Figures \ref{fig:exple-1} to \ref{fig:exple-15} are organised in two series, one in which the bulge-disc decompositions are successful  (\sct\ref{sct-appendix-successful}), and the other containing those that are problematic (\sct\ref{sct-appendix-problematic}) and serve as illustration of the outliers identified in the various comparisons of the structural parameters between the single-S\'ersic model and the bulge-disc decomposition (\sct\ref{sct-quality-check}). The $x$ and $y$ axes of the CCD image of the ERO-Perseus field, hence of the vignettes shown in \fg\ref{fig:exple-1} to \ref{fig:exple-15}, are rotated by \ang{38} from the North-South and East-West directions, respectively, of the equatorial system. The parameters of all displayed galaxies are listed in \tab\ref{tab-values-model-exples}.

\subsection{Successful bulge-disc fits\label{sct-appendix-successful}}

The images displayed in \fgs\ref{fig:exple-1} to \ref{fig:exple-7} represent the most common cases of both the single-S\'ersic and bulge-disc models fitting galaxies well and providing structural parameters that are consistent with each other, as discussed in \sct\ref{sct-quality-check}. These images also highlight the higher degree of complexity reached by the larger number of parameters in the bulge-disc decomposition compared to single-S\'ersic fits, leading to better simultaneous modelling of the bulge and the disc components. In several objects, the disc might be better modelled with a S\'ersic index smaller than unity, but using a variable index for the disc in addition to that for the bulge would cause more degeneracies. The last four objects illustrate how bars are modelled by the bulge components, biasing both the axis ratio and $B/T$ of the bulge towards high values (see \tab\ref{tab-values-model-exples}).

\begin{table*}[ht]
\caption{Parameters of the single-S\'ersic profile fits and bulge-disc decompositions in the VIS \IE band, complementing the coordinates, magnitudes and bulge-to-total light ratio of the example galaxies shown in \fg\ref{fig:exple-1} to \ref{fig:exple-15}.}
\resizebox{\linewidth}{!}{%
\begin{tabular}{ c c c l l l l l l l l r r r l l l }
\hline\hline
Obj. & RA & Dec & $m_\mathrm{auto}$ & $m_\mathrm{1p}$ & $m_\mathrm{B+D}$ & $m_\mathrm{B}$ & $m_\mathrm{D}$ & $B/T$ & $n$ & $n_\mathrm{B}$ & $R_\mathrm{e,1p}$ & $R_\mathrm{e,B}$ & $R_\mathrm{e,D}$ & $(b/a)_\mathrm{1p}$ & $(b/a)_\mathrm{B}$ & $(b/a)_\mathrm{D}$ \\
& J2000 deg & J2000 deg & mag & mag & mag & mag & mag &  & &  & pixels & pixels & pixels &  &  & \\
\hline
1 & 49.33030 & 41.53805 & $18.25$ & $18.29$ & $18.21$ & $21.80{\pm0.02}$ & $18.25$ & $0.04$ & $0.84$ & $0.82{\pm0.04}$ & $29.5{\pm0.1}$ & $2.2{\pm0.0}$ & $31.1{\pm0.1}$ & $0.69$ & $0.73$ & $0.70$ \\
2 & 50.23211 & 41.71729 & $19.62$ & $19.57$ & $19.48$ & $22.98{\pm0.03}$ & $19.52$ & $0.04$ & $0.88{\pm0.02}$ & $0.50$ & $20.3{\pm0.2}$ & $1.6{\pm0.0}$ & $22.7{\pm0.3}$ & $0.88$ & $0.89{\pm0.04}$ & $0.89$ \\
3 & 49.06566 & 41.63792 & $18.62$ & $18.98$ & $19.03$ & $21.64{\pm0.04}$ & $19.13$ & $0.09$ & $1.49{\pm0.02}$ & $9.98{\pm0.27}$ & $14.4{\pm0.1}$ & $1.2{\pm0.1}$ & $14.3{\pm0.1}$ & $0.85$ & $0.60{\pm0.03}$ & $0.82$ \\
4 & 49.38678 & 41.61667 & $20.31$ & $20.15$ & $20.24$ & $21.41{\pm0.04}$ & $20.70{\pm0.02}$ & $0.34$ & $4.13{\pm0.10}$ & $2.65{\pm0.12}$ & $21.2{\pm0.7}$ & $3.6{\pm0.2}$ & $31.6{\pm0.6}$ & $0.27$ & $0.67{\pm0.02}$ & $0.13$ \\
5 & 49.78521 & 41.58108 & $16.80$ & $16.78$ & $16.77$ & $18.04{\pm0.02}$ & $17.18$ & $0.31$ & $1.31$ & $1.08$ & $41.3{\pm0.1}$ & $19.0{\pm0.2}$ & $60.6{\pm0.4}$ & $0.47$ & $0.75$ & $0.34$ \\
6 & 50.13557 & 41.36430 & $18.49$ & $18.23$ & $18.43$ & $20.02$ & $18.72$ & $0.23$ & $6.90{\pm0.06}$ & $2.41{\pm0.05}$ & $59.4{\pm1.4}$ & $3.1{\pm0.1}$ & $31.2{\pm0.2}$ & $0.69$ & $0.76$ & $0.61$ \\
7 & 49.99226 & 41.42617 & $19.93$ & $19.86$ & $19.88$ & $21.25{\pm0.05}$ & $20.24{\pm0.02}$ & $0.28$ & $2.68{\pm0.04}$ & $7.52{\pm0.55}$ & $22.7{\pm0.3}$ & $8.3{\pm1.4}$ & $20.2{\pm0.2}$ & $0.65$ & $0.42$ & $0.87$ \\
8 & 49.22191 & 41.49665 & $17.93$ & $17.99$ & $17.87$ & $20.68$ & $17.95$ & $0.07$ & $0.78$ & $0.94{\pm0.02}$ & $39.1{\pm0.1}$ & $5.2{\pm0.1}$ & $43.3{\pm0.2}$ & $0.36$ & $0.39$ & $0.35$ \\
9 & 49.69704 & 41.98058 & $15.90$ & $15.87$ & $15.87$ & $16.66$ & $16.58$ & $0.48$ & $2.00$ & $1.77$ & $41.1{\pm0.1}$ & $23.7{\pm0.1}$ & $77.8{\pm0.3}$ & $0.67$ & $0.60$ & $0.44$ \\
10 & 49.37662 & 41.53632 & $17.58$ & $17.68$ & $17.61$ & $21.09{\pm0.02}$ & $17.66$ & $0.04$ & $0.51$ & $1.21{\pm0.06}$ & $29.1{\pm0.1}$ & $7.3{\pm0.1}$ & $29.8{\pm0.1}$ & $0.75$ & $0.12$ & $0.75$ \\
\hline
11 & 50.00209 & 41.68845 & $20.01$ & $20.07$ & $20.04$ & $20.59{\pm0.02}$ & $21.04{\pm0.03}$ & $0.60$ & $0.56$ & $0.78{\pm0.02}$ & $10.7{\pm0.0}$ & $8.4{\pm0.1}$ & $9.7{\pm0.1}$ & $0.25$ & $0.29$ & $0.31$ \\
12 & 49.37197 & 41.91597 & $20.50$ & $20.30$ & $20.34$ & $20.38$ & $23.85{\pm0.25}$ & $0.96$ & $8.56{\pm0.34}$ & $8.30{\pm0.20}$ & $8.0{\pm0.5}$ & $7.0{\pm0.1}$ & $2.3{\pm0.3}$ & $0.34$ & $0.31$ & $0.90{\pm0.17}$ \\
13 & 50.04730 & 41.64479 & $19.40$ & $19.47$ & $19.43$ & $20.19$ & $20.18$ & $0.50$ & $0.65$ & $0.76$ & $11.7{\pm0.0}$ & $9.6{\pm0.0}$ & $10.8{\pm0.1}$ & $0.66$ & $0.64$ & $0.64$ \\
14 & 49.78091 & 42.09539 & $20.93$ & $20.95$ & $20.92$ & $21.64$ & $21.71$ & $0.51$ & $0.70$ & $1.15{\pm0.02}$ & $8.4{\pm0.0}$ & $7.2{\pm0.0}$ & $6.9{\pm0.1}$ & $0.51$ & $0.56$ & $0.54$ \\
15 & 50.19885 & 41.66254 & $20.44$ & $20.53$ & $20.47$ & $21.38$ & $21.08$ & $0.43$ & $0.62$ & $0.50$ & $9.7{\pm0.1}$ & $5.2{\pm0.0}$ & $10.7{\pm0.1}$ & $0.32$ & $0.57$ & $0.30$ \\
16 & 49.43170 & 41.28917 & $20.46$ & $20.49$ & $20.48$ & $21.64{\pm0.02}$ & $20.93$ & $0.34$ & $0.74$ & $0.69{\pm0.02}$ & $14.1{\pm0.1}$ & $11.2{\pm0.0}$ & $10.7{\pm0.1}$ & $0.54$ & $0.52$ & $0.84$\\
17 & 49.49079 & 41.81583 & $20.78$ & $20.77$ & $20.68$ & $22.03{\pm0.06}$ & $21.06{\pm0.03}$ & $0.29$ & $1.44{\pm0.03}$ & $0.97{\pm0.04}$ & $16.5{\pm0.2}$ & $9.7{\pm0.3}$ & $20.6{\pm0.5}$ & $0.54$ & $0.33$ & $0.77{\pm0.02}$ \\
18a & 49.22173 & 41.44524 & $20.35$ & $20.40$ & $20.30$ & $21.20{\pm0.02}$ & $20.93{\pm0.02}$ & $0.44$ & $1.16{\pm0.02}$ & $0.82{\pm0.02}$ & $9.7{\pm0.1}$ & $5.7{\pm0.1}$ & $17.2{\pm0.3}$ & $0.89$ & $0.92$ & $0.68{\pm0.02}$ \\
18b & 49.22135 & 41.44690 & $20.12$ & $20.32$ & $20.43$ & $20.73{\pm0.02}$ & $21.97{\pm0.06}$ & $0.76$ & $2.38{\pm0.04}$ & $1.97{\pm0.04}$ & $9.8{\pm0.1}$ & $6.5{\pm0.1}$ & $12.6{\pm0.4}$ & $0.72$ & $0.74$ & $0.51{\pm0.02}$ \\
\hline
\vspace{1pt}
\end{tabular}}
{\small\textbf{Note.} All values have a calculated standard error, only those $\ge0.015$ are displayed.}
\label{tab-values-model-exples}
\end{table*}

\subsection{Problematic bulge-disc fits\label{sct-appendix-problematic}}

\begin{figure*}
    \includegraphics[width=0.163\textwidth]{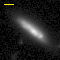}
    \includegraphics[width=0.163\textwidth]{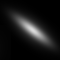}
    \includegraphics[width=0.163\textwidth]{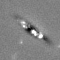}
    \includegraphics[width=0.163\textwidth]{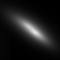}
    \includegraphics[width=0.163\textwidth]{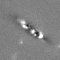}
    \includegraphics[width=0.163\textwidth]{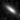}
    \caption{Object 11: this highly inclined strongly flocculent and poorly resolved disc galaxy, with no visible bulge, has its bulge and disc model components split in flux at $B/T=0.60$, with similar values of structural parameters for both components as well as for the single-S\'ersic profile fit. The disc and bulge models fit the top-left and bottom-right parts of the galaxy, respectively, with a 9.8 pixels ($1.0\,R_\mathrm{e,disc}$) separation in their centring. The flocculence cannot be identified from the \YE lower resolution image, and the light excess near the centre of the object could be erroneously identified as a bulge.}
    \label{fig:exple-8}
\end{figure*}

\begin{figure*}
    \includegraphics[width=0.163\textwidth]{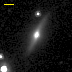}
    \includegraphics[width=0.163\textwidth]{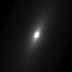}
    \includegraphics[width=0.163\textwidth]{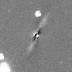}
    \includegraphics[width=0.163\textwidth]{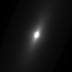}
    \includegraphics[width=0.163\textwidth]{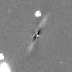}
    \includegraphics[width=0.163\textwidth]{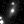}
    \caption{Object 12: because of an edge-on and probably warped disc, most of the flux of this galaxy is included in the bulge component ($B/T=0.96$), with similar parameters as for the single-S\'ersic fit, whereas the 4\% disc model component is nearly round with a 3 to 4 times smaller effective radius, hence an inverted bulge and disc modelling. Still, the bulge effective radius is too small to model the outer extensions of the disc. The peculiar morphology of this object and the disc warp cannot be identified from the \YE lower resolution image.}
    \label{fig:exple-9}
\end{figure*}

\begin{figure*}
    \includegraphics[width=0.163\textwidth]{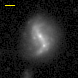}
    \includegraphics[width=0.163\textwidth]{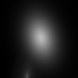}
    \includegraphics[width=0.163\textwidth]{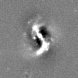}
    \includegraphics[width=0.163\textwidth]{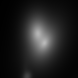}
    \includegraphics[width=0.163\textwidth]{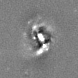}
    \includegraphics[width=0.163\textwidth]{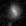}
    \caption{Object 13: this object might be a merger of two galaxies with the appearance of what could be two bulges, and a central bar. The single-S\'ersic fit underestimates the bulge flux, and the bulge-disc decomposition is split with the bulge and disc models at the top and bottom galaxy, respectively, leading to a $1.00\,R_\mathrm{e,1p}= 1.08\,R_\mathrm{e,disc}$ offset in their centring and a non-physical $B/T=0.50$. The bar and asymmetry remain visible in the \YE image, but the flocculence and other structures are not resolved.}
    \label{fig:exple-10}
\end{figure*}

\begin{figure*}
    \includegraphics[width=0.163\textwidth]{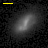}
    \includegraphics[width=0.163\textwidth]{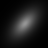}
    \includegraphics[width=0.163\textwidth]{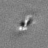}
    \includegraphics[width=0.163\textwidth]{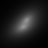}
    \includegraphics[width=0.163\textwidth]{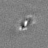}
    \includegraphics[width=0.163\textwidth]{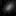}
    \caption{Object 14: because no central bulge is visible, this flocculent and poorly resolved galaxy with a possible strong bar gets its bulge and disc model components split in flux ($B/T=0.51$) and in their centring by $7.3$ pixels ($1.1\,R_\mathrm{e,disc}$) onto opposite sides of the disc (with the bulge model to the upper-right and the disc model to the lower-left of the object), and without modelling the bar. The bar and flocculence cannot be identified from the \YE lower resolution image.}
    \label{fig:exple-11}
\end{figure*}

\begin{figure*}
    \includegraphics[width=0.163\textwidth]{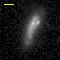}
    \includegraphics[width=0.163\textwidth]{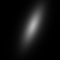}
    \includegraphics[width=0.163\textwidth]{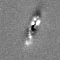}
    \includegraphics[width=0.163\textwidth]{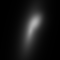}
    \includegraphics[width=0.163\textwidth]{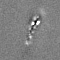}
    \includegraphics[width=0.163\textwidth]{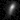}
    \caption{Object 15: because no central bulge is visible, this highly inclined, flocculent, and warped disc galaxy gets its bulge and disc model components split in centring by 9.7 pixels ($1.0\,R_\mathrm{e,1p}$ or $0.9\,R_\mathrm{e,disc}$) and flux ($B/T=0.43$). The elongated disc model component is to the lower-left, whereas the rounder bulge model component is to upper-right, with its effective radius about half the nearly equal values for the disc component and the single-S\'ersic fit. The flocculence cannot be identified from the lower resolution \YE image, and the light excess near the centre of the object could be erroneously identified as a significant bulge.}
    \label{fig:exple-12}
\end{figure*}

\begin{figure*}
    \includegraphics[width=0.163\textwidth]{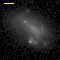}
    \includegraphics[width=0.163\textwidth]{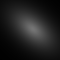}
    \includegraphics[width=0.163\textwidth]{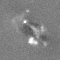}
    \includegraphics[width=0.163\textwidth]{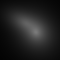}
    \includegraphics[width=0.163\textwidth]{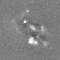}
    \includegraphics[width=0.163\textwidth]{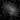}
    \caption{Object 16: an irregular (bulge-less) barred galaxy, whose single-S\'ersic model focuses on the brightest part of the galaxy and has a low S\'ersic index of $n=0.74$. The bulge and disc decomposition leads to bulge and disc models separated by 13.5 pixels ($0.95\,R_\mathrm{e,1p}$ or $1.3\,R_\mathrm{e,disc}$), with the bulge at the top-left and disc at the bottom-right and characterised by non-physical parameters (for instance $ R_\mathrm{e,bulge} > R_\mathrm{e,disc}$). The flocculence and the bar can be roughly identified from the \YE lower resolution image.}
    \label{fig:exple-13}
\end{figure*}

\begin{figure*}
    \includegraphics[width=0.163\textwidth]{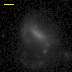}
    \includegraphics[width=0.163\textwidth]{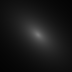}
    \includegraphics[width=0.163\textwidth]{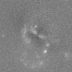}
    \includegraphics[width=0.163\textwidth]{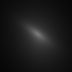}
    \includegraphics[width=0.163\textwidth]{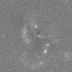}
    \includegraphics[width=0.163\textwidth]{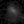}
    \caption{Object 17: an irregular (bulgeless) barred galaxy, for which both the single-S\'ersic and bulge model are biased by the bar, with respective axis ratios of 0.54 and 0.33, against 0.77 for the disc. There is also a \ang{25.5} difference between the position angles of the disc and single-S\'ersic profiles. The flocculence cannot be identified from the lower resolution \YE image, and the well defined bar in the \IE image could be erroneously modelled as a bulge.}
    \label{fig:exple-14}
\end{figure*}

\begin{figure*}
    \includegraphics[width=0.163\textwidth]{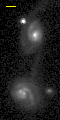}
    \includegraphics[width=0.163\textwidth]{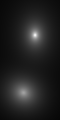}
    \includegraphics[width=0.163\textwidth]{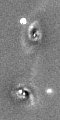}
    \includegraphics[width=0.163\textwidth]{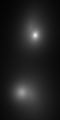}
    \includegraphics[width=0.163\textwidth]{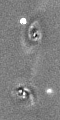}
    \includegraphics[width=0.163\textwidth]{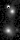}
    \caption{Objects 18a and b: a pair of spiral galaxies probably undergoing an interaction, since  a `plume' of matter appears to connect them and their discs are asymmetric, hence their being displayed together, with Object 18a at the bottom and Object 18b at the top. Although the single-S\'ersic fit residuals appear similar to those from the bulge-disc decompositions, the latter models for both galaxies have a disc and bulge offset in their centring due to the disc asymmetries: in Object 18a, the disc model is positioned to the right of the bulge model, while in Object 18b, it is to the upper left. For both objects, the effective radius of the single-S\'ersic fit is nearly the average of those of the bulge and the disc, which could give the false impression of two successful bulge-disc decompositions -- it is the split between the disc and bulge centres that indicates that these bulges and discs are not physical. Moreover, in Object 18a, the central part of the galaxy hosts two bright spots, one of which may be a bulge and the other a bright \ion{H}{ii} region (perhaps triggered by the interaction). None of these spots in Object 18a are modelled by the bulge component, which has too low a S\'ersic index ($n_\mathrm{B}=0.82$). In contrast, the bulge of Object 18b is well defined but is as much underestimated in flux by the bulge model as by the single-S\'ersic fit. However, the disc model being offset towards the spiral arm (to the upper-left) leads the bulge model to also include the inner disc around the bulge, yielding an overestimated $B/T=0.76$ and a likely underestimated $n_\mathrm{B}=1.97$. Lastly, for both galaxies, the single-S\'ersic fit indices are similar to those of the bulge component, leaving significant residuals, comparable to those for the bulge-disc decomposition, and also caused by the disc asymmetry. The bulge-disc decomposition would be largely unconstrained with the \YE lower angular resolution, in which the well-designed spiral arms seen in the \IE band are undetectable.}
    \label{fig:exple-15}
\end{figure*}

The images displayed in \fgs\ref{fig:exple-8} to \ref{fig:exple-15} were identified by examining the outliers in the various figures of \sct\ref{sct-quality-check}, and represent cases in which the bulge-disc decomposition reaches the limits of its capabilities, or a peculiarity of the image prevents any attempt at modelling the galaxy with the chosen profiles, resulting in non-physical bulges and discs. In these objects, the effective radius of the single-S\'ersic fit is the best available approximation of the object size.

\newpage

\section{Impact of a common profile modelled over multiple bands \label{appendix-common-profile}}

The parametric fits implemented within the processing function of the Organizational Unit OU-MER\footnote{OU-MER aims at producing object catalogues from the merging by cross-identification of all the multi-wavelength data, both from \Euclid and ground-based complementary observations.} will provide structural parameters for all \Euclid galaxies, therefore creating the largest catalogue of S\'ersic parameters ever built \citep{Q1-TP004, Q1-SP040}. In the current state of the tools developed for OU-MER, these fits are also performed with \texttt{SourceXtractor++}, but with two models, one for the VIS image and another for all three NISP images, contrary to the fully independent fits in all four bands mainly used here (see \sct\ref{sct-model-fitting}). For comparison, we therefore performed a run in a similar configuration (see approach 2 in \sct\ref{sct-model-fitting}) to see how the single set of NISP parameters would compare to those fitted independently in the individual \YE, \JE, and \HE NISP bands. 

\begin{figure}
\centering
\includegraphics[width=0.8\columnwidth]{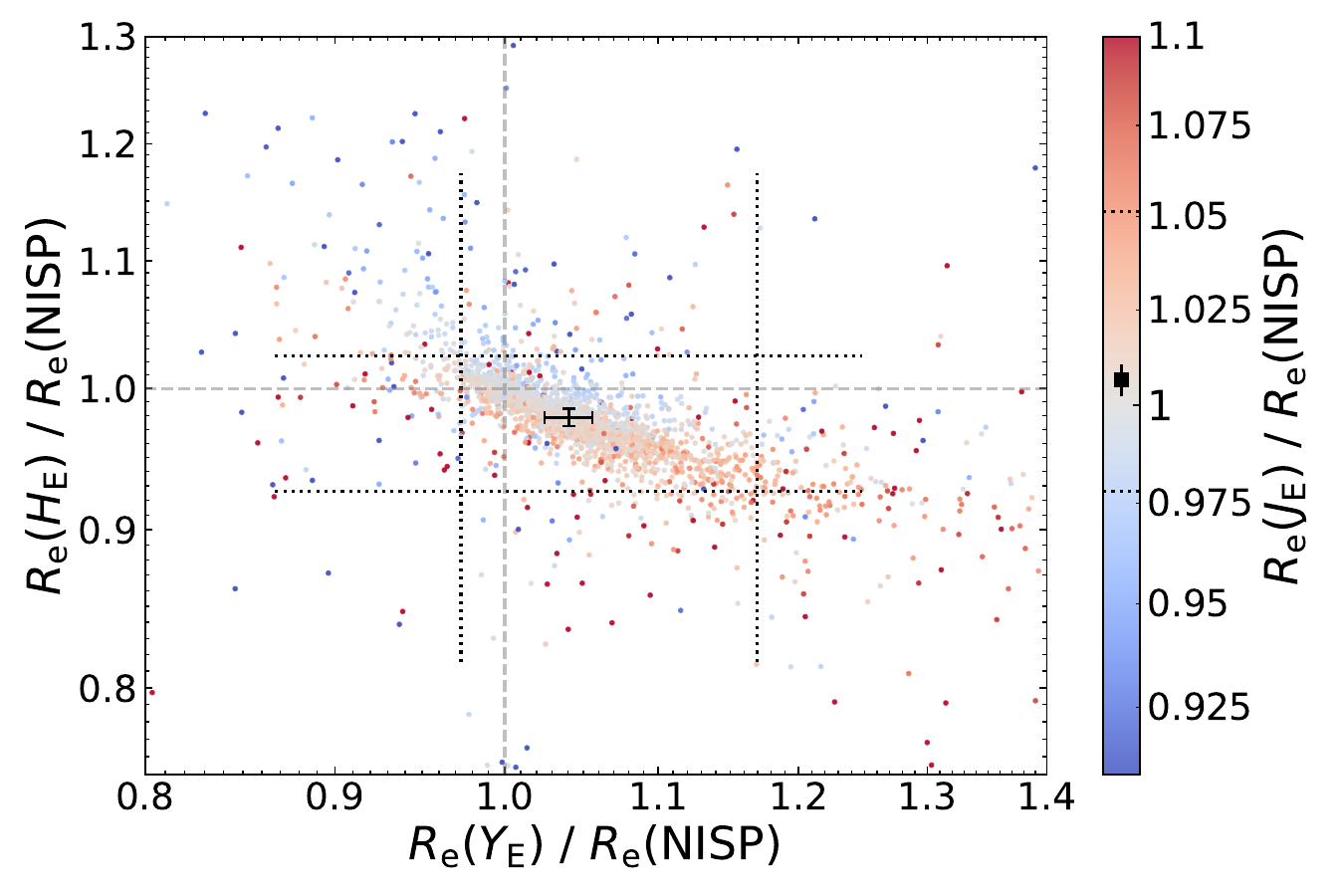}
\caption{Ratios of the single-S\'ersic profile effective radii $R_\mathrm{e}(\YE)$, $R_\mathrm{e}(\JE)$ and $R_\mathrm{e}(\HE)$ derived for the individual NISP bands, and $R_\mathrm{e}(\mathrm{NISP})$ derived by fitting a unique set of structural parameters on the three NISP images for the $2445$ galaxies with $\IE\le21$. The ratios for the \YE, \JE, and \HE bands are shown as the $x$-axis, colour map and $y$-axis, respectively. The cross in the plot and the square in the colour bar indicate the median values of the ratio distributions as well as bootstrap errors multiplied by 10 for visibility, whereas the dotted black lines indicate the 0.1 and 0.9 quantiles. The dashed grey lines correspond to the identity ratios on both axes. Because the wavelength range of \JE is intermediate between those of \YE and \HE, the common $R_\mathrm{e}(\mathrm{NISP})$ is consistent with $R_\mathrm{e}(\JE)$ and exhibits similar colour effects with respect to the \JE and \HE bands.}
\label{color-NISP}
\end{figure}

In \fg\ref{color-NISP} we plot for each galaxy with $\IE\le21$ the effective radii for the independent single-S\'ersic fits to each of the three NISP bands analysed so far, denoted $R_\mathrm{e}(\YE)$, $R_\mathrm{e}(\JE)$, and $R_\mathrm{e}(\HE)$, divided by the single-S\'ersic effective radii $R_\mathrm{e}(\mathrm{NISP})$ fitted commonly to the three NISP images. For these various distributions of $R_\mathrm{e}$ ratios, the medians and bootstrap errors (multiplied by 10 for visibility) are shown as the black cross and square, respectively, whereas the 0.1 and 0.9 quantiles are represented as black dotted lines, both inside the graph and along the colour bar. 

Figure \ref{color-NISP} indicates that on average the $R_\mathrm{e}(\mathrm{NISP})$ effective radius is smaller, larger, and similar to the effective radii in the \YE, \HE, and \JE band, respectively, with median values and associated bootstrap errors of $1.041 \pm 0.0016$, $0.978\pm 0.0006$, and $1.007 \pm 0.0004$, respectively, and dispersions of 0.06, 0.03, and 0.02, respectively (estimated as half the 16--84th percentile range). The ratio between $R_\mathrm{e}(\JE)$ and $R_\mathrm{e}(\mathrm{NISP})$ shown along the colour bar also exhibits a distribution more concentrated around 1 (the scale of the colour map is stretched compared to the axis labels), with $80\%$ of the galaxies having a ratio in the interval $[0.98, 1.05]$ compared with  $[0.97, 1.17]$ and $[0.93, 1.02]$ for the ratios with \YE and \HE, respectively (with the boundaries of the intervals corresponding to the 0.1 and 0.9 quantiles). Overall, this indicates that the single-S\'ersic $R_\mathrm{e}(\mathrm{NISP})$ fitted on all three NISP bands does not deviate significantly from $R_\mathrm{e}(\JE)$ and is intermediate between the smaller $R_\mathrm{e}(\HE)$ and larger $R_\mathrm{e}(\YE)$ values. This results from the fact that the effective wavelength of the \JE band is close to the average over those in the \YE and \HE bands. The anti-correlation between the $R_\mathrm{e}(\YE)/R_\mathrm{e}(\mathrm{NISP})$ and $R_\mathrm{e}(\HE)/R_\mathrm{e}(\mathrm{NISP})$ ratios seen in \fg\ref{color-NISP} results directly from the correlation between $R_\mathrm{e}(\JE)/R_\mathrm{e}(\YE)$ and $R_\mathrm{e}(\HE)/R_\mathrm{e}(\JE)$, already discussed in \sct\ref{sct-bulge-disk-gradient}.

Finally, a similar approach on the differences between the single-S\'ersic indices was performed and led to median values and standard errors of $-0.019 \pm 0.018$, $0.005 \pm 0.017$, and $0.033 \pm 0.015$, for $n(\YE) - n(\mathrm{NISP})$, $n(\JE) - n(\mathrm{NISP})$, and $n(\HE) - n(\mathrm{NISP})$, respectively, with again the \JE band providing similar S\'ersic index values as for the joint fit to all three NISP filters. This confirms the fact that the unique set of structural parameters fitted on all three NISP images is well approximated by the \JE structural parameters, but slightly differ from the values in the \YE and \HE bands, because of the colour gradients detected in \sct\ref{sct-bulge-disk-gradient}.

\end{appendix}

\end{document}